\documentclass[11pt]{article}
\usepackage{eurosym}
\usepackage{amsfonts}
\usepackage{amssymb}
\usepackage{graphicx}
\usepackage{amsmath}
\usepackage{makeidx}
\usepackage{indentfirst}
\usepackage[T1]{fontenc}
\usepackage[utf8]{inputenc}

\newcounter{resultnum}[section]
\newcounter{conclusionnum}[section]
\newcounter{conditionnum}[section]
\newcounter{conjecturenum}[section]
\newcounter{examplenum}[section]
\newcounter{exercisenum}[section]
\newcounter{lemmanum}[section]
\newcounter{notationnum}[section]
\newcounter{theoremnum}[section]
\newcounter{definitionnum}[section]
\newcounter{corollarynum}[section]
\newcounter{remarknum}[section]
\newcounter{propositionnum}[section]
\newcounter{acknowledgementnum}[section]
\newcounter{algorithmnum}[section]
\newtheorem{axiom}{Axiom}[section]

\newcounter{axiomnum}[section]
\newcounter{casenum}[section]
\newcounter{claimnum}[section]
\newcounter{summarynum}[section]
\newcounter{problemnum}[section]
\begin{document}

\title{Constantin Carath\'{e}odory axiomatic approach and Grigory Perelman
thermodynamics for geometric flows and cosmological solitonic solutions}
\date{April 18, 2021}

\author{
\vspace{.1 in} {\ \textbf{Iuliana Bubuianu}}\thanks{%
email: iulia.bubu@gmail.com } \\
{\small \textit{Radio Ia\c{s}i, \ 44 Lasc\v{a}r Catargi, Ia\c{s}i, \ 700107, Romania}}  
\ and \vspace{.1 in} \\
 \textbf{Sergiu I. Vacaru}
\thanks{
emails: sergiu.vacaru@gmail.com and sergiuvacaru@mail.fresnostate.edu ;\newline
\textit{Address for post correspondence as a visitor senior researcher at YF CNU Ukraine is:\ } 37 Yu. Gagarina street, ap. 3, Chernivtsi, Ukraine, 58008}  \\
{\small \textit{Physics Department, California State University at Fresno,
Fresno, CA 93740, USA; and }}\\
 {\small \textit{Dep. Theoretical Physics and Computer Modelling,
 Yu. Fedkovych Chernivtsi National University,}}\\
{\small  101 Storozhynetska street, Chernivtsi, 58029, Ukraine}
}

\maketitle

\begin{abstract}
We elaborate on statistical thermodynamics models of relativistic geometric flows as generalizations of G. Perelman and R. Hamilton theory centred around C. Carath\'{e}odory axiomatic approach to thermodynamics with Pfaffian differential equations. The anholonomic frame deformation method, AFDM, for constructing generic off--diagonal and locally anisotropic cosmological solitonic solutions in the theory of relativistic geometric flows and general relativity is developed. We conclude that such solutions can not be described in terms of the Hawking--Bekenstein thermodynamics for hypersurface, holographic, (anti) de Sitter and similar configurations. The geometric thermodynamic values are defined and computed for nonholonomic Ricci flows, (modified) Einstein equations, and new classes of locally anisotropic cosmological solutions encoding solitonic hierarchies.

\vskip3pt


\vskip3pt


PACS 2010:\ 02.40.Vh, 02.90.+p, 04.20.Cv, 04.20.Jb, 04.90.+e, 05.90.+m
\end{abstract}

\tableofcontents




\section{Introduction}

Thermodynamics is a fundamental physical theory with various branches and
applications in modern physics, engineering, biology, chemistry, information
theory and mathematics, see reviews \cite%
{sommerfeld55,zemansky68,pippard97,callen60,landsberg61} and references
therein. The equilibrium thermodynamics originated from the study of heat
engines when the combination of mechanical and thermal concepts was done in
an empirical way with further essential developments and contributions to
statistical physics and ergodic theory, modern gravity, cosmology etc.

Thermodynamic ideas and methods were developed and applied in the black hole,
BH, physics \cite{bek72,bek73,haw73,haw75}, using the Bekenstein--Hawking
entropy, and (using different concepts and constructions) in the proof of
the Thurston--Poincar\'{e} conjecture due to Grigory Perelman \cite%
{perelman1}, see original fundamental physical and mathematical works and
reviews of geometric analysis and topological results in \cite%
{friedan80,hamilton82,cao06,morgan07,kleiner06,vacaru08}. In a series of our
and co-authors works, we studied possible implications of the approach
elaborated by G. Perelman for geometric flow statistical thermodynamics in
certain directions of modified and Einstein gravity and cosmology and
astrophysics \cite{vacaru13,vacaru16,vacaru17,bub19} and classical and
quantum geometric information flow theory \cite{vacaru19,vacaru19a,vacaru19b}%
. It was exploited the idea that the concept of Perelman W-entropy and
associated statistical models presents more general mathematical and
physical possibilities comparing to those elaborated for theories and
solutions with Bekenstein--Hawking and another area--holographic type
entropies.

Constantin Carath\'{e}ory formulated the first systematic and axiomatic
formulation of equilibrium thermodynamics \cite{carath09,carath25}. In such
an approach (see \cite{landsberg61,redlich68,gian74,gurtin86} on further
extensions), the geometry of thermodynamics is symplectic and analogous to
the structure of Hamilton mechanics and can be expressed through Pfaff forms
and related systems of first-order partial differential equations. The
axiomatic treatment of thermodynamics caught the attention of a number of
famous and well-known scientists \cite%
{born21,lande26,chand58,buchdahl49,buchdahl66,pauli58,landsberg61,landsberg91}
who recognized and in one case criticized \cite{planck26} Carath\'{e}odory's
papers; for brief reviews, we cite \cite{pog00,ant02,bel02,bel02a}.\footnote{%
M. Planck and some other authors criticism "targeting quick results" was
about the difficulty to provide a simple physical picture of the Carath\'{e}%
odory method and the concept of entropy together with sophisticate geometric
methods unknown at that time to the bulk of physicists and mathematicians.
At present, the functional analysis, measure theory and topology techniques
are familiar to researchers publishing works in mathematical physics and
geometry and physics.}

In the present work, we will not get into the details of C. Carath\'{e}odory
and G. Perelman achievements; see above-cited works and \cite%
{rass91,carath99,carrod00,span00,pog00,ant02}, on life and contributions in
mathematics, physics, and education. We shall study only how the mathematic
tools of axiomatic thermodynamics can be applied to relativistic
generalizations of geometric flow theory and compute geometric thermodynamic
values for locally anisotropic cosmological solutions. There will be
considered also certain applications in modern cosmology. There are three
main purposes of this article: 1) To show how the C. Carath\'{e}ory
axiomatic approach to thermodynamics in the language of Pfaff forms can be
extended in order to include in the scheme generalizations of the G.\
Perelman thermodynamics for relativistic geometric flows; 2) To consider
possible applications of the anholonomic frame deformation method, AFDM, and
study main properties of geometric evolution flows of locally anisotropic
cosmological models (in particular with generic off-diagonal solitonic
deformations of the Friedman-Lema\^{i}tre-Robertson-Walker,\ FLRW, metrics;
3) To provide explicit examples of how geometric flow thermodynamic values are
computed for cosmological solitonic solutions which can not be described by
thermodynamic concepts elaborated in the framework of Bekenstein--Hawking
entropy and generalizations.

This paper is organized as follows: In section \ref{sec2}, we present an
introduction into the theory of relativistic nonholonomic flows with
modified F- and W--functionals and elaborate on respective statistical
thermodynamic models. How the Carath\'{e}odory axiomatic approach can be
extended in order to include some classes of generalized Ricci flows and
solitons is considered. Then, in section \ref{sec3}, we develop the AFDM and
show that the important general decoupling and integration properties of
geometric evolution flow and Ricci soliton equations are preserved for
cosmological solitonic spaces. Possible locally anisotropic cosmological
parameterizations are summarized in Table 1. We provide Table 2 summarizing
the AFDM for generating such locally anisotropic cosmological solutions. In
section \ref{sec4}, we show how exact and parametric cosmological solitonic
solutions can be constructed for relativistic geometric flow evolution
equations. There are analyzed explicit examples of computing respective
W-entropy, thermodynamic values and Pfaffians. Finally, conclusions and
perspectives are considered in section \ref{sec5}. In Appendix \ref%
{appendixa}, we outline some results on Pfaffian differential equations.
Appendix \ref{appendixb} contains necessary parameterizations for flows of
cosmological solitonic metrics.

\section{Generalized Carath\'{e}ory--Perelman thermodynamics and Ricci flows}

\label{sec2} In this section, we provide a brief introduction to the theory
of relativistic geometric flows and analogous statistical thermodynamics for
nonholonomic Einstein systems, NESs, see details in \cite%
{vacaru13,vacaru16,vacaru17,bub19} and references therein. How such
constructions can be formalized following the Carath\'{e}odory axiomatic
approach is analyzed.

\subsection{Relativistic models of geometric flow thermodynamics}

We consider a relativistic spacetime as in general relativity, GR.
Geometrically, it is defined by a (pseudo) Riemannian manifold $\mathbf{V}$
with a conventional splitting of dimension, $\dim \mathbf{V}=4=2+2,$ and two
dimensional horizontal, h, and two dimensional vertical, v, components (such 
decomposition will be useful for constructing exact solutions of systems
of important physical equations). This induces diadic decompositions of
local bases and corresponding tangent bundles $T\mathbf{V}$ and, its dual, $%
\ T^{\ast }\mathbf{V.}$ Being enabled with a metric $\mathbf{g}=(h\mathbf{g}%
,v\mathbf{g})$ of a local pseudo-Euclidean signature $(+++-)$ and
postulating local causality conditions as in special relativity theory, we
model a curved spacetime as a Lorentzian manifold. We can always consider
that such spacetimes are endowed with a double nonholonomic 2+2 and 3+1
splitting (the first splitting will be used for elaborating new methods of
constructing exact solutions and the second splitting will be necessary for
elaborating thermodynamical models).

In this work, we say that a Lorentz manifold $\mathbf{V}$ is nonholonomic
(in literature, there are also used equivalent terms like anholonomic, or
non-integrable) if it is endowed with a h- and/or v-splitting defined by a
Whitney sum defining a nonlinear connection, N-connection, structure $%
\mathbf{N}:\ T\mathbf{V}=h\mathbf{\mathbf{V\oplus }}v\mathbf{V}$, where $T%
\mathbf{V}$ is the tangent bundle on $\mathbf{V.}$ Such a geometric
structure is a fundamental one for elaborating various models of
Finsler--Lagrange--Hamilton geometry which are determined in complete form
if there are prescribed three fundamental geometric objects/structures (a
nonlinear quadratic line element, a nonlinear connection and a distinguished
connection which is adapted to a h-v-splitting). N--connections can be
introduced also in (pseudo) Riemannian geometry when (in local form) $%
\mathbf{N}=N_{i}^{a}(u)dx^{i}\otimes \partial _{a}$ is determined by for a
corresponding set of coefficients $\{N_{i}^{a}\} $ which can be related to
certain off-diagonal terms of metrics in certain local frames of coordinates.%
\footnote{%
We parameterize the coordinates as $u^{\mu }=(x^{i},y^{a}),$ in brief, $%
u=(x,y),$ where $i,j,...=1,2$ and $a,b=3,4$, with small Greek indices $%
\alpha ,\beta ,...=1,2,3,4,$ when $u^{4}=y^{4}=t$ is the time like
coordinate. We shall summarise on "up-low" repeating indices and use
boldface symbols for spaces and geometric objects adapted to a N-connection
splitting. For a double 2+2 and 3+1 splitting, the local coordinates are
labeled $u^{\alpha }=(x^{i},y^{a})=(x^{\grave{\imath}},u^{4}=t)$ for $\grave{%
\imath},\grave{j},\grave{k}=1,2,3.$ The nonholonomic distributions can be
N-adapted form for any open region $U\subset $ $\mathbf{V}$ covered by a
family of 3-d spacelike hypersurfaces $\Xi _{t}$ with a time like parameter $%
t.$} Corresponding subclasses of N-adapted (co) frames allow, for instance,
nonholonomic diadic decompositions of geometric and physical objects.
Together with a so-called canonical nonholonomic deformations of linear
connection structures (we shall use "hats" on geometric and physical objects
adapted to such canonical nonholonomic frames) this allows to integrate
(modified) Einstein and geometric flow equations in very general forms
depending, in principle, on all spacetime coordinates and a geometric
evolution parameter.

We shall use two important linear connections which can be constructed using
the same metric structure:
\begin{equation}
\mathbf{g}\rightarrow \left\{
\begin{array}{ccccc}
\nabla : &  & \nabla \mathbf{g}=0;\ ^{\nabla }\mathbf{T}=0, &  &
\mbox{ the
Levi--Civita, LC, connection;} \\
\widehat{\mathbf{D}}: &  & \widehat{\mathbf{D}}\ \mathbf{g}=0;\ h\widehat{%
\mathbf{T}}=0,\ v\widehat{\mathbf{T}}=0. &  &
\mbox{ the canonical
d--connection.}%
\end{array}%
\right.  \label{lcconcdcon}
\end{equation}%
In these formulas (see \cite{vacaru13,vacaru16,vacaru17,bub19} for details
on computing coefficients with respect to N-adapted and/or coordinate
frames), the distinguished connection (d--connection) $\widehat{\mathbf{D}}%
=(h\widehat{\mathbf{D}},v\widehat{\mathbf{D}})$ preserves under parallelism
the decomposition $\mathbf{N}$ and $\widehat{\mathbf{T}}\mathbf{[g,N]}$ is
the corresponding torsion d-tensor.{\ We use the terms d-tensor,
d-connection etc. for geometric objects adapted to an N--connection
h-v-splitting. The LC--connection }$\nabla $ can be introduced without any
N--connection structure but the zero torsion condition in the case of
generic off-diagonal metrics do not allow to prove a decoupling property
and explicit integration of physically important systems of nonlinear
partial differential equations, PDEs. Nevertheless, any geometric data $(%
\mathbf{g,}\nabla )$ can be distorted to some canonical ones, $(\mathbf{g,}%
\widehat{\mathbf{D}}),$ with decoupling of (modified) Einstein equations and
encoding in general form various classes of physically important solutions)
\begin{equation}
\widehat{\mathbf{D}}\mathbf{[g,N]}=\nabla \mathbf{[g,N]}+\widehat{\mathbf{Z}}%
\mathbf{[g,N],}  \label{distr}
\end{equation}%
where $\widehat{\mathbf{Z}}$ is the distortion d-tensor determined in
standard algebraic form by the torsion tensor $\widehat{\mathbf{T}}\mathbf{%
[g,N]}$ of $\widehat{\mathbf{D}}.$ These values are completely defined by
the metric d-tensor $\mathbf{g}=(h\mathbf{g},v\mathbf{g})$ adapted to a
prescribed $\mathbf{N.}$ The values $h$ $\widehat{\mathbf{T}}$ and $v%
\widehat{\mathbf{T}}$ denote respective torsion components which vanish on
conventional h- and v--subspaces. There are also nontrivial components $hv$ $%
\widehat{\mathbf{T}}$ defined by certain anholonomy (equivalently,
nonholonomic/non-integrable) relations. All geometric constructions on a
Lorentz manifold $\mathbf{V}$ can be performed in a not adapted N-connection
form with $\nabla $ and/or in N-adapted form using $\widehat{\mathbf{D}}$
from (\ref{lcconcdcon}) or other type d-connections. The corresponding
torsion d-tensor, $\widehat{\mathbf{T}}_{\ \alpha \beta }^{\gamma }$; Ricci
d-tensor, $\widehat{\mathbf{R}}_{\ \beta \gamma };$ scalar curvature $\ ^{s}%
\widehat{R}:=\mathbf{g}^{\alpha \beta }\widehat{\mathbf{R}}_{\ \beta \gamma
};$ and Einstein d-tensor, $\widehat{\mathbf{E}}_{\ \beta \gamma }:=\widehat{%
\mathbf{R}}_{\ \beta \gamma }-\frac{1}{2}\mathbf{g}_{\ \beta \gamma }$ $\
^{s}\widehat{R},$ are defined and computed in standard forms as in
metric--affine geometry and related via distortion formulas to respective
values determined by $\nabla .$

In the theory of Ricci flows of geometric objects on $\mathbf{V,}$ it is
considered an evolution positive parameter $\tau ,0\leq \tau \leq \tau _{0},$
which for thermodynamic models can identified with the temperature, or
chosen to be proportional to a temperature parameter. For the geometric flow
evolution of Riemannian metrics and respective statistical thermodynamic
models, this was considered in G. Perelman's famous preprint \cite{perelman1}%
. For evolution of (generalized) pseudo--Riemannian configurations we can
elaborate on two classes of (effective) geometric theories, when families of
metrics 1] $\mathbf{g}(\tau ):=\mathbf{g}(\tau ,u)$ are labelled by a
conventional evolution (relativistic temperature parameter) or 2] $\mathbf{g}%
(\tau ):=\mathbf{g}(\tau ,x^{\grave{\imath}})$ for an imaginary time like
coordinate $u^{\alpha }=(x^{i},y^{a})=(x^{\grave{\imath}},u^{4}=ict),$ where
$i^{2}=-1$ and, for simplicity, there used unities when the fundamental
speed of light is $c=1.$ Hereafter, we shall write in brief only the
dependence on evolution parameter, without spacetime or space coordinates if
that will not result in ambiguities. In this work, we study only theories of
class 1] when the evolution models are relativistic, encode solitonic waves
for pseudo-Riemannian metric signatures and can be characterized by
relativistic thermodynamic models, see details in \cite%
{vacaru13,vacaru16,vacaru17} and references therein. In such geometric and
thermodynamic theories, we consider also flows of N--connections $\mathbf{N}%
(\tau )=\mathbf{N}(\tau ,u),$ canonical d-connections $\widehat{\mathbf{D}}%
(\tau )=\widehat{\mathbf{D}}(\tau ,u).$ On $\mathbf{V,}$ we can introduce
also families of Lagrange densities $\ ^{g}\mathcal{L}(\tau ),$ for
gravitational fields in a MGT or GR (when $\ ^{g}\mathcal{L}(\tau )=$ $\
^{s}R[\nabla (\tau )]$ ), and $\ ^{tot}\mathcal{L}(\tau )=\ ^{tot}\mathcal{L}%
[\mathbf{g}(\tau ),\widehat{\mathbf{D}}(\tau ),\varphi (\tau )],$ as total
Lagrangians for effective and matter fields which will be defined below for
certain cosmological models with scalar fields $\varphi (\tau )=\varphi
(\tau ,u).$

For any region $U\subset $ $\mathbf{V}$ with a 2+2 splitting $(\mathbf{N,g}%
), $ we consider an additional structure of 3-d hypersurfaces $\Xi _{t}$
parameterized by time like coordinate $y^{4}=t$ for coordinates $u^{\alpha
}=(x^{i},y^{a})=(x^{\grave{\imath}},t).$ The families of metrics can be
represented as d-metrics with 3+1 splitting and N--adapted geometric
evolution,%
\begin{eqnarray}
\mathbf{g}(\tau ) &=&\mathbf{g}_{\alpha ^{\prime }\beta ^{\prime }}(\tau ,\
u)d\ \mathbf{e}^{\alpha ^{\prime }}(\tau )\otimes d\mathbf{e}^{\beta
^{\prime }}(\tau )  \label{decomp31} \\
&=&q_{i}(\tau ,x^{k})dx^{i}\otimes dx^{i}+\mathbf{q}_{3}(\tau ,x^{k},y^{a})%
\mathbf{e}^{3}(\tau )\otimes \mathbf{e}^{3}(\tau )-[\ _{q}N(\tau
,x^{k},y^{a})]^{2}\mathbf{e}^{4}(\tau )\otimes \mathbf{e}^{4}(\tau ),  \notag
\\
&& \mbox{for\ } \mathbf{e}^{\mu }(\tau ) =(e^{i}=dx^{i},\mathbf{e}^{a}(\tau
)=dy^{a}+\ N_{i}^{a}(\tau )dx^{i}).  \label{frameflows}
\end{eqnarray}%
In (\ref{decomp31}), there are considered geometric flows of "shift"
coefficients $\mathbf{q}_{\grave{\imath}}(\tau )=(q_{i}(\tau ),\mathbf{q}%
_{3}(\tau ))$ related to flows of a 3-d metric $\mathbf{q}_{ij}(\tau )=diag(%
\mathbf{q}_{\grave{\imath}}(\tau ))=(q_{i}(\tau ),\mathbf{q}_{3}(\tau ))$ on
a hypersurface $\Xi _{t}$ if $\mathbf{q}_{3}(\tau )=\mathbf{g}_{3}(\tau )$
and $[\ _{q}N(\tau )]^{2}=-\mathbf{g}_{4}(\tau ),$ where $\ _{q}N(\tau )$ is
a family of lapse functions. Here, it should be noted that we follow
notations which are different from those in \cite{misner} for GR. In this
work, it is used a left label $q$ in order to avoid ambiguities with the
notations for the coefficients $N_{i}^{a}$ of a N-connection. There are
considered flows of N-adapted frames (\ref{frameflows}) determined by the
flow evolution of N-connection coefficients.

In nonholonomic canonical variables, the relativistic versions of G.
Perelman functionals (originally defined in \cite{perelman1} for flows of
Riemannian metrics), in this work encoding also the geometric evolution of
matter fields, are postulated \cite{vacaru13,vacaru16,vacaru17} in the form
\begin{eqnarray}
\widehat{\mathcal{F}}(\tau ) &=&\int \left( 4\pi \tau \right) ^{-2}e^{-%
\widehat{f}}\sqrt{|\mathbf{g}|}d^{4}u(\ _{s}\widehat{R}+\ ^{tot}\mathcal{L}+|%
\widehat{\mathbf{D}}\widehat{f}|^{2})\mbox{
and }  \label{ffcand} \\
\widehat{\mathcal{W}}(\tau ) &=&\int \widehat{\mu }\sqrt{|\mathbf{g}|}%
d^{4}u[\tau (\ _{s}\widehat{R}+\ ^{tot}\mathcal{L}+|\ \ _{h}\widehat{\mathbf{%
D}}\widehat{f}|+|\ \ _{v}\widehat{\mathbf{D}}\widehat{f}|)^{2}+\widehat{f}%
-4].  \label{wfcand}
\end{eqnarray}%
In such formulas, a normalizing function $\widehat{f}(\tau ,u)$ can be a
convenient one for elaborating certain topological/ geometric / physical
models or subjected to the conditions
\begin{equation}
\widehat{\mathcal{V}}(\tau )=\int \widehat{\mu }\sqrt{|\mathbf{g}|}%
d^{4}u=\int_{t_{1}}^{t_{2}}\int_{\Xi _{t}}\widehat{\mu }\sqrt{|\mathbf{g}|}%
d^{4}u=1,  \label{normalmes}
\end{equation}%
for a classical integration measure$\ \widehat{\mu }=\left( 4\pi \tau
\right) ^{-2}e^{-\widehat{f}}$ (a version of Carath\'{e}odory measure); and
the Ricci scalar $\ _{s}\widehat{R}$ is taken for the Ricci d-tensor $%
\widehat{\mathbf{R}}_{\alpha \beta }$ of a d-connection $\widehat{\mathbf{D}}%
.$

There is a series of arguments for writing the $\widehat{\mathcal{F}}$%
-functional (\ref{ffcand}) and $\widehat{\mathcal{W}}$-functional\ (\ref%
{wfcand}) in above forms:

\begin{enumerate}
\item Fixing variations of such functionals on a d-metric and respective
matter field evolution scenarios, and considering self-similar
configurations for a $\tau =\tau _{0},$ we obtain systems of nonlinear PDEs
for relativistic Ricci solitons, which are equivalent to the gravitational
field equations in nonholonomic variables,
\begin{equation}
\widehat{\mathbf{R}}_{\alpha \beta }=\widehat{\mathbf{\Upsilon }}_{\alpha
\beta }  \label{einstnes}
\end{equation}%
for nonholonomic Einstein systems, NESs, if the normalizing function $%
\widehat{f}$ is correspondingly chosen and the nonholonomic constraints for
extracting Levi--Civita configurations are imposed to extract
LC--configurations $\widehat{\mathbf{D}}_{\mid \widehat{\mathbf{T}}%
=0}=\nabla $. The sources $\ \widehat{\mathbf{\Upsilon }}_{\alpha \beta
}(\tau )=[\widehat{\mathbf{\Upsilon }}_{ij}(\tau ),\widehat{\mathbf{\Upsilon
}}_{ab}(\tau )]$ with coefficients defined with respect to N-adapted frames
in (\ref{einstnes}) are of type $\widehat{\mathbf{\Upsilon }}_{\mu \nu }=\
^{e}\widehat{\mathbf{\Upsilon }}_{\mu \nu }+\ ^{m}\widehat{\mathbf{\Upsilon }%
}_{\mu \nu },$ where $\ ^{e}\widehat{\mathbf{\Upsilon }}_{\mu \nu }$ are
effective sources determined by distortions of the linear connections and
effective Lagrangians for gravitational fields. Such a source is not zero
even in GR if there are nonzero distortions (\ref{distr}) from $\widehat{%
\mathbf{D}}$ to $\nabla .$ A source for matter field, $\ ^{m}\widehat{%
\mathbf{\Upsilon }}_{\mu \nu },$ can be constructed using a N--adapted
variational calculus for a Lagrande density $\ ^{m}\mathcal{L}(\mathbf{g,}%
\widehat{\mathbf{D}},\ \ ^{A}\varphi ), $ when
\begin{equation*}
\ ^{m}\widehat{\mathbf{\Upsilon }}_{\mu \nu }=\varkappa (\ ^{m}\widehat{%
\mathbf{T}}_{\mu \nu }-\frac{1}{2}\mathbf{g}_{\mu \nu }\ ^{m}\widehat{%
\mathbf{T}})\rightarrow \varkappa (\ ^{m}T_{\mu \nu }-\frac{1}{2}\mathbf{g}%
_{\mu \nu }\ ^{m}T)
\end{equation*}%
for [coefficients of $\ \widehat{\mathbf{D}}]$ $\rightarrow $ [coefficients
of $\nabla $]. In such formulas, we consider $\ ^{m}\widehat{\mathbf{T}}=%
\mathbf{g}^{\mu \nu }\ ^{m}\widehat{\mathbf{T}}_{\mu \nu }$ for the
N-adapted energy--momentum tensor
\begin{equation}
\ ^{m}\widehat{\mathbf{T}}_{\alpha \beta }:=-\frac{2}{\sqrt{|\mathbf{g}_{\mu
\nu }|}}\frac{\delta (\sqrt{|\mathbf{g}_{\mu \nu }|}\ \ ^{m}\mathcal{L})}{%
\delta \mathbf{g}^{\alpha \beta }}.  \label{emten}
\end{equation}%
For simplicity, we shall consider only Lagrange densities$\ ^{m}\mathcal{L}%
=\ ^{\phi }\mathcal{L}(\mathbf{g,}\widehat{\mathbf{D}},\ \phi )$ determined
by a scalar field $\phi (x,u)$ and/or geometric evolution of scalar fields $%
\phi (\tau )=\phi (\tau ,x,u),$ when $\ ^{m}\widehat{\mathbf{T}}_{\alpha
\beta }=\ ^{\phi }\widehat{\mathbf{T}}_{\alpha \beta }$.

\item For three dimensional, 3-d, Riemannian metrics, there are obtained
respective Lyapunov type functionals as it was postulated in \cite{perelman1}
and used for the proof of the Thurston--Poincar\'{e} conjecture.

\item The functional $\widehat{\mathcal{W}}$ (\ref{wfcand}) defines a
nonholonomic canonical and relativistic generalization of the so-called
W-entropy introduced in \cite{perelman1}. Various types of 4-d - 10-d $%
\mathcal{W}$--entropies and associated statistical and quantum
thermodynamics values are used for elaborating models of classical and
(commutative and noncommutative/ supersymmetric) quantum geometric flows and
geometric information flows, see \cite%
{vacaru08,vacaru13,vacaru16,vacaru17,bub19,vacaru19,vacaru19a,vacaru19b} and
references therein.

\item The functionals $\widehat{\mathcal{F}}$ and $\widehat{\mathcal{W}}$
result in generalized R. Hamilton equations \cite{hamilton82} considered
earlier in physics by D. Friedan \cite{friedan80} (respective proofs for and
N-adapted variational calculus are presented in \cite%
{vacaru13,vacaru16,vacaru17,bub19}):
\begin{eqnarray}
\frac{\partial \mathbf{g}_{ij}}{\partial \tau } &=&-2\left( \widehat{\mathbf{%
R}}_{ij}-\widehat{\mathbf{\Upsilon }}_{ij}\right) ;\ \frac{\partial \mathbf{g%
}_{ab}}{\partial \tau }=-2\left( \widehat{\mathbf{R}}_{ab}-\widehat{\mathbf{%
\Upsilon }}_{ab}\right) ;  \label{canhamiltevol} \\
\widehat{\mathbf{R}}_{ia} &=&\widehat{\mathbf{R}}_{ai}=0;\ \widehat{\mathbf{R%
}}_{ij}=\widehat{\mathbf{R}}_{ji};\ \widehat{\mathbf{R}}_{ab}=\widehat{%
\mathbf{R}}_{ba};  \notag \\
\partial _{\tau }\ \widehat{f} &=&-\widehat{\square }\widehat{f}+\left\vert
\ \widehat{\mathbf{D}}\ \widehat{f}\right\vert ^{2}-\ \ _{s}\widehat{R}+%
\widehat{\mathbf{\Upsilon }}_{a}^{a},  \label{normalizfunct}
\end{eqnarray}%
where $\ \widehat{\square }(\tau )=\widehat{\mathbf{D}}^{\alpha }(\tau )\
\widehat{\mathbf{D}}_{\alpha }(\tau )$ is used for the geometric flows of
the d'Alambert operator. In nonholonomic canonical variables with $\widehat{%
\mathbf{D}}$, such systems of nonlinear PDEs can be integrated in very
general forms and restricted to describe the geometric evolution and Ricci
soliton configurations of NESs, see details and proofs in references from
the previous paragraph (point).

\item The measure $\ \widehat{\mu }\sqrt{|\mathbf{g}|}d^{4}u=\left( 4\pi
\tau \right) ^{-2}e^{-\widehat{f}}\sqrt{|\mathbf{g}|}d^{4}u$ consists an
explicit example of a Carath\'{e}odory type measure which allows to
construct geometric and statistical thermodynamic models. Such a model was
elaborated for the flow evolution of 3-d Riemannian metrics by G. Perelman
and considered in the proof of the Thurston--Poincar\'{e} conjecture.
Thermodynamic measures have more rich implications in various branches of
topology and applications and provide a natural tool to understand the
difficulties (ergodicity, approach to equilibrium, irreversibility etc.) in
the foundations of statistical physic and non--equilibrium thermodynamics,
see discussions and references in \cite%
{misra79,misra83,ant00,ant02,bel02,bel02a}. Fixing a normalizing function $%
\widehat{f},$ we prescribe an evolution scenarios with respective scales and
phase space integral properties determined by geometric and physical data $(%
\mathbf{g,}\widehat{\mathbf{D}},\ ^{A}\varphi ).$ We can consider a
geometric evolution model without $\ ^{m}\mathcal{L}$ but with a re-defined
functional measure $\mu \ ^{\bigtriangledown }\sqrt{|\mathbf{g}|}%
d^{4}u=\left( 4\pi \tau \right) ^{-2}e^{-\widehat{f}(f\ ^{\bigtriangledown
})}\sqrt{|\mathbf{g}|}d^{4}u,$ where $f^{\bigtriangledown }$ is chosen to be
a solution of this system of PDEs:%
\begin{equation}
\ ^{tot}\mathcal{L}+|\widehat{\mathbf{D}}\widehat{f}|^{2} =|\widehat{\mathbf{%
D}}(f^{\bigtriangledown })|^{2} \mbox{ and } \tau (\ ^{tot}\mathcal{L}+|\ \
_{h}\widehat{\mathbf{D}}\widehat{f}|+|\ \ _{v}\widehat{\mathbf{D}}\widehat{f}%
|)^{2}+\widehat{f} =\tau (|\ \ _{h}\widehat{\mathbf{D}}(f^{\bigtriangledown
})|+|\ \ _{v}\widehat{\mathbf{D}}(f^{\bigtriangledown
})|)^{2}+f^{\bigtriangledown }.  \label{normalfunct}
\end{equation}%
The solutions for such a $\ f^{\bigtriangledown }(\widehat{f})$ and/or $%
\widehat{f}(f^{\bigtriangledown })$ can be found in an explicit form
(usually, such a normalizing function can be approximated to a constant) for
a large class of generic off-diagonal or diagonal solutions of systems (\ref%
{canhamiltevol}) and (\ref{normalizfunct}). In such constructions, usually
there are prescribed respective values of some generating functions and
sources and effective cosmological constants subjected to certain nonlinear
symmetry conditions, see details in \cite%
{vacaru08,vacaru13,vacaru16,vacaru17,bub19,vacaru19,vacaru19a,vacaru19b} and
section \ref{sec5a}. For various physical applications, it is enough to find
a class of solutions of generalized R. Hamilton equations (\ref%
{canhamiltevol}) and to consider that the geometric evolution is normalized
by some functions a $\ f^{\bigtriangledown }(\widehat{f})$ and/or $\widehat{f%
}(f^{\bigtriangledown })$ subjected to conditions \ (\ref{normalizfunct})
and (\ref{normalfunct}). In many cases, such normalizations can be performed
with certain integration constants or for series expansions on a small
parameter.
\end{enumerate}

Using formulas (\ref{ffcand}),\ (\ref{wfcand}), (\ref{canhamiltevol}), (\ref%
{normalizfunct}), we can elaborate on statistical thermodynamic models for
geometric flows determined by data $(\mathbf{g,}\widehat{\mathbf{D}},\ \
^{A}\varphi ,f^{\bigtriangledown })$ and applying concepts and formulas from
statistical thermodynamics. It is considered a canonical ensemble at
temperature $\beta ^{-1}=T$, in this work $T$ is proportional to $\tau$,
with partition function $Z=\int \exp (-\beta E)d\omega (E),$ where a measure
$\omega (E)$ is defined as a density of states. In standard form, there are
computed such important thermodynamical values: average flow energy, $%
\mathcal{E}=\ \left\langle E\right\rangle :=-\partial \log Z/\partial \beta
; $\ flow entropy, $\mathcal{S}:=\beta \left\langle E\right\rangle +\log Z;$%
\ flow fluctuation, $\eta :=\left\langle \left( E-\left\langle
E\right\rangle \right) ^{2}\right\rangle =\partial ^{2}\log Z/\partial \beta
^{2}.$ A geometric thermodynamic model can be constructed if we associate to
(\ref{wfcand}) a respective thermodynamic generating functions (in this
work, flows of $\ ^{tot}\mathcal{L}$ are encoded into $f^{\bigtriangledown }(%
\widehat{f},\ ^{tot}\mathcal{L})$ subjected to the conditions (\ref%
{normalfunct})),
\begin{equation}
\widehat{\mathcal{Z}}[\mathbf{g}(\tau ),f^{\bigtriangledown }]=\int (4\pi
\tau )^{-2}e^{-\widehat{f}(f^{\bigtriangledown })}\sqrt{|\mathbf{g}|}%
d^{4}u(-f^{\bigtriangledown }+2),\mbox{ for }\mathbf{V.}  \label{genfcanv}
\end{equation}%
Hereafter we shall not write functional dependencies on $\mathbf{g}(\tau )$
and $f^{\bigtriangledown }$ if it will not result in ambiguities.

Applying a similar variational calculus similar to that presented in details
in \cite{perelman1,morgan07,kleiner06,vacaru08} (in N-adapted form for
frames (\ref{frameflows}) and d-connections $\widehat{\mathbf{D}})$ to (\ref%
{genfcanv}) and (\ref{wfcand}) and respective 3+1 parameterizations of
d-metrics (\ref{decomp31}), we define and compute analogous thermodynamic
values for geometric evolution flows of NES,%
\begin{eqnarray}
\widehat{\mathcal{E}}\ (\tau ) &=&-\tau ^{2}\int (4\pi \tau )^{-2}e^{-%
\widehat{f}(f^{\bigtriangledown })}\sqrt{|q_{1}q_{2}\mathbf{q}_{3}(_{q}N)|}%
\delta ^{4}u(\ \ _{s}\widehat{R}+|\widehat{\mathbf{D}}f^{\bigtriangledown
}|^{2}\mathbf{\ }-\frac{2}{\tau }),  \label{thvcanon} \\
\widehat{\mathcal{S}}\ (\tau ) &=&-\int (4\pi \tau )^{-2}e^{-\widehat{f}%
(f^{\bigtriangledown })}\sqrt{|q_{1}q_{2}\mathbf{q}_{3}(_{q}N)|}\delta ^{4}u%
\left[ \tau \left( \ _{s}\widehat{R}+|\widehat{\mathbf{D}}%
f^{\bigtriangledown }|^{2}\right) +f^{\bigtriangledown }-4\right] ,  \notag
\\
\widehat{\eta }\ (\tau ) &=&-2\tau ^{4}\int (4\pi \tau )^{-2}e^{-\widehat{f}%
(f^{\bigtriangledown })}\sqrt{|q_{1}q_{2}\mathbf{q}_{3}(_{q}N)|}\delta
^{4}u[|\ \widehat{\mathbf{R}}_{\alpha \beta }+\widehat{\mathbf{D}}_{\alpha
}\ \widehat{\mathbf{D}}_{\beta }\ f^{\bigtriangledown }-\frac{1}{2\tau }%
\mathbf{g}_{\alpha \beta }|^{2}].  \notag
\end{eqnarray}%
In these formulas, $\delta ^{4}u$ contains N-elongated differentials and the
data on matter fields and nonlinear symmetries are encoded in $\
f^{\bigtriangledown }(\widehat{f},\ ^{tot}\mathcal{L}).$ For fixed
self-similar Riemannian configurations, the values $\widehat{\mathcal{E}}$
and $\widehat{\mathcal{S}}$ provide an equilibrium thermodynamic description
of Ricci solitons. Such concepts can be considered along any causal curve on
a Lorentz manifold when fixing certain normalization functions and nonlinear
symmetries the conventional thermodynamic description holds true for
evolution models of NESs. The fluctuation $\widehat{\eta }\ (\tau )$ allows
to include into consideration small perturbations of metrics and
corresponding distortion values. Such a description defines a relativistic
thermodynamic model which is irreversible and describes various types of
nonlinear self-organizing, pattern forming, kinetic and/or stochastic
processes, see examples and references in \cite%
{vacaru08,vacaru13,vacaru16,vacaru17,bub19,vacaru19,vacaru19a,vacaru19b}.

After Carath\'{e}odory had completed the proof of Poincar\'{e} recurrence
theorem \cite{carath19}, he was the first who saw that measure theory is the
natural language to discuss the problems of statistical physics and
thermodynamics. The Poincar\'{e} hypothesis was formulated on topological
properties of three dimensional hypersurfaces endowed with Riemannian
metrics. The proof of its generalized form as the Thurston--Poincar\'{e}
conjecture was possible by introducing measures of type%
\begin{equation}
\widehat{M}=\widehat{\mu }\sqrt{|\mathbf{g}|}d^{4}u=\left( 4\pi \tau \right)
^{-2}e^{-\widehat{f}}\sqrt{|\mathbf{g}|}d^{4}u  \label{measure}
\end{equation}%
for Riemannian versions of functionals $\widehat{\mathcal{F}}$ (\ref{ffcand}%
) $\widehat{\mathcal{W}}$\ (\ref{wfcand}), when with $\widehat{\mathbf{D}}%
_{\mid \widehat{\mathbf{T}}=0}=\nabla ,$ and a respective thermodynamic
values (\ref{thvcanon}). If $\sqrt{|\mathbf{g}|}$ is considered for
pseudo-Riemannian metrics (and various, for instance,
Finsler-Lagrange-Hamilton generalizations \cite{vacaru08,bub19}), we can
speculate on respective generalized Poincar\'{e} recurrence theorem which
can be reformulated in this form: \textit{The volume preserving dynamical
transformations of an (effective) phase space with a measure }$\widehat{\mu }%
\sqrt{|\mathbf{g}|}d^{4}u$\textit{\ have the \ property that almost all
points in any region of positive volume (excepting possible subsets of zero
volume) will return back into their region after some finite time. }For
relativistic configurations, we consider 3+1 splitting and a Lorentz type
causality on respective spacetime and/or phase space (co) tangent Lorentz
bundles. Of course, the return time for each point connected by a causal
curve, is different but can be computed for a respective exact/parametric
solution of the relativistic geometric flows and/or (modified) gravity
theory (i.e. nonholonomic Ricci soliton configuration).

Following the Carath\'{e}odory measure theoretic idea \cite{carath19} and
Birkhoff's approach to ergodicity \cite{birkhoff31,confeld82}, we can
clarify the relation between ergodic and recurrent systems. In the case of
geometric flows, we can define ergodicity by replacing Boltzmann's sets with
sets of non zero volume measure which was generalized for nonholonomic
manifolds and generalized Finsler spaces in \cite%
{vacaru00,vacaru10,vacaru12,vacaru13}. Here we note that geometric flows as
ergodic systems are recurrent but not vice versa. Non--ergodic systems
decompose into time-invariant ergodic sub-systems and this property can be
extended to relativistic flows defined along causal curves.

Mixing (it means that the statistical correlations decay and results in
statistical regularity) of geometric flows implies ergodicity and is
compatible with recurrence. Let us explain how it characterizes geometric
evolution of NESs relating certain subsets $A,B,Y\subset \mathbf{V,}$ when $%
S_{t}:Y\rightarrow Y$ is the geometric (gradient) flow evolution on the
phase space $Y$ and there are satisfied non-zero measure conditions, $%
\widehat{M}[A]\neq 0,\widehat{M}[B]\neq 0,$ for $\widehat{M}[Y]=1$
considered as a probability measure determined by (\ref{measure}), when $%
\lim_{t\rightarrow \pm \infty }\frac{\widehat{M}[A\cap S_{t}B]}{\widehat{M}%
[A]}=\frac{\widehat{M}[B]}{\widehat{M}[Y]}$. Such conditions for geometric
evolution of NES determined by on families of 3-d hypersurfaces $\Xi _{t}$
mean that any set $B$ spreads over the phase space $Y$ so that for any
(fixed set/hypersurface/window) the fraction of $B$ in $A$ approaches the
fraction of $B$ over the whole phase space $Y$ (this is an uniform mixture
of d-metrics).

Relativistic thermodynamic systems (\ref{thvcanon}) with measure theoretic
definitions of ergodicity and mixing for corresponding classes of solutions
of (\ref{canhamiltevol}) possess such properties on an open spacetime region
$U\subset \mathbf{V},$

\begin{enumerate}
\item as ergodic systems they have an unique equilibrium distribution
defined by $\mathbf{g}(\tau _{0});$

\item as mixing systems they approach an equilibrium state by $\mathbf{g}%
(\tau _{0});$

\item the rates of approach to equilibrium is determined by the rates of
decay of correlations which can be computed, for instance, for locally
anisotropic cosmological solutions;

\item using exact solutions, we can study irreversibility as an
unidirectional spontaneous evolution from present to future; this issue can
addressed using operator theory and functional analysis (in this article, we
do not consider such issues in Hilbert space and convex spaces, see
references in \cite{ant02}).
\end{enumerate}

Nonholonomically deformed G. Perelman functionals (\ref{ffcand}) and (\ref%
{wfcand}) determine both the relativistic dynamical and thermodynamic (in
general, with irreversible and non-equilibrium configurations) properties of
NESs.

\subsection{Carath\'{e}odory axiomatic thermodynamics and Ricci flows}

Following the first seminal Carath\'{e}odory's work \cite{carath09}, we
state the main definitions of the concepts of states, equilibrium, energy
and entropy, and thermodynamic coordinates. A geometric flow state is given
by any data $\{\mathbf{g}_{\alpha \beta }=[q_{1}q_{2}\mathbf{q}_{3}(_{q}N)],%
\mathbf{N,}f^{\bigtriangledown }\}$ defining a solution of the nonholonomic
Ricci flow equations (\ref{canhamiltevol}) for a fixed normalizing function $%
\ f^{\bigtriangledown }(\widehat{f})$ and/or $\widehat{f}(f^{%
\bigtriangledown })$ subjected to some conditions \ (\ref{normalizfunct})
and (\ref{normalfunct}). The evolution parameter $\tau $ can be identified
with the temperature $\mathcal{T}$ in some conventional systems of
references and chosen physical unities (in general, we can consider any
convenient $\mathcal{T}(\tau )$; we use a "cal" symbol in order to avoid
ambiguities when the capital letter $T$ is used for torsion (\ref{lcconcdcon}%
) or the energy-momentum tensor (\ref{emten})). For such geometric data, we
can always compute the statistical thermodynamic values $\widehat{\mathcal{E}%
}\ (\tau )$ and $\widehat{\mathcal{S}}\ (\tau ),$ see (\ref{thvcanon}). To
elaborate on analogous thermodynamic models we can consider a family of
volumes $\widehat{\mathcal{V}}(\tau )\neq 1$ (\ref{normalmes}) when the
condition $\widehat{\mathcal{V}}(\tau )=1$ can be imposed by a corresponding
$\ f^{\bigtriangledown }(\widehat{f})$ but $d\widehat{\mathcal{V}}(\tau
)\neq 0. $ This allows us to introduce a conventional pressure $P$ and
external work $A,$ and postulate that for any fixed $\tau _{0}$ the first
law of thermodynamics for geometric flows%
\begin{equation}
d\widehat{\mathcal{E}}=-Pd\widehat{\mathcal{V}}(\tau ).  \label{therm1law}
\end{equation}%
Stating $\tau _{i}$ and $\tau _{f}$ for respective initial, and final
states, we formulate

\begin{axiom}
\label{axiom1}For any geometric flow thermodynamics of NESs, $\widehat{%
\mathcal{E}}\ (\tau _{f})-\widehat{\mathcal{E}}\ (\tau _{i})+A=0.$
\end{axiom}

This axiom\footnote{%
A mathematical project usually starts as an axiomatic system starting with
an ensemble of declarations/ statements. This contains certain
constructions, solutions of equations, and proofs of theorems. In the case
of Euclidean geometry, the axioms are considered to be self-evident but
various motivations and fundamental/ experimental arguments are put forward
for advanced theories related to physics and applications. As a typical
axiomatic approach to modern thermodynamics can be considered \cite%
{gurtin86,lieb97}. The axioms and certain definitions and "rules of
interference" provide the basis for proving theorems. The word "postulate"
is used in many cases instead of "axiom". Here we explain that in
mathematics and logics the axioms are considered as general statements
accepted without proofs. In their turns, postulates are used for some
specific cases and can not be considered as "very general" statements. In
many papers in non-mathematical journals oriented to mathematical physics
and applications the axioms, definitions and rules of interference are not
cite and related rules of interference are not sited but certain proofs and
solutions are provided using corresponding mathematical tools. Such a
geometric and PDE theory style will be used in this work.} can be considered
as the fist postulate of the relativistic thermodynamics of Ricci flows.
Reversibility for such systems can be introduced for self-similar
configurations for a fixed $\tau _{0},$ i.e. for relativistic Ricci solitons
defined equivalently by generalized Einstein equations (\ref{einstnes}). As
dynamical equations, such (modified) gravitational and matter field
equations possess reversible (at least in certain regions) solutions.%
\footnote{%
For standard thermodynamic systems, i.e. not for the Ricci flows, this is
just the internal energy and external work conservation law, i.e. the first
postulate of thermodynamics.} Nevertheless, a general geometric flow
evolution is described by irreversible equations (\ref{canhamiltevol}) and (%
\ref{normalizfunct}).

After that we can state the second axiom for relativistic geometric flows%
\footnote{%
Following Carath\'{e}odory (see also discussions and references in \cite%
{pog00}), for standard thermodynamic systems the English version of such a
famous second axiom is "In the neighborhood of any equilibrium state of a
system (of any number of thermodynamic coordinates), there exists states
that are inaccessible by reversible adiabatic processes". This axiom is
better understood if it is used the Kelvin's formulation of the second law
of (standard, not geometric) thermodynamics "no cycle can exist whose net
effect is a total conversion of heat into work".}:

\begin{axiom}
\label{axiom2}In neighborhood of any self-similar configurations for a fixed
$\tau _{0},$ i.e. of a nonholonomic Ricci soliton, there exists
states/geometric data $\{\mathbf{g}_{\alpha \beta }=[1_{1}q_{2}\mathbf{q}%
_{3}(_{q}N)],\mathbf{N,}f^{\bigtriangledown }\}$ which are inaccessible as
nonholonomic Ricci soliton systems for a fixed $f^{\bigtriangledown }(%
\widehat{f})$ but as accessible for some $\tau \geq \tau _{0}$ if there are
nontrivial solutions of generalized Hamilton equations (\ref{canhamiltevol})
and (\ref{normalizfunct}) relating $\tau _{0}$ as an initial state and a
final state with $\tau$ .
\end{axiom}

In principle, one could be certain "un-physical" processes which may connect
two geometric thermodynamic NES or general d-metric systems when $\widehat{%
\mathcal{E}}\ (\tau )$ and $\widehat{\mathcal{S}}\ (\tau )$ are computed for
data $\{\mathbf{g}_{\alpha \beta }\mathbf{=[}q_{1}q_{2}\mathbf{q}%
_{3}(_{q}N)],\mathbf{N,}f^{\bigtriangledown }\}$ which are not solutions of (%
\ref{canhamiltevol}). Here we note that in the axioms and definitions of the
Carath\'{e}odory method there is no mention of heat, temperature or entropy
because heat is regarded as derived value (and not a fundamental quantity)
that appears as soon as the adiabatic restriction is removed. That was the
weakness and strength of the approach to standard thermodynamics developed
by the aid of the theory of Pfaffian equations. M. Born \cite{born21} was
the first who centered the attention to the elegance of the new method but M
. Planck \cite{planck26} sharply criticized Carath\'{e}odory's method
considering that the Thomson-Clausious treatment was more reliable being
much nearer to experimental evidence, i.e. to natural processes. For
details, discussions and main references, we cite \cite{pog00} and \cite%
{ant02}.

Theories of geometric Ricci flows are not elaborated similarly to typical
thermodynamical systems characterizing engines and cycles with heat, or
chemical reactions and can not be studied in a phenomenological manner with
engineering methods. The strength of axiomatic methods is that introducing
in terms of geometric objects (measure defined by a metric, linear
connection and corresponding curvature scalar and Ricci tensor) the concept
of W-entropy, and related functionals and statistical thermodynamic
constructions, G.\ Perelman was able to prove the Thurston--Poincar\'{e}
conjecture. Such geometric methods are even more advanced and sophisticated
than those used by C. Carath\'{e}odory and M. Planck's criticism on
mathematical "harshness" is not relevant for such geometric thermodynamic
theories. Fundamental values similar to energy and entropy of thermodynamic
systems can be defined and computed in rigorous mathematical forms (\ref%
{thvcanon}) and one of the main goals of this paper is to show that the
Carath\'{e}odory approach with Pfaff forms can be naturally extended (even
in an almost phenomenological manner) to geometric flow models of NES and
generalized gravity systems. We shall show that the Carath\'{e}odory method
can be extended in such forms that to include in a relativistic form the
original constructions with Riemannian metrics and statistical thermodynamic
ideas.

\subsubsection{Adiabatic approximation for nonholnomic Ricci solitons}

The axiomatic thermodynamics of Carath\'{e}odory is based on the theory of
Pfaffian differential equations (fist studied by J. J. Pfaff, who proposed a
general method of integrating PDEs of first order in 1814-1815), see \cite%
{pog00}, for a brief review and references, and Appendix \ref{appendixa}. We
can elaborate on analogous "adiabatic" transformation of an "ideal" gas of
evolution flows of relativistic systems $\{\mathbf{g}_{\alpha \beta
}=[1_{1}q_{2}\mathbf{q}_{3}(_{q}N)],\mathbf{N,}f^{\bigtriangledown }\}$
defining a solution for respective nonholonomic deformations of Einstein
equations (\ref{einstnes}).

In the approximation of ideal gas of NESs, for a nonholonomic Ricci soliton
in a $\tau _{0},$ we state an equation of thermodynamic state,
\begin{equation}
P\widehat{\mathcal{V}}(\tau _{0})=\rho \tau _{0}\mbox{ for }\widehat{%
\mathcal{V}}(\tau _{0})=const,  \label{idealgas}
\end{equation}%
when the normalization (\ref{normalmes}) is not imposed, and (\ref{therm1law}%
), and write $d\widehat{\mathcal{E}}=-Pd\widehat{\mathcal{V}}(\tau
_{0})=C_{v}d\tau =dA$. Considering above formulas in local form, all values
like $R,P,C_{v},$ etc. depend also on spacetime coordinates $u^{\alpha }$ or
on space like coordinates $x^{\acute{\imath}}$ for cosmological configurations
if we do not perform integration as in (\ref{thvcanon}). Using the last two
equations, we obtain
\begin{equation}
\frac{C_{v}}{\tau _{0}}d\tau +\frac{\rho }{\widehat{\mathcal{V}}(\tau _{0})}d%
\widehat{\mathcal{V}}=0.  \label{pfaff01}
\end{equation}%
For constant coefficients, such a Pfaff equation is exact (see Appendix \ref%
{appendixa}), $\frac{\partial }{\partial \widehat{\mathcal{V}}}(\frac{C_{v}}{%
\tau })\mid _{\tau =\tau _{0}}=\frac{\partial }{\partial \tau }(\frac{\rho }{%
\widehat{\mathcal{V}}})\mid _{\tau =\tau _{0}}=0$, i.e. the Schwarz equation
(\ref{schw1}) is satisfied. For such geometric flow and thermodynamic
configurations, there is a solution of (\ref{pfaff01}) as a function
\begin{equation*}
\phi (\tau _{0},\widehat{\mathcal{V}})=\int \frac{C_{v}}{\tau }d\tau \mid
_{\tau =\tau _{0}}+\int \frac{\rho }{\widehat{\mathcal{V}}(\tau _{0})}d%
\widehat{\mathcal{V}}=const.
\end{equation*}

Above formulas are similar to the well-known ones used for adiabatic
transformations of an ideal gas when $\tau _{0}\widehat{\mathcal{V}}^{\gamma
_{0}-1}=const,\mbox{ \ for }\gamma _{0}=C_{p}/C_{v}$ and $\rho =C_{p}-C_{v}$%
. This is not surprising because all constructions are derived for
corresponding approximations in the statistical thermodynamic energy of
relativistic Ricci flows $\widehat{\mathcal{E}}$ (\ref{thvcanon}) computed
for nonholonomic Ricci soliton configurations.

\subsubsection{General transforms \& integrating factors for ideal gases of
nonholonomic Ricci solitons}

For general geometric flows, the analogous systems became asymmetric which
can be characterized by an equation of type
\begin{equation*}
\delta Q=d\widehat{\mathcal{E}}-\delta A,
\end{equation*}%
where $Q$ is a conventional "heat" related to nonholonomic Ricci flow
evolution (the term is introduced as for the usual thermodynamic systems
which are not adiabatic). In such cases, not each term of this equation can
be a state function. If we consider the approximation of ideal gas (\ref%
{idealgas}) for NESs, we can write
\begin{equation}
\delta Q=C_{v}d\tau +\frac{\rho \tau _{0}}{\widehat{\mathcal{V}}(\tau _{0})}d%
\widehat{\mathcal{V}}.  \label{heatgeom}
\end{equation}%
Because for this Pfaff form the Schwarz condition (\ref{schw1}) is not
satisfied, we conclude that $Q$ is not a state thermodynamic function.

The value $\delta A$ is not a state function because the equations (\ref%
{schw1}) are not satisfied for $\delta A=\frac{\rho \tau _{0}}{\widehat{%
\mathcal{V}}(\tau _{0})}d\widehat{\mathcal{V}}+0\cdot d\tau $. Nevertheless,
$\widehat{\mathcal{E}}$ is a state function because the Schwarz relation
holds for
\begin{equation*}
d\widehat{\mathcal{E}}=C_{v}d\tau +0\cdot d\widehat{\mathcal{V}}.
\end{equation*}

The Schwartz condition (\ref{schw1}) fails for $\delta \phi =\frac{\rho \tau
_{0}}{P}dP-\rho d\tau $. Using an integrating factor $K=-1/P,$ we satisfy
the condition (\ref{schw2}) which allows to find a solution (with possible
total differential) of $\widehat{\mathcal{V}}=\widehat{\mathcal{V}}(\tau
_{0},P)=\rho \tau _{0}/P.$

In a similar manner, we can find an integrating factor $\tau _{0}=\tau _{0}(%
\mathcal{T})$ for (\ref{heatgeom}) when the Schwartz condition (\ref{schw2})
is fulfilled. This allows us to define a new state function, the
thermodynamic entropy $S(\mathcal{T}),$ when%
\begin{equation*}
dS(\mathcal{T}):=\frac{\delta Q}{\tau _{0}(\mathcal{T})}=\frac{d\widehat{%
\mathcal{E}}+Pd\widehat{\mathcal{V}}}{\tau _{0}(\mathcal{T})}.
\end{equation*}%
In the ideal gas approximation for NESs, we obtain an exact differential $dS(%
\mathcal{T}):=\frac{C_{v}}{\tau _{0}(\mathcal{T})}d\tau +\frac{\rho }{%
\widehat{\mathcal{V}}(\mathcal{T})}d\widehat{\mathcal{V}}$.

It should be emphasized that the thermodynamic entropy $S(\mathcal{T})$, in
general, is different from the statistical geometric thermodynamic one $%
\widehat{\mathcal{S}}\ (\tau )$ (\ref{thvcanon}). For certain nonholonomic
Ricci soliton configuration and flow evolution of NESs, we can chose such
effective values $\tau _{0}(\mathcal{T}),C_{v},\rho $ etc. in order to have $%
S(\mathcal{T})=\widehat{\mathcal{S}}\ (\tau )$ for certain well defined
models with respective normalization functions and relativistic causal
structures. The values $\widehat{\mathcal{E}}\ (\tau )$ and $\widehat{%
\mathcal{S}}\ (\tau )$ are defined by a generalized W-entropy from an
axiomatic approach to Ricci flows but extended to relativistic geometric
flow. The method with Pfaff forms can be applied to such
statistical/geometric thermodynamic models (which are different from
standard thermodynamic ones) which allow to state the conditions when
generalizations of G. Perelman thermodynamics can be described following the
Carath\'{e}odory axiomatic approach.

Finally, we note that using homogeneous Pfaff forms as in \cite{bel02} we
can elaborate on relativistic models of Carath\'{e}odory--Gibbs--Perelman
thermodynamics. In \cite{bel02a}, a study of the Bekenstein--Hawking black
hole thermodynamics \cite{bek72,bek73,haw73,haw75} in Carath\'{e}odory
approach was performed (such constructions can be extended to and black
hole/cosmological/ holographic models with conventional horizons). The
geometric and statistical thermodynamic methods involving G. Perelman
W--entropy are more general ones \cite%
{vacaru13,vacaru16,vacaru17,bub19,vacaru19b} because provide an unified
approach to various classes of flow evolution and dynamical field theories
when the thermodynamic ideas are not limited to horizon type configurations
of some exact and parametric solutions.

\section{Decoupling and integrability of cosmological solitonic flow
equations}

\label{sec3} The goal of this section is to apply the anholonomic frame
deformation method, AFDM, in order to show a general decoupling and
integration property of the system of nonlinear PDEs (\ref{canhamiltevol})
for locally anisotropic cosmological configurations encoding solitonic
hierarchies. The geometric/ physical objects for such effective statistical
thermodynamical systems and corresponding Pfaff equations are determined by
generating and integration functions and (effective) matter sources encoding
geometric flow evolutions and solitonic configurations, see proofs and
examples in \cite{vacaru01,vacaru10a,anco06,vacaru15}. Locally anisotropic
cosmological solutions in gravity theories and corresponding inflation and
dark matter and dark energy models were studied in \cite%
{vacaru18tc,bubuianu18,vacaru19b}.

\subsection{How geometric cosmological flows can be encoded into solitonic
hierarchies?}

We model geometric evolution of a a 'prime' cosmological metric, $\mathbf{%
\mathring{g}}$, into a family 'target' d-metrics $\mathbf{g}(\tau )$ (\ref%
{decomp31}), when the nonholonomic deformations $\mathbf{\mathring{g}%
\rightarrow g}(\tau )$ a modelled by $\eta $-polarization functions,
\begin{eqnarray}
\mathbf{g}(\tau ) &=&\eta _{\alpha }(\tau ,x^{k},t)\mathring{g}_{\alpha }%
\mathbf{e}^{\alpha }[\eta ]\otimes \mathbf{e}^{\alpha }[\eta ]=\eta
_{i}(\tau ,x^{k})\mathring{g}_{i}dx^{i}\otimes dx^{i}+\eta _{a}(\tau
,x^{k},t)\mathring{h}_{a}e^{a}[\eta ]\otimes e^{a}[\eta ],  \notag \\
\mathbf{e}^{\alpha }[\eta ] &=&(dx^{i},\mathbf{e}^{a}=dy^{a}+\eta _{i}^{a}%
\mathring{N}_{i}^{a}dx^{i}).  \label{dme}
\end{eqnarray}%
The target N-connection coefficients are parameterized in the form $%
N_{i}^{a}(\tau ,x^{k},t)=\eta _{i}^{a}(\tau ,x^{k},t)\mathring{N}%
_{i}^{a}(\tau ,x^{k},t).$ The values $\eta _{i}(\tau )=\eta _{i}(\tau
,x^{k}),\eta _{a}(\tau )=\eta _{a}(\tau ,x^{k},t)$ and $\eta _{i}^{a}(\tau
)=\eta _{i}^{a}(\tau ,x^{k},t)$ are gravitational polarization functions, or
$\eta $-polarizations. Any target d-metric $\mathbf{g}(\tau )$ is defines a
solution of the N-adapted Hamilton equations in canonical variables (\ref%
{canhamiltevol}), or for relativistic nonholonomic Ricci soliton equations (%
\ref{einstnes}) with $\tau =\tau _{0}$ which are equivalent to the canonical
nonholonomic deformations of Einstein equations.

A cosmological prime metric $\mathbf{\mathring{g}}=\mathring{g}_{\alpha
\beta }(x^{i},y^{a})du^{\alpha }\otimes du^{\beta }$ is parameterized in a
general coordinate form with off-diagonal N-coefficients and/or represented
equivalently in N-adapted form
\begin{eqnarray}
\mathbf{\mathring{g}} &=&\mathring{g}_{\alpha }(u)\mathbf{\mathring{e}}%
^{\alpha }\otimes \mathbf{\mathring{e}}^{\beta }=\mathring{g}%
_{i}(x)dx^{i}\otimes dx^{i}+\mathring{g}_{a}(x,y)\mathbf{\mathring{e}}%
^{a}\otimes \mathbf{\mathring{e}}^{a},  \label{primedm} \\
&& \mbox{for } \mathbf{\mathring{e}}^{\alpha }=(dx^{i},\mathbf{e}^{a}=dy^{a}+%
\mathring{N}_{i}^{a}(u)dx^{i}),\mbox{ and }\mathbf{\mathring{e}}_{\alpha }=(%
\mathbf{\mathring{e}}_{i}=\partial /\partial y^{a}-\mathring{N}%
_{i}^{b}(u)\partial /\partial y^{b},\ {e}_{a}=\partial /\partial y^{a}).
\notag
\end{eqnarray}%
In general, such a d-metric $\mathbf{\mathring{g}(\tau }_{0}\mathbf{)=}%
\mathring{g}_{\alpha }(u)$ can be, or not, a cosmological solution of
gravitational field equations in GR but we impose the condition that under
geometric evolution it transforms into a target metric (\ref{dme}) which must
be an exact or parametric solution.

\subsubsection{Generating cosmological solitonic hierarchies}

Let us associate a non--stretching curve $\gamma (\tau ,\mathbf{l})$ on a
Einstein manifold $\mathbf{V}$ to geometric evolution of a d-metric $\mathbf{%
g}(\tau ),$ where $\tau $ can be identified with the geometric flow
parameter of temperature type and $\mathbf{l}$ is the arclength of the
curve, see details in \cite{vacaru10,anco06,vacaru15}. Such a curve is
characterized by an evolution d--vector $\mathbf{Y}=\gamma _{\tau }$ and
tangent d--vector $\mathbf{X}=\gamma _{\mathbf{l}}$ for which $\mathbf{%
g(X,X)=}1.$ It also swept out $\gamma (\tau ,\mathbf{l})$ as a
two--dimensional surface in $T_{\gamma (\tau ,\mathbf{l})}\mathbf{V}\subset T%
\mathbf{V.}$

On such nonholonomic configurations, we consider a coframe $\mathbf{e}\in
T_{\gamma }^{\ast }\mathbf{V}_{\mathbf{N}}\otimes (h\mathfrak{p\oplus }v%
\mathfrak{p})$ constructed as a N--adapted $\left( SO(n)\mathfrak{\oplus }%
SO(m)\right) $--parallel basis along curve $\gamma .$ For 4-d nonholonomic
Lorentz manifolds and their N-adapted flow evolution models, we consider
that $n=m=4$ and model the evolution of 4-d Lorentzian d-metrics. The labels
for respective dimension as $n$ and $m$ can be used in order to distinguish
N-adapted decompositions into $h$- and $v$-, or $cv$-components.

Families of canonical d-connections $\widehat{\mathbf{D}}(\tau )$ can be
associated with respective families of linear d-connection 1--forms $%
\widehat{\mathbf{\Gamma }}(\tau )\in T_{\gamma }^{\ast }\mathbf{V}_{\mathbf{N%
}}\otimes (\mathfrak{so}(n)\mathfrak{\oplus so}(m)).$ Similar families of
1-forms can be introduced for other types of d-connections or for a
LC-connection. We parameterize frame bases by 1-forms $\mathbf{e}_{\mathbf{X}%
}(\tau )=\mathbf{e}_{h\mathbf{X}}(\tau )+\mathbf{e}_{v\mathbf{X}}(\tau ).$
In any point $\tau _{0}$ and for $(1,\overrightarrow{0})\in \mathbb{R}^{n},%
\overrightarrow{0}\in \mathbb{R}^{n-1}$ and $(1,\overleftarrow{0})\in
\mathbb{R}^{m},\overleftarrow{0}\in \mathbb{R}^{m-1},$ for
\begin{equation*}
\mathbf{e}_{h\mathbf{X}}=\gamma _{h\mathbf{X}}\rfloor h\mathbf{e=}\left[
\begin{array}{cc}
0 & (1,\overrightarrow{0}) \\
-(1,\overrightarrow{0})^{T} & h\mathbf{0}%
\end{array}%
\right] ,\mathbf{e}_{v\mathbf{X}}=\gamma _{v\mathbf{X}}\rfloor v\mathbf{e=}%
\left[
\begin{array}{cc}
0 & (1,\overleftarrow{0}) \\
-(1,\overleftarrow{0})^{T} & v\mathbf{0}%
\end{array}%
\right] .
\end{equation*}%
For a $n+m$ splitting, $\widehat{\mathbf{\Gamma }}(\tau )=\left[ \widehat{%
\mathbf{\Gamma }}_{h\mathbf{X}}(\tau ),\widehat{\mathbf{\Gamma }}_{v\mathbf{X%
}}(\tau )\right] ,$ with%
\begin{eqnarray*}
\widehat{\mathbf{\Gamma }}_{h\mathbf{X}}(\tau ) &=&\gamma _{h\mathbf{X}%
}\rfloor \widehat{\mathbf{L}}\mathbf{=}\left[
\begin{array}{cc}
0 & (0,\overrightarrow{0}) \\
-(0,\overrightarrow{0})^{T} & \widehat{\mathbf{L}}%
\end{array}%
\right] \in \mathfrak{so}(n+1), \\
\mbox{ where }\widehat{\mathbf{L}} &=&\left[
\begin{array}{cc}
0 & \overrightarrow{v} \\
-\overrightarrow{v}^{T} & h\mathbf{0}%
\end{array}%
\right] \in \mathfrak{so}(n),~\overrightarrow{v}\in \mathbb{R}^{n-1},~h%
\mathbf{0\in }\mathfrak{so}(n-1);
\end{eqnarray*}
\begin{eqnarray*}
\mbox{ and }\widehat{\mathbf{\Gamma }}_{v\mathbf{X}}(\tau ) &=&\gamma _{v%
\mathbf{X}}\rfloor \widehat{\mathbf{C}}=\left[
\begin{array}{cc}
0 & (0,\overleftarrow{0}) \\
-(0,\overleftarrow{0})^{T} & \widehat{\mathbf{C}}%
\end{array}%
\right] \in \mathfrak{so}(m+1), \\
\mbox{ where }\widehat{\mathbf{C}} &=&\left[
\begin{array}{cc}
0 & \overleftarrow{v}) \\
-\overleftarrow{v}^{T} & v\mathbf{0}%
\end{array}%
\right] \in \mathfrak{so}(m),~\overleftarrow{v}\in \mathbb{R}^{m-1},~v%
\mathbf{0\in }\mathfrak{so}(m-1).
\end{eqnarray*}

Using a family of canonical d--connections $\widehat{\mathbf{D}}(\tau ),$ we
can define certain families of N-adapted matrices which are decomposed with
respect to the flow direction:\ in the h--direction,
\begin{equation*}
\mathbf{e}_{h\mathbf{Y}}=\gamma _{\tau }\rfloor h\mathbf{e}=\left[
\begin{array}{cc}
0 & \left( h\mathbf{e}_{\parallel },h\overrightarrow{\mathbf{e}}_{\perp
}\right) \\
-\left( h\mathbf{e}_{\parallel },h\overrightarrow{\mathbf{e}}_{\perp
}\right) ^{T} & h\mathbf{0}%
\end{array}%
\right] ,\mbox{ when }\newline
\mathbf{e}_{h\mathbf{Y}}\in h\mathfrak{p,}\left( h\mathbf{e}_{\parallel },h%
\overrightarrow{\mathbf{e}}_{\perp }\right) \in \mathbb{R}^{n},h%
\overrightarrow{\mathbf{e}}_{\perp }\in \mathbb{R}^{n-1},
\end{equation*}
\begin{eqnarray*}
\mbox{ and }\widehat{\mathbf{\Gamma }}_{h\mathbf{Y}}(\tau ) &\mathbf{=}%
&\gamma _{h\mathbf{Y}}\rfloor \widehat{\mathbf{L}}\mathbf{=}\left[
\begin{array}{cc}
0 & (0,\overrightarrow{0}) \\
-(0,\overrightarrow{0})^{T} & h\mathbf{\varpi }_{\tau }%
\end{array}%
\right] \in \mathfrak{so}(n+1), \\
\mbox{ where }h\mathbf{\varpi }_{\tau } &\mathbf{=}&\left[
\begin{array}{cc}
0 & \overrightarrow{\varpi } \\
-\overrightarrow{\varpi }^{T} & h\widehat{\mathbf{\Theta }}%
\end{array}%
\right] \in \mathfrak{so}(n),~\overrightarrow{\varpi }\in \mathbb{R}^{n-1},~h%
\widehat{\mathbf{\Theta }}\mathbf{\in }\mathfrak{so}(n-1).
\end{eqnarray*}%
Similar families of geometric objects and parameterizations can be
constructed for the v--direction,
\begin{equation*}
\mathbf{e}_{v\mathbf{Y}}=\gamma _{\tau }\rfloor v\mathbf{e=}\left[
\begin{array}{cc}
0 & \left( v\mathbf{e}_{\parallel },v\overleftarrow{\mathbf{e}}_{\perp
}\right) \\
-\left( v\mathbf{e}_{\parallel },v\overleftarrow{\mathbf{e}}_{\perp }\right)
^{T} & v\mathbf{0}%
\end{array}%
\right] ,\mbox{ when }\mathbf{e}_{v\mathbf{Y}}\in v\mathfrak{p,}\left( v%
\mathbf{e}_{\parallel },v\overleftarrow{\mathbf{e}}_{\perp }\right) \in
\mathbb{R}^{m},v\overleftarrow{\mathbf{e}}_{\perp }\in \mathbb{R}^{m-1},
\end{equation*}
\begin{eqnarray*}
\mbox{ and }\widehat{{\mathbf{\Gamma }}}_{v\mathbf{Y}} &\mathbf{=}&\gamma _{v%
\mathbf{Y}}\rfloor \widehat{\mathbf{C}}\mathbf{=}\left[
\begin{array}{cc}
0 & (0,\overleftarrow{0}) \\
-(0,\overleftarrow{0})^{T} & v\widehat{\mathbf{\varpi }}_{\tau }%
\end{array}%
\right] \in \mathfrak{so}(m+1), \\
\mbox{ where }v\mathbf{\varpi }_{\tau } &\mathbf{=}&\left[
\begin{array}{cc}
0 & \overleftarrow{\varpi } \\
-\overleftarrow{\varpi }^{T} & v\widehat{\mathbf{\Theta }}%
\end{array}%
\right] \in \mathfrak{so}(m),~\overleftarrow{\varpi }\in \mathbb{R}^{m-1},~v%
\widehat{\mathbf{\Theta }}\mathbf{\in }\mathfrak{so}(m-1).
\end{eqnarray*}

Adapting for general cosmological metrics the results proven in \cite%
{vacaru10,anco06,vacaru15} for parameterizations related to geometric flows
of 4-d Lorentzian metrics, we formulate such possibilities for generating
solitonic hierarchies with explicit dependence on geometric flow parameter
and on a time like coordinate (and, for locally anisotropic cases, on space
like coordinates):

\begin{itemize}
\item The $0$ flows locally anisotropic cosmological spaces are convective
(travelling wave) maps ${\gamma }_{\tau }={\gamma }_{\mathbf{l}}$
distinguished as $\left( h{\gamma }\right) _{\tau }=\left( h{\gamma }\right)
_{h\mathbf{X}}$ and $\left( v{\gamma }\right) _{\tau }=\left( v{\gamma }%
\right) _{v\mathbf{X}}$. The classification of such maps depend on the type
of cosmological d-metrics and d-connection structures.

\item There are +1 flows defined as non--stretching mKdV maps describing
geometric cosmological flows
\begin{eqnarray*}
-\left( h{\gamma }\right) _{\tau } &=&\widehat{\mathbf{D}}_{h\mathbf{X}%
}^{2}\left( \tau ,h{\gamma }\right) _{h\mathbf{X}}+\frac{3}{2}|\widehat{%
\mathbf{D}}_{h\mathbf{X}}\left( \tau ,h{\gamma }\right) _{h\mathbf{X}}|_{h%
\mathbf{g}}^{2}~\left( h{\gamma }\right) _{h\mathbf{X}}, \\
-\left( v{\gamma }\right) _{\tau } &=&\widehat{\mathbf{D}}_{v\mathbf{X}%
}^{2}\left( \tau ,v{\gamma }\right) _{v\mathbf{X}}+\frac{3}{2}|\widehat{%
\mathbf{D}}_{v\mathbf{X}}\left( \tau ,v{\gamma }\right) _{v\mathbf{X}}|_{v%
\mathbf{g}}^{2}~\left( v{\gamma }\right) _{v\mathbf{X}},
\end{eqnarray*}%
and the +2,... flows as higher order analogs.

\item Finally, the -1 flows are defined by the kernels of families of
canonical recursion h--operator,
\begin{equation*}
h\widehat{{\mathfrak{R}}}(\tau )=\widehat{\mathbf{D}}_{h\mathbf{X}}(\tau
)\left( \widehat{\mathbf{D}}_{h\mathbf{X}}(\tau )+\widehat{\mathbf{D}}_{h%
\mathbf{X}}^{-1}(\tau )\left( \overrightarrow{v}\cdot \right)
\overrightarrow{v}\right) +\overrightarrow{v}\rfloor \widehat{\mathbf{D}}_{h%
\mathbf{X}}^{-1}(\tau )\left( \overrightarrow{v}\wedge \widehat{\mathbf{D}}%
_{h\mathbf{X}}(\tau )\right) ,
\end{equation*}%
and families of canonical recursion v--operator,
\begin{equation*}
v\widehat{{\mathfrak{R}}}(\tau )=\widehat{\mathbf{D}}_{v\mathbf{X}}(\tau
)\left( \widehat{\mathbf{D}}_{v\mathbf{X}}(\tau )+\widehat{\mathbf{D}}_{v%
\mathbf{X}}^{-1}(\tau )\left( \overleftarrow{v}\cdot \right) \overleftarrow{v%
}\right) +\overleftarrow{v}\rfloor \widehat{\mathbf{D}}_{v\mathbf{X}%
}^{-1}(\tau )\left( \overleftarrow{v}\wedge \widehat{\mathbf{D}}_{v\mathbf{X}%
}(\tau )\right) ,
\end{equation*}%
inducing non--stretching maps $\widehat{\mathbf{D}}_{h\mathbf{Y}}\left( \tau
,h{\gamma }\right) _{h\mathbf{X}}=0$ and $\widehat{\mathbf{D}}_{v\mathbf{Y}%
}\left( \tau ,v{\gamma }\right) _{v\mathbf{X}}=0$.
\end{itemize}

The families of canonical recursion d-operator $\widehat{{\mathfrak{R}}}%
(\tau )=(h\widehat{{\mathfrak{R}}}(\tau ),v\widehat{{\mathfrak{R}}}(\tau ))$
are respectively related to bi-Hamiltonian structures for families of
cosmological solitonic configurations. Such configurations are characterized
also by respective Carath\'{e}odory and/or Perelman thermodynamic values
with explicit dependence on a temperature parameter and a time like
coordinate.

\subsubsection{Examples of solitonic cosmological distributions and
nonlinear waves}

The geometric flow evolution of any cosmological d-metric on a Lorentz
manifold can be encoded into solitonic hierarchies. In this work, the
geometric cosmological flow evolution is described by exact and parametric
solutions of type $\mathbf{g}(\tau )=\mathbf{g}(\tau ,x^{i},t)=[h\mathbf{g}%
(\tau ,x^{i}),v\mathbf{g}(\tau ,x^{i},t)],$ with Killing symmetry on $%
\partial _{3}$ when in adapted coordinates the coefficients of such
d-metrics do not depend on a space like coordinate $y^{3}.$ In principle, it
is possible to construct more general classes of solutions with dependence
on all spacetime coordinates and a temperature like parameter but general
formulas and classification are technically cumbersome and we omit such
considerations. For certain cosmological configurations, we shall consider
d-metrics of type $\mathbf{g}(\tau )=\mathbf{g}(\tau ,t)=[h\mathbf{g}(\tau
),v\mathbf{g}(\tau ,t)],$ see Appendix \ref{appendixb}.

\paragraph{Cosmological solitonic waves: \newline
}

We can consider nonlinear waves $\ \iota =\iota (x^{1},x^{2},y^{4}=t)$ as
solutions of solitonic 3-d equations
\begin{eqnarray}
\partial _{11}^{2}\iota +\epsilon \partial _{4}(\partial _{2}\iota +6\iota
\partial _{4}\iota +\partial _{444}^{3}\iota )=0,\ \partial _{11}^{2}\iota
+\epsilon \partial _{2}(\partial _{4}\iota +6\iota \partial _{2}\iota
+\partial _{222}^{3}\iota ) &=&0,  \label{solitdistr} \\
\partial _{22}^{2}\iota +\epsilon \partial _{4}(\partial _{1}\iota +6\iota
\partial _{4}\iota +\partial _{444}^{3}\iota )=0,\partial _{22}^{2}\iota
+\epsilon \partial _{1}(\partial _{4}\iota +6\iota \partial _{1}\iota
+\partial _{111}^{3}\iota ) &=&0,  \notag \\
\partial _{11}^{2}\iota +\epsilon \partial _{1}(\partial _{2}\iota +6\iota
\partial _{1}\iota +\partial _{111}^{3}\iota )=0,\ \partial _{44}^{2}\iota
+\epsilon \partial _{2}(\partial _{1}\iota +6\iota \partial _{2}\iota
+\partial _{222}^{3}\iota ) &=&0,  \notag
\end{eqnarray}%
for $\epsilon =\pm 1.$ These equations and their solutions can be redefined
via frame/coordinate transforms for temperature-time cosmological generating
functions.

\paragraph{Generating nonlinear temperature-time solitonic waves: \newline
}

Geometric flows of cosmological metrics for a Lorentz manifold can be
characterized by 3-d solitonic waves with explicit dependence flow parameter
$\tau $ defined by functions $\iota (\tau ,u)$ as solutions of such
nonlinear PDEs:
\begin{equation}
\ \ \iota =\left\{
\begin{array}{ccc}
\iota (\tau ,x^{2},t) & \mbox{ as a solution of } & \partial _{\tau \tau
}^{2}\ \ \iota +\epsilon \partial _{4}[\partial _{2}\ \ \iota +6\ \ \iota
\partial _{4}\iota +\partial _{444}^{3}\iota ]=0; \\
\iota (x^{2},\tau ,t) & \mbox{ as a solution of } & \partial _{22}^{2}\
\iota +\epsilon \partial _{4}[\partial _{\tau }\ \iota +6\ \ \iota \partial
_{4}\iota +\partial _{444}^{3}\iota ]=0; \\
\iota (\tau ,x^{1},t) & \mbox{ as a solution of } & \partial _{\tau \tau
}^{2}\ \ \iota +\epsilon \partial _{4}[\partial _{1}\ \ \iota +6\ \ \iota
\partial _{4}\iota +\partial _{444}^{3}\iota ]=0; \\
\iota (x^{1},\tau ,t) & \mbox{ as a solution of } & \partial _{11}^{2}\
\iota +\epsilon \partial _{4}[\partial _{\tau }\ \iota +6\ \ \iota \partial
_{4}\iota +\partial _{444}^{3}\iota ]=0; \\
\iota (\tau ,t,x^{2}) & \mbox{ as a solution of } & \partial _{\tau \tau
}^{2}\ \ \iota +\epsilon \partial _{2}[\partial _{4}\ \ \iota +6\ \ \iota
\partial _{2}\iota +\partial _{222}^{3}\iota ]=0; \\
\ \iota (t,\tau ,x^{2}) & \mbox{ as a solution of } & \partial _{44}^{2}\ \
\iota +\epsilon \partial _{2}[\partial _{\tau }\ \ \iota +6\ \ \iota
\partial _{2}\iota +\partial _{222}^{3}\iota ]=0.%
\end{array}%
\right.  \label{swaves}
\end{equation}%
Applying general frame/coordinate transforms on respective solutions of
equations (\ref{swaves}), we construct cosmological solitonic waves
parameterized by functions labeled in the form $\ \iota =\iota (\tau
,x^{i}), $ $=\iota (\tau ,x^{1},t),$ or $=\iota (\tau ,x^{2},t).$

In a similar form, we can consider other types of cosmologic solitonic
configurations determined, for instance, by sine-Gordon and various types of
nonlinear wave configurations characterized by geometric curve flows, see
details in \cite{vacaru10,anco06,vacaru15} and references therein. Any
solitonic hierarchy configuration with nonlinear waves of type $\iota (\tau
,u)$ (\ref{swaves}) or $\iota =\iota (x^{i},t)$ (\ref{solitdistr}) \ can be
can be used as generating functions for certain classes of nonholonomic
deformations of cosmological metrics. In this work, d-metrics of type $%
\mathbf{g}(\tau )$ (\ref{decomp31}) and/or (\ref{dme}) are d-tensor
functionals of type
\begin{equation}
\mathbf{g}(\tau )=\mathbf{g[}\iota (\tau ,u\mathbf{)]=g[}\iota \mathbf{]=}%
(g_{i}[\iota ],g_{a}[\iota ])  \label{solitondm}
\end{equation}%
with polarization functions $\eta _{i}(\tau )=\eta _{i}(\tau ,x^{k})=\eta
_{i}[\iota ],\eta _{a}(\tau )=\eta _{a}(\tau ,x^{k},y^{b})=\eta _{a}[\iota ]$
and $\eta _{i}^{a}(\tau )=\eta _{i}^{a}(\tau ,x^{k},y^{b})=\eta
_{i}^{a}[\iota ].$ A functional dependence $\mathbf{[}\iota \mathbf{]}$ can
be considered for multiple solitonic hierarchies with mixing (for instance,
on some different solutions of equations of type (\ref{swaves}) and/or (\ref%
{solitdistr})). This can be written conventionally in the form $[\iota ]=[\
_{1}\iota ,\ _{2}\iota ,...]$ where the left label states the type of
solitonic hierarchies. We shall construct in explicit form such cosmological
solutions for geometric flows of NES in next section.

\subsubsection{Table 1 with ansatz for cosmological geometric flows and
solitonic hierarchies}

In this work, we use brief notations of partial derivatives $\partial
_{\alpha }q=\partial q/\partial u^{\alpha }$ of a function $q(x^{k},y^{a}),$
for instance, $\partial _{1}q=q^{\bullet }=\partial q/\partial
x^{1},\partial _{2}q=q^{\prime }=\partial q/\partial x^{2},\partial
_{3}q=\partial q/\partial y^{3}=\partial q/\partial \varphi =q^{\diamond
},\partial _{4}q=\partial q/\partial t=\partial _{t}q=q^{\ast }.$Second
order derivatives are written in the form $\partial _{33}^{2}q=q^{\diamond
\diamond },\partial _{44}^{2}=\partial ^{2}q/\partial t^{2}=\partial
_{tt}^{2}q=q^{\ast \ast }.$ Partial derivatives on a flow parameter will be
written in the form $\partial _{\tau }=\partial /\partial \tau .$ The $\tau $%
-evolution of any d--metric $\mathbf{g}(\tau )$ of type (\ref{decomp31}), (%
\ref{dme}) and (\ref{solitondm}) \ can be parameterized (using respective
frame transforms) for respective local coordinates $(x^{k},y^{4}=t)$ and a
common geometric flow evolution and/or curve flows temperature like
parameter $\tau ,$
\begin{eqnarray}
g_{i}(\tau ) &=&e^{\psi {(\tau ,x}^{i}{)}}=e^{\psi \lbrack \ _{1}\iota
]},\,\,g_{a}(\tau )=\omega ({\tau ,x}^{i},y^{b})h_{a}({\tau ,x}%
^{i},t)=\omega \lbrack \ _{3}\iota ]h_{a}[\ _{2}\iota ],  \notag \\
\,\,N_{i}^{3}(\tau ) &=&n_{i}({\tau ,x}^{i},t)=n_{i}[\ _{4}\iota ,\
_{2}\iota ],\ N_{i}^{4}(\tau )=w_{i}({\tau ,x}^{i},t)=w_{i}[\ _{2}\iota ]\,.
\label{cosmf}
\end{eqnarray}%
For simplicity, we can consider $\omega =1$ for a large class of
cosmological models with at least one Killing symmetry, for instance, on $%
\partial _{3}.$

We can introduce effective sources for geometric flows of NES (\ref%
{canhamiltevol}) which by corresponding nonholonomic frame transforms and
tetradic (vierbein) fields are parameterized in N--adapted form,
\begin{equation*}
\ ^{eff}\widehat{\mathbf{\Upsilon }}_{\mu \nu }(\tau )=\mathbf{e}_{\ \mu
}^{\mu ^{\prime }}(\tau )\mathbf{e}_{\nu }^{\ \nu ^{\prime }}(\tau )[~\
\widehat{\mathbf{\Upsilon }}_{\mu ^{\prime }\nu ^{\prime }}(\tau )+\frac{1}{2%
}~\partial _{\tau }\mathbf{g}_{\mu ^{\prime }\nu ^{\prime }}(\tau )]=[~\ _{h}%
\widehat{\mathbf{\Upsilon }}(\tau ,{x}^{k})\delta _{j}^{i},\widehat{\mathbf{%
\Upsilon }}(\tau ,x^{k},y^{c})\delta _{b}^{a}].
\end{equation*}%
The values $\ _{h}\widehat{\mathbf{\Upsilon }}(\tau )$ and $\widehat{\mathbf{%
\Upsilon }}(\tau )$\ can be taken as functionals of certain solutions of
nonlinear solitonic equations and then considered as generating data for
(effective) matter sources and certain forms compatible with solitonic
hierarchies for d-metrics (\ref{cosmf}). We write
\begin{equation}
\widehat{\Im }[\iota ]=\ \ ^{eff}\widehat{\mathbf{\Upsilon }}_{\ \nu }^{\mu
}(\tau )=[~\ _{h}\widehat{\Im }[\ _{1}\iota ]=~\ _{h}\widehat{\mathbf{%
\Upsilon }}(\tau ,{x}^{i})\delta _{j}^{i},~\ _{v}\widehat{\Im }[\ _{2}\iota
]=\widehat{\mathbf{\Upsilon }}(\tau ,{x}^{i},t)\delta _{b}^{a}].
\label{dsourcparam}
\end{equation}%
There are used "hat" symbols in order to emphasize that such values are
considered for systems of nonlinear PDEs involving a canonical d-connection.

We can work using canonical nonholonomic variables with functional
dependence of d-metrics and prescribed effective sources on some
cosmological solitonic hierarchies. In such cases, the system of
nonholonomic entropic R. Hamilton equations (\ref{canhamiltevol}) can be
written in a formal nonholonomic Ricci soliton form (equivalently, as a
nonholonomic deformation of Einstein equations when the geometric objects
depend additionally on a temperature like parameter $\tau $ and for
effective source (\ref{dsourcparam})),%
\begin{equation}
\widehat{\mathbf{R}}_{\alpha \beta }[\iota ]=\widehat{\Im }_{\alpha \beta
}[\iota ].  \label{solitonhierarcheq}
\end{equation}

Table 1 (see below) summarize the geometric data on nonholonomic 2+2
variables and corresponding ansatz which allow us to transform relativistic
geometric flow equations and/or nonholonomic Ricci solitons into respective
systems of nonlinear ordinary differential equations, ODEs, and partial
differential equations, PDEs, determined by cosmological solitonic
hierarchies.


{\scriptsize
\begin{eqnarray*}
&&%
\begin{tabular}{l}
\hline\hline
\begin{tabular}{lll}
& {\ \textsf{Table 1:\ Geometric cosmological solitonic flows and modified
Einstein eqs as systems of nonlinear PDEs}} &  \\
& and the Anholonomic Frame Deformation Method, \textbf{AFDM}, &  \\
& \textit{for constructing generic off-diagonal exact and/or parametric
cosmological solutions} &
\end{tabular}%
\end{tabular}
\\
&&{%
\begin{tabular}{lll}
\hline
diagonal ansatz: PDEs $\rightarrow $ \textbf{ODE}s &  & AFDM: \textbf{PDE}s
\textbf{with decoupling; \ generating functions} \\
radial coordinates $u^{\alpha }=(r,\theta ,\varphi ,t)$ & $u=(x,y):$ &
\mbox{  2+2 splitting, } $u^{\alpha }=(x^{1},x^{2},y^{3},y^{4}=t);$%
\mbox{  flow parameter  }$\tau $ \\
LC-connection $\mathring{\nabla}$ & [connections] & $%
\begin{array}{c}
\mathbf{N}:T\mathbf{V}=hT\mathbf{V}\oplus vT\mathbf{V,}\mbox{ locally }%
\mathbf{N}=\{N_{i}^{a}(x,y)\} \\
\mbox{ canonical connection distortion }\widehat{\mathbf{D}}=\nabla +%
\widehat{\mathbf{Z}}%
\end{array}%
$ \\
$%
\begin{array}{c}
\mbox{ diagonal ansatz  }g_{\alpha \beta }(u) \\
=\left(
\begin{array}{cccc}
\mathring{g}_{1} &  &  &  \\
& \mathring{g}_{2} &  &  \\
&  & \mathring{g}_{3} &  \\
&  &  & \mathring{g}_{4}%
\end{array}%
\right)%
\end{array}%
$ & $\mathbf{\mathring{g}}\Leftrightarrow \mathbf{g}(\tau )$ & $%
\begin{array}{c}
g_{\alpha \beta }(\tau )=%
\begin{array}{c}
g_{\alpha \beta }(\tau ,x^{i},y^{a})\mbox{ general frames / coordinates} \\
\left[
\begin{array}{cc}
g_{ij}+N_{i}^{a}N_{j}^{b}h_{ab} & N_{i}^{b}h_{cb} \\
N_{j}^{a}h_{ab} & h_{ac}%
\end{array}%
\right] ,\mbox{ 2 x 2 blocks }%
\end{array}
\\
\mathbf{g}_{\alpha \beta }(\tau )=[g_{ij}(\tau ),h_{ab}(\tau )], \\
\mathbf{g}(\tau )=\mathbf{g}_{i}(\tau ,x^{k})dx^{i}\otimes dx^{i}+\mathbf{g}%
_{a}(\tau ,x^{k},y^{b})\mathbf{e}^{a}\otimes \mathbf{e}^{b}%
\end{array}%
$ \\
$\mathring{g}_{\alpha \beta }=\mathring{g}_{\alpha }(t)\mbox{ for FLRW} $ &
[coord.frames] & $g_{\alpha \beta }(\tau )=g_{\alpha \beta }(\tau
,x^{i},y^{4}=t)\mbox{ cosmological configurations}$ \\
$%
\begin{array}{c}
\mbox{coord. transforms }e_{\alpha }=e_{\ \alpha }^{\alpha ^{\prime
}}\partial _{\alpha ^{\prime }}, \\
e^{\beta }=e_{\beta ^{\prime }}^{\ \beta }du^{\beta ^{\prime }},\mathring{g}%
_{\alpha \beta }=\mathring{g}_{\alpha ^{\prime }\beta ^{\prime }}e_{\ \alpha
}^{\alpha ^{\prime }}e_{\ \beta }^{\beta ^{\prime }} \\
\begin{array}{c}
\mathbf{\mathring{g}}_{\alpha }(x^{k},y^{a})\rightarrow \mathring{g}_{\alpha }(t), \\
\mathring{N}_{i}^{a}(x^{k},y^{a})\rightarrow 0.%
\end{array}%
\end{array}%
$ & [N-adapt. fr.] & $\left\{
\begin{array}{cc}
\mathbf{g}_{i}(\tau , x^i)=\mathbf{g}_{i}[\ _{1}\iota ],\mathbf{g}%
_{a}(\tau ,x^i,t)=\mathbf{g}_{a}[\ _{2}\iota ], &
\mbox{
d-metrics } \\
N_{i}^{4}(\tau )=w_{i}[\ _{2}\iota ],N_{i}^{3}=n_{i}[\ _{4}\iota ,\
_{2}\iota ], & \mbox{N-connections}%
\end{array}%
\right. $ \\
$\mathring{\nabla},$ $Ric=\{\mathring{R}_{\ \beta \gamma }\}$ & Ricci tensors
& $\widehat{\mathbf{D}},\ \widehat{\mathcal{R}}ic=\{\widehat{\mathbf{R}}_{\
\beta \gamma }\}$ \\
$~^{m}\mathcal{L[\mathbf{\phi }]\rightarrow }\ ^{m}\mathbf{T}_{\alpha \beta }%
\mathcal{[\mathbf{\phi }]}$ & sources & $%
\begin{array}{cc}
\widehat{\Im }[\iota ]=\widehat{\mathbf{\Upsilon }}_{\ \nu }^{\mu }(\tau )=%
\mathbf{e}_{\ \mu ^{\prime }}^{\mu }\mathbf{e}_{\nu }^{\ \nu ^{\prime }}%
\widehat{\mathbf{\Upsilon }}_{\ \nu ^{\prime }}^{\mu ^{\prime }} &  \\
=[~\ _{h}\widehat{\Im }[\ _{1}\iota ]\delta _{j}^{i},\widehat{\Im }[\
_{2}\iota ]\delta _{b}^{a}], & \mbox{ cosmological conf.}%
\end{array}%
$ \\
trivial equations for $\mathring{\nabla}$-torsion & LC-conditions & $%
\widehat{\mathbf{D}}_{\mid \widehat{\mathcal{T}}\rightarrow 0}=\mathbf{%
\nabla }\mbox{
extracting new classes of solutions in GR}$ \\ \hline\hline
\end{tabular}%
}
\end{eqnarray*}%
}

In this paper, we study physically important cases when $\mathbf{\mathring{g}%
}$ defines a cosmological metric in GR. For diagonalizable via coordinate
transforms prime metrics, we can always find a coordinate system when $%
\mathring{N}_{i}^{b}=0.$ Non-singular nonholonomic deformations can be
constructed for cosmological solutions with nontrivial functions $\eta
_{\alpha }=(\eta _{i},\eta _{a}),\eta _{i}^{a},$ and nonzero coefficients $%
\mathring{N}_{i}^{b}(u)$. For a d-metric (\ref{dme}), we can analyze the
conditions of existence and properties of some target and/or prime
cosmological solutions with solitonic waves and structure formation when,
for instance, $\eta _{\alpha }\rightarrow 1$ and $N_{i}^{a}\rightarrow
\mathring{N}_{i}^{a}$. The values $\eta _{\alpha }=1$ and/or $\mathring{N}%
_{i}^{a}=0$ can be imposed as some special nonholonomic constraints on
temperature-time cosmological flows.

\subsection{Decoupling of geometric flow equations into cosmological
solitonic hierarchies}

In this subsection, we prove that the system of nonlinear PDEs (\ref%
{solitonhierarcheq}) describing geometric flow evolution of cosmological NES
encoding solitonic hierarchies can be decoupled in general form.

\subsubsection{Canonical Ricci d-tensors for cosmological flows encoding
solitonic hierarchies}

\label{ssdecst}We can chose certain systems of reference/ coordinates when
coefficients of a d-metric (\ref{cosmf}) and derived geometric objects like
the canonical d-connection and corresponding curvature and torsion d-tensors
do not depend on $y^{4}=t$ with respect to a class of N-adapted frames.
Using a d-metric ansatz with $\omega =1$ and a source $\widehat{\Im }[\iota
]=[~\ _{h}\widehat{\Im }[\ _{1}\iota ],~\ _{v}\widehat{\Im }[\ _{2}\iota ]]$
(\ref{dsourcparam}) , we compute the coefficients of the Ricci d-tenso and
write the geometric flow modified Einstein equations (\ref{solitonhierarcheq}%
) in the form
\begin{eqnarray}
\widehat{\mathbf{R}}_{1}^{1}[\ \iota ] &=&\widehat{\mathbf{R}}_{2}^{2}[\
\iota ]=-\ _{h}\widehat{\Im }[\ _{1}\iota ]\mbox{ i.e.}\frac{g_{1}^{\bullet
}g_{2}^{\bullet }}{2g_{1}}+\frac{\left( g_{2}^{\bullet }\right) ^{2}}{2g_{2}}%
-g_{2}^{\bullet \bullet }+\frac{g_{1}^{\prime }g_{2}^{\prime }}{2g_{2}}+%
\frac{(g_{1}^{\prime })^{2}}{2g_{1}}-g_{1}^{\prime \prime }=-2g_{1}g_{2}\
_{h}\widehat{\Im };  \label{eq1a} \\
\widehat{\mathbf{R}}_{3}^{3}[\ \iota ] &=&\widehat{\mathbf{R}}_{4}^{4}[\
\iota ]=-~\ _{v}\widehat{\Im }[\ _{2}\iota ]\mbox{
i.e. }\frac{\left( h_{3}^{\ast }\right) ^{2}}{2h_{3}}+\frac{h_{3}^{\ast }\
h_{4}^{\ast }}{2h_{4}}-h_{3}^{\ast \ast }=-2h_{3}h_{4}~\ _{v}\widehat{\Im };
\label{eq2a} \\
\widehat{\mathbf{R}}_{3k}(\tau ) &=&\frac{h_{3}}{2h_{3}}n_{k}^{\ast \ast }+(%
\frac{3}{2}h_{3}^{\ast }-\frac{h_{3}}{h_{4}}h_{4}^{\ast })\frac{n_{k}^{\ast }%
}{2h_{4}}=0;  \label{eq3a} \\
\widehat{\mathbf{R}}_{4k}(\tau ) &=&-w_{k}\left[ \left( \frac{h_{3}^{\ast }}{%
2h_{3}}\right) ^{2}+\frac{h_{3}^{\ast }\ }{2h_{3}}\frac{h_{4}^{\ast }}{2h_{4}%
}-\frac{h_{3}^{\ast \ast }}{2h_{3}}\right] +\frac{h_{3}^{\ast }}{2h_{3}}(%
\frac{\partial _{k}h_{3}}{2h_{3}}+\frac{\partial _{k}h_{4}}{2h_{4}})-\frac{%
\partial _{k}h_{3}^{\ast }}{2h_{3}}=0.  \label{eq4a}
\end{eqnarray}%
Let us explain the decoupling property of this system of nonlinear PDEs:
From equation (\ref{eq1a}), we can find $g_{1}$ (or, inversely, $g_{2}$) for
any prescribed functional of solitonic hierarchies encoded into a h-source $%
\ _{h}\widehat{\Im }[\ _{1}\iota ]$ and any given coefficient $g_{2}(\tau
,x^{i})=g_{2}[\ \iota ]$ (or, inversely, $g_{1}(\tau ,x^{i})=g_{1}[\ \iota ]$%
) when the solitonic hierarchies for the coefficients of a h-metric in such
coordinates depend on a temperature like parameter but not on time like
coordinates. We can integrate on time like coordinate $y^{4}=t$ in (\ref%
{eq2a}) and define $h_{4}(\tau ,x^{i},t)$ as a solution of first order PDE
for any prescribed v-source $~\ _{v}\widehat{\Im }[\ _{2}\iota ]$ and given
coefficient $h_{3}(\tau ,{x}^{i},t)=h_{3}[\ _{2}\iota ].$ Inversely, it is
possible to define $h_{3}(\tau ,{x}^{i},t)$ if $h_{4}(\tau ,{x}%
^{i},t)=h_{4}[\ _{2}\iota ]$ is given for such solutions we have to solve a
second order PDE. The coefficients of v-metrics involve, in general,
different types of solitonic hierarchies even they are related via
corresponding formulas to another classes of solitonic hierarchies
prescribed for the effective v-source. We have to integrate two times on $t$
in (\ref{eq3a}) in order to compute $n_{k}(\tau ,{x}^{i},t)=n_{k}[\ \iota ]$
for any defined $h_{3}$ and $h_{4}$. Introducing certain values of $h_{3}$
and $h_{4}$ in equations (\ref{eq4a}), we obtain a system of algebraic
linear equations for $w_{k}(\tau ,{x}^{i},t)=w_{k}[\ \iota ].$ Here it
should be also emphasized that the cosmological solitonic hierarchies
encoded in the coefficients of a N-connection are different (in general)
from those encoded in the coefficients of d-metric and nontrivial effective
sources.

Using the decoupling property of nonlinear off-diagoan cosmological
solitonic systems (\ref{eq1a})--(\ref{eq4a}), we can integrate such PDEs
step by step by prescribing respectively the effective sources, the
h-coefficients, $g_{i},$ and v--coefficients, $h_{a},$ for geometric flowd
of d-metrics. The geometric evolution of such solutions involves a
prescribed nonholonomic constraint on $~\partial _{\tau }\mathbf{g}_{\mu
^{\prime }\nu ^{\prime }}(\tau )$ included in $\widehat{\Im }[\iota ].$

\subsubsection{Nonlinear symmetries for cosmological solitonic generating
functions and sources}

Let us define the the coefficients $\alpha _{i}=h_{4}^{\ast }\partial
_{i}\varpi ,\ \beta =h_{4}^{\ast }\varpi ^{\ast },\ \gamma =\left( \ln
|h_{3}|^{3/2}/|h_{4}|\right) ^{\ast },$ where
\begin{equation}
\varpi {=\ln |h_{3}^{\ast }/\sqrt{|h_{3}h_{4}|}|}  \label{aux02}
\end{equation}%
for nonsingular values for $h_{a}^{\ast }\neq 0$ and $\partial _{t}\varpi
\neq 0$ (we have to elaborate other methods if such conditions are not
satisfied), we transform the system of solitonic nonlinear PDEs (\ref{eq1a}%
)--(\ref{eq4a}) into
\begin{equation}
\psi ^{\bullet \bullet }+\psi ^{\prime \prime }=2~\ \ \ _{h}\widehat{\Im }[\
_{1}\iota ],\quad \varpi ^{\ast }\ h_{3}^{\ast }=2h_{3}h_{4}\ \ _{v}\widehat{%
\Im }[\ _{2}\iota ],n_{k}^{\ast \ast }+\gamma n_{k}^{\ast }=0,\ \beta
w_{i}-\alpha _{i}=0.  \label{estatsimpl}
\end{equation}%
Such a system can be integrated in explicit and general forms (depending on
the type of parameterizations, \ see details in \cite{vacaru10,vacaru13} and
some examples of cosmological solutions will be provided in next sections)
if there are prescribed a generating function $\Psi (\tau )=\Psi (\tau
,x^{i},t)=\Psi \lbrack \ \iota ]:=e^{\varpi }$ and generating sources $\ _{h}%
\widehat{\Im }$ and $\ \ _{v}\widehat{\Im }.$ Here, we note that
Levi-Civita, LC, conditions for extracting cosmological solitonic solution
with zero torsion, can be transformed into a system of 1st order PDEs,
\begin{equation}
(\partial _{i}-w_{i}\partial _{t})\ln \sqrt{|h_{3}|}=0,\partial
_{t}w_{i}=(\partial _{i}-w_{i}\partial _{t})\ln \sqrt{|h_{4}|},\partial
_{t}n_{i}=0,\partial _{i}n_{k}=\partial _{k}n_{i},\partial
_{k}w_{i}=\partial _{i}w_{k},  \label{lccond}
\end{equation}%
which are considered as additional constraints on off-diagonal coefficients
of metrics of type (\ref{cosmf}).

To generate exact and parametric solutions we have to solve a system of two
equations for $\varpi $ in (\ref{aux02}) and (\ref{estatsimpl}) involving
four functions ($h_{3},h_{4},\ _{v}\widehat{\Im },$ and $\Psi ).$ We can
check by corresponding computations that there is an important nonlinear
symmetry which allows to redefine the generating function and the effective
source and to introduce a family of effective cosmological constants $%
\Lambda (\tau )\neq 0,\Lambda (\tau _{0})=const,$ not depending on spacetime
coordinates $u^{\alpha }.$ Such nonlinear transforms $(\Psi (\tau ),\ _{v}%
\widehat{\Im }(\tau ))\iff (\Phi (\tau ),\Lambda (\tau ))$ are defined by
formulas
\begin{equation}
\Lambda (\ \Psi ^{2}[\ _{1}\iota ])^{\ast }=|\ _{v}\widehat{\Im }[\ \
_{2}\iota ]|(\Phi ^{2}[\ \iota ])^{\ast },\mbox{
or  }\Lambda \ \Psi ^{2}[\ _{1}\iota ]=\Phi ^{2}[\ \iota ]|\ _{v}\widehat{%
\Im }[\ \ _{2}\iota ]|-\int dt\ \Phi ^{2}[\ \iota ]|\ _{v}\widehat{\Im }[\
_{2}\iota ]|^{\ast }  \label{nsym1a}
\end{equation}%
and allow us to introduce families of new generating functions $\Phi (\tau
,x^{i},t)=\Phi \lbrack \ \iota ]$ and families of (effective) cosmological
constants$.$ The families of constants $\Lambda (\tau )$ can be chosen for
certain physical models and the geometric/physical data for $\Phi $ encode
nonlinear symmetries both for the generating functions and sources and
respective cosmological solitonic hierarchies for $_{v}\widehat{\Im },$ and $%
\Psi .$ In result of nonlinear symmetries, we can describe nonlinear systems
of PDEs by two equivalent sets of generating data $(\Psi ,\Upsilon )$ or $%
(\Phi ,\Lambda ).$ But such symmetries of solitonic cosmological hierarchies
are encoded into functionals with respective partial derivations $\partial
_{t}$ and/or integration on $dt.$ To generate certain classes of solutions,
we can work with effective cosmological constants but for other ones we have
to consider generating sources. Finally, we note that modules in formulas (%
\ref{nsym1a}) should be chosen in certain forms resulting in physically
motivated nonlinear symmetries, relativistic causal models and thermodynamic
values which are compatible with observational data in modern cosmology.

\subsection{Integrability of geometric cosmological flow equations with
solitonic hierarchies}

We study properties of some classes of generic off-diagonal cosmological
solutions of (\ref{solitonhierarcheq}) determined by generated functions and
sources with solitonic hierarchies.

\subsubsection{Cosmological solutions for off-diagonal metrics and
N--coefficients}

\label{ssintst}By straightforward computations we can verify that the system
(\ref{eq1a})--(\ref{eq4a}) \ represented in the form (\ref{estatsimpl}) (see
similar details in \cite{gheorghiu14, bubuianu18}) can be solved by if the
coefficients of a d--metric and N--connection are computed {\small
\begin{eqnarray}
\ g_{i}(\tau ) &=&e^{\ \psi (\tau ,x^{k})}%
\mbox{ as a solution of 2-d
Poisson eqs. }\psi ^{\bullet \bullet }+\psi ^{\prime \prime }=2~~\ _{h}%
\widehat{\Im }[\ _{1}\iota ];  \notag \\
g_{3}[\ _{3}\iota ] &=&h_{3}(\tau ,{x}^{i},t)=h_{3}^{[0]}(\tau ,x^{k})-\int
dt\frac{(\Psi ^{2})^{\ast }}{4\ _{v}\widehat{\Im }}=h_{3}^{[0]}(\tau
,x^{k})-\Phi ^{2}/4\Lambda (\tau );  \label{offcosm1} \\
g_{4}[\ _{4}\iota ] &=&h_{4}(\tau ,{x}^{i},t)=-\frac{(\Psi ^{\ast }[\iota
])^{2}}{4(\ _{v}\widehat{\Im }[\ _{2}\iota ])^{2}h_{3}[\ _{3}\iota ]}=-\frac{%
(\Psi ^{\ast })^{2}}{4(\ _{v}\widehat{\Im })^{2}\left( h_{3}^{[0]}(\tau
,x^{k})-\int dt(\Psi ^{2})^{\ast }/4\ _{v}\widehat{\Im }\right) }  \notag \\
&=&-\frac{(\Phi ^{2})(\Phi ^{2})^{\ast }}{h_{3}|\Lambda (\tau )\int dt\ _{v}%
\widehat{\Im }[\Phi ^{2}]^{\ast }|}=-\frac{[(\Phi ^{2})^{\ast }]^{2}}{%
4[h_{3}^{[0]}(\tau ,x^{k})-\Phi ^{2}/4\Lambda (\tau )]|\int dt\ \ _{v}%
\widehat{\Im }(\Phi ^{2})^{\ast }|};  \notag
\end{eqnarray}%
\begin{eqnarray*}
N_{k}^{3}[\ _{5}\iota ] &=&n_{k}(\tau ,{x}^{i},t)=\ _{1}n_{k}(\tau ,x^{i})+\
_{2}n_{k}(\tau ,x^{i})\int dt\frac{(\Psi ^{\ast })^{2}}{\ _{v}\widehat{\Im }%
^{2}|h_{3}^{[0]}(\tau ,x^{i})-\int dt(\Psi ^{2})^{\ast }/4\ _{v}\widehat{\Im
}|^{5/2}} \\
&=&\ _{1}n_{k}(\tau ,x^{i})+\ _{2}n_{k}(\tau ,x^{i})\int dt\frac{(\Phi
^{\ast })^{2}}{4|\Lambda (\tau )\int dt\ _{v}\widehat{\Im }(\Phi ^{2})^{\ast
}||h_{3}|^{5/2}}; \\
\ N_{i}^{4}[\iota ] &=&w_{i}(\tau ,{x}^{i},t)=\frac{\partial _{i}\ \Psi }{%
\Psi ^{\ast }}\ =\frac{\partial _{i}\ \Psi ^{2}}{(\Psi ^{2})^{\ast }}\ =%
\frac{\partial _{i}[\int dt\ \ _{v}\widehat{\Im }(\Phi ^{2})^{\ast }]}{\ _{v}%
\widehat{\Im }(\Phi ^{2})^{\ast }}.
\end{eqnarray*}%
} In these formulas, there are considered different sets of solitonic
hierarchies and respective integration functions $h_{3}^{[0]}(\tau ,x^{k}),$
$\ _{1}n_{k}(\tau ,x^{i}),$ and $\ _{2}n_{k}(\tau ,x^{i})$ encoding (non)
commutative parameters and integration constants but also nonlinear
geometric flow scenarios on $\tau $ and cosmological evolution. These data
and symmetries of solitonic hierarchies for generating geometric evolution
data $(\Psi ,\Upsilon ),$ or $(\Phi ,\Lambda ),$ (all related by nonlinear
differential / integral transforms (\ref{nsym1a})) can be prescribed in
explicit form following certain topology/ symmetry / asymptotic conditions
and compatibility with observational data. The coefficients (\ref{offcosm1})
define generic off-diagonal cosmological solitonic solutions with associated
bi Hamilton structures if the corresponding anholonomy coefficients are not
trivial. In general, such geometric flow cosmological solutions are with
nontrivial nonholonomically induced d-torsions and solitonic hierarchies
determined by evolution of N-adapted coefficients of d-metric structures. We
can impose additional nonholonomic constraints (\ref{lccond}) in order to
extract LC-configurations for cosmological metrics under geometric flow
evolution.

\subsubsection{Quadratic line elements for off-diagonal cosmological
solitonic hierarchies}

Instead of $\Psi $ and/or $\Phi ,$ we can consider as a generating function
any coefficient $h_{3}[\ _{3}\iota ]=h_{3}^{[0]}-\Phi ^{2}/4\Lambda $,
\newline
$h_{3}^{\ast }(\tau )\neq 0,$ and write formulas
\begin{equation*}
\Phi ^{2}(\tau )=4\Lambda \left( h_{3}[\ _{3}\iota ]-h_{3}^{[0]}\right)
,(\Phi ^{2})^{\ast }=4\Lambda (h_{3})^{\ast }\mbox{ and }(\Phi ^{\ast
})^{2}=\Lambda (h_{3})^{\ast }(\frac{h_{3}}{h_{3}^{[0]}}-1).
\end{equation*}%
Using nonlinear symmetries (\ref{nsym1a}), we find $\ $%
\begin{equation*}
(\Psi ^{2})^{\ast }=4\ \left\vert \ _{v}\widehat{\Im }[\ _{2}\iota
]\right\vert (h_{3})^{\ast }\mbox{ and }\Psi ^{2}=4\left\vert \ _{v}\widehat{%
\Im }\right\vert h_{3}-4\int dt\left\vert \ _{v}\widehat{\Im }\right\vert
^{\ast }h_{3}.
\end{equation*}
Such formulas determine corresponding functionals $\Psi \lbrack \ _{v}%
\widehat{\Im },h_{3},h_{3}^{[0]}]$ and $\Phi \lbrack \Lambda
,h_{3},h_{3}^{[0]}].$ Introducing these values into respective formulas for $%
h_{a},N_{i}^{b}$ and $\ _{v}\widehat{\Im }$ in (\ref{offcosm1}) and
expressing the generating functions and the \ d--metric (\ref{dme}) with
cosmological data (\ref{cosmf}) in terms of $h_{3},$ for respective
integration functions and effective sources for geometric evolution, we
compute {\small
\begin{eqnarray*}
g_{3}[\ _{3}\iota ] &=&h_{3}(\tau ,{x}^{i},t)=h_{3}^{[0]}(\tau ,x^{k})-\int
dt\frac{(\Psi ^{2})^{\ast }}{4\ _{v}\widehat{\Im }}=h_{3}^{[0]}(\tau
,x^{k})-\Phi ^{2}/4\Lambda (\tau ); \\
g_{4}[\ _{4}\iota ] &=&h_{4}(\tau ,{x}^{i},t)=-\frac{(\Phi ^{2})(\Phi
^{2})^{\ast }}{h_{3}|\Lambda (\tau )\int dt\ _{v}\widehat{\Im }(\Phi
^{2})^{\ast }|}=\frac{4|(h_{3})^{\ast }|}{|\int dt\ _{v}\widehat{\Im }%
(h_{3})^{\ast }|}; \\
\ \ N_{k}^{3}[\ _{5}\iota ] &=&n_{k}(\tau ,{x}^{i},t)=\ _{1}n_{k}(\tau
,x^{i})+\ _{2}n_{k}(\tau ,x^{i})\int dt\frac{(\Phi ^{\ast })^{2}}{4|\Lambda
(\tau )\int dt\ _{v}\widehat{\Im }(\Phi ^{2})^{\ast }||h_{3}|^{5/2}}= \\
&=&\ _{1}n_{k}(\tau ,x^{i})+\ _{2}\widetilde{n}_{k}(\tau ,x^{i})\int dt\frac{%
(h_{3})^{\ast }(1-h_{3}/h_{3}^{[0]})}{|\Lambda \int dt\ _{v}\widehat{\Im }%
(h_{3})^{\ast }||h_{3}|^{5/2}}; \\
\ N_{i}^{4}[\iota ] &=&w_{i}(\tau ,{x}^{i},t)=\frac{\partial _{i}\ \Psi }{%
(\Psi )^{\ast }}\ =\frac{\partial _{i}\ \Psi ^{2}}{(\Psi ^{2})^{\ast }}\ =%
\frac{\partial _{i}[\int dt\ \ _{v}\widehat{\Im }(\Phi ^{2})^{\ast }]}{\ _{v}%
\widehat{\Im }(\Phi ^{2})^{\ast }}=\frac{\partial _{i}\left( \left\vert \
_{v}\widehat{\Im }\right\vert h_{3}-\int dt\left\vert \ _{v}\widehat{\Im }%
\right\vert ^{\ast }h_{3}\right) }{\ \left\vert \ _{v}\widehat{\Im }[\
_{2}\iota ]\right\vert h_{3}^{\ast }}.
\end{eqnarray*}%
}

In result, we can express the quadratic line element corresponding to this
class of cosmological flow solutions in three equivalent forms: {\small
\begin{eqnarray}
ds^{2} &=&e^{\ \psi (\tau ,x^{k})}[(dx^{1})^{2}+(dx^{2})^{2}]
\label{gencosm1} \\
&&\left\{
\begin{array}{cc}
\begin{array}{c}
-h_{3}[dy^{3}+(\ \ _{1}n_{k}(\tau ,x^{i})+\ _{2}\widetilde{n}_{k}(\tau
,x^{i})\int dt\frac{(h_{3})^{\ast }(1-h_{3}/h_{3}^{[0]})}{|\Lambda \int dt\
_{v}\widehat{\Im }(h_{3})^{\ast }||h_{3}|^{5/2}})dx^{k}] \\
-\frac{4|(h_{3})^{\ast }|}{|\int dt\ _{v}\widehat{\Im }(h_{3})^{\ast }|}[dt+%
\frac{\partial _{i}\left( \left\vert \ _{v}\widehat{\Im }\right\vert
h_{3}-\int dt\left\vert \ _{v}\widehat{\Im }\right\vert ^{\ast }h_{3}\right)
}{\ \left\vert \ _{v}\widehat{\Im }[\ _{2}\iota ]\right\vert h_{3}^{\ast }}%
dx^{i}], \\
\mbox{ or }%
\end{array}
&
\begin{array}{c}
\mbox{gener.  funct.}h_{3}, \\
\mbox{ source }_{v}\widehat{\Im },\mbox{ or }\Lambda ;%
\end{array}
\\
\begin{array}{c}
-[dy^{3}+(_{1}n_{k}+\ _{2}n_{k}\int dt\frac{(\Psi ^{\ast })^{2}}{4(\ _{v}%
\widehat{\Im })^{2}|h_{3}^{[0]}-\int dt\frac{(\Psi ^{2})^{\ast }}{4\ \ _{v}%
\widehat{\Im }}|^{5/2}})dx^{k}] \\
(h_{3}^{[0]}-\int dt\frac{(\Psi ^{2})^{\ast }}{4\ \ _{v}\widehat{\Im }})+%
\frac{(\Psi ^{2})^{\ast }}{4(\ _{v}\widehat{\Im })^{2}(h_{3}^{[0]}-\int dt%
\frac{(\Psi ^{2})^{\ast }}{4\ _{v}\widehat{\Im }})}[dt+\frac{\partial _{i}\
\Psi }{\ \partial _{3}\Psi }dx^{i}] \\
\mbox{ or }%
\end{array}
&
\begin{array}{c}
\mbox{gener.  funct.}\Psi , \\
\mbox{source }\ _{v}\widehat{\Im };%
\end{array}
\\
\begin{array}{c}
-[dy^{3}+(_{1}n_{k}+\ _{2}n_{k}\int dt\frac{[(\Phi ^{2})^{\ast }]^{2}}{|\
4\Lambda \int dt\ \ _{v}\widehat{\Im }[(\Phi )^{2}]^{\ast }|}|h_{3}^{[0]}-%
\frac{\Phi ^{2}}{4\Lambda }|^{-5/2})dx^{k}] \\
(h_{3}^{[0]}-\frac{\Phi ^{2}}{4\Lambda })-\frac{[(\Phi ^{2})^{\ast }]^{2}}{%
4|\Lambda \int dt\ \ _{v}\widehat{\Im }[(\Phi )^{2}]^{\ast }|\ (h_{3}^{[0]}-%
\frac{\Phi ^{2}}{4\Lambda })}[dt+\frac{\partial _{i}[\int dt\ \ _{v}\widehat{%
\Im }(\Phi ^{2})^{\ast }]}{\ \ \ _{v}\widehat{\Im }\ (\Phi ^{2})^{\ast }}%
dx^{i}],%
\end{array}
&
\begin{array}{c}
\mbox{gener.  funct.}\Phi \\
\mbox{effective }\Lambda \mbox{ for }\ _{v}\widehat{\Im }.%
\end{array}%
\end{array}%
\right.  \notag
\end{eqnarray}%
} Formulas (\ref{offcosm1}) \ and (\ref{gencosm1}) encode cosmological
solitonic hierarchies determined by generating functions. A generating
source $\ _{v}\widehat{\Im }$ and effective cosmological constant $\Lambda $
do not involve (in general) any solitonic behaviour. Nonlinear symmetries (%
\ref{nsym1a}) mix different cosmological solitonic structures of generating
functions and any cosmological functional for sources.

\subsubsection{Off-diagonal Levi-Civita cosmological solitonic hierarchies}

We can solve the equations (\ref{lccond}) for zero torsion conditions
considering special classes of generating functions and sources. For
instance, we prescribe a $\Psi (\tau )=\check{\Psi}(\tau ,x^{i},t)$ for
which $(\partial _{i}\check{\Psi})^{\ast }=\partial _{i}(\check{\Psi}^{\ast
})$ and fix a $\ _{v}\widehat{\Im }(\tau ,x^{i},t)=\ _{v}\widehat{\Im }[%
\check{\Psi}]=\ _{v}\check{\Im}(\tau ),$ or $\ _{v}\widehat{\Im }=const.$
The nonlinear symmetries (\ref{nsym1a}) transforms into
\begin{equation*}
\Lambda \ \check{\Psi}^{2}=\check{\Phi}^{2}|\ _{v}\widehat{\Im }|-\int dt\
\Phi ^{2}|\ _{v}\widehat{\Im }|^{\ast }\mbox{ and }\check{\Phi}%
^{2}=-4\Lambda \check{h}_{3}(\tau ,x^{i},t),\check{\Psi}^{2}=\int dt\ \ _{v}%
\widehat{\Im }\check{h}_{3}^{\ast }.
\end{equation*}%
In the second case, the coefficient $h_{3}(\tau )=\check{h}_{3}(\tau
,x^{i},t)$ can be considered also as generating function when $h_{4}$ and
the N-connection coefficients are computed using corresponding formulas an
certain nonlinear symmetries and nonholonomic constraints. To generate zero
torsion cosmologic solitonic hierarchies, we find some functions $\check{A}%
(\tau )=\check{A}(\tau ,x^{i},t)$ and $n(\tau )=n(\tau ,x^{i})$ when the
coefficients of N-connection are
\begin{equation*}
n_{k}(\tau )=\check{n}_{k}(\tau )=\partial _{k}n(\tau ,x^{i})\mbox{ and
}w_{i}(\tau )=\partial _{i}\check{A}(\tau )=\frac{\partial _{i}(\int dt\ _{v}%
\check{\Im}\ \check{h}_{3}^{\ast }{}])}{\ \ _{v}\check{\Im}\ \check{h}%
_{3}^{\ast }{}}=\frac{\partial _{i}\check{\Psi}}{\check{\Psi}^{\ast }}=\frac{%
\partial _{i}[\int dt\ \ \ _{v}\check{\Im}(\check{\Phi}^{2})^{\ast }]}{\ \
_{v}\check{\Im}(\check{\Phi}^{2})^{\ast }}.
\end{equation*}%
In result, the quadratic line elements for new classes of off-diagonal zero
torsion locally anisotropic cosmologic solutions encoding solitonic
hierarchies and defined as subclasses of solutions (\ref{gencosm1}), {\small
\begin{equation}
ds^{2}=e^{\ \psi (\tau ,x^{k})}[(dx^{1})^{2}+(dx^{2})^{2}]-\left\{
\begin{array}{cc}
\begin{array}{c}
\check{h}_{3}\left[ dy^{3}+(\partial _{k}n)dx^{k}\right] +\frac{(\check{h}%
_{3}^{\ast }{})^{2}}{|\int dt\ _{v}\check{\Im}\check{h}_{3}^{\ast }|\ \check{%
h}_{3}}[dt+(\partial _{i}\check{A})dx^{i}], \\
\mbox{ or }%
\end{array}
&
\begin{array}{c}
\mbox{gener.  funct.}\check{h}_{3}, \\
\mbox{ source }\ \ _{v}\check{\Im},\mbox{ or }\Lambda ;%
\end{array}
\\
\begin{array}{c}
(h_{3}^{[0]}-\int dt\frac{(\check{\Psi}^{2})^{\ast }}{4\ \ _{v}\check{\Im}})%
\left[ dy^{3}+(\partial _{k}n)dx^{k}\right] + \\
\frac{(\check{\Psi}^{2})^{\ast }}{4\ (\ _{v}\check{\Im})^{2}(h_{3}^{[0]}-%
\int dt\frac{(\check{\Psi}^{2})^{\ast }}{4\ \ _{v}\check{\Im}})}%
[dt+(\partial _{i}\check{A})dx^{i}] \\
\mbox{ or }%
\end{array}
&
\begin{array}{c}
\mbox{gener.  funct.}\check{\Psi}, \\
\mbox{source }\ \ _{v}\check{\Im};%
\end{array}
\\
\begin{array}{c}
(h_{3}^{[0]}-\frac{\check{\Phi}^{2}}{4\Lambda })\left[ dy^{3}+(\partial
_{k}n)dx^{k}\right] + \\
\frac{\lbrack (\check{\Phi}^{2})^{\ast }]^{2}}{4|\Lambda \int dt\ \ _{v}%
\check{\Im}(\check{\Phi}^{2})^{\ast }|\ (h_{3}^{[0]}-\frac{\check{\Phi}^{2}}{%
4\Lambda })}[dt+(\partial _{i}\check{A})dx^{i}],%
\end{array}
&
\begin{array}{c}
\mbox{gener.  funct.}\ \check{\Phi} \\
\mbox{effective }\Lambda \mbox{ for }\ \ _{v}\check{\Im}.%
\end{array}%
\end{array}%
\right.  \label{lcsolstat}
\end{equation}%
} For any value of flow parameter $\tau ,$ such cosmological solitonic
metrics are generic off-diagonal and define new classes of solutions which
are different, for instance, from the FLRW metric. We may check if the
anholonomy coefficients $C_{\alpha \beta }^{\gamma }=\{C_{ia}^{b}=\partial
_{a}N_{i}^{b},C_{ji}^{a}=\mathbf{e}_{j}N_{i}^{a}-\mathbf{e}_{i}N_{j}^{a}\}$
are not zero for solitonic values of $N_{i}^{3}=\partial _{i}\check{A}$ and $%
N_{k}^{4}=\partial _{k}n$ and conclude if certain metrics are or not generic
off-diagonal. We can \ study certain nonholonomic cosmologic solitonic
configurations determined, for instance, by data $(\ _{v}\check{\Im},\check{%
\Psi},h_{3}^{[0]},\check{n}_{k}),$ with $w_{i}=\partial _{i}\check{A}%
\rightarrow 0$ and $\partial _{k}n\rightarrow 0,$ see Appendix \ref%
{appendixb}.

\subsection{Table 2: AFDM for constructing cosmological solitonic flows}

We consider the coefficient $h_{3}(\tau )=h_{3}(\tau ,x^{i},t)$ in (\ref%
{gencosm1}) as a generating function. Such a value can be determined by a
family of solitonic hierarchies, $h_{3}(\tau )=h_{3}[\ _{3}\iota ]$ with
explicit dependence on a time like coordinate $t.$ We can perform a
deformation procedure for constructing a class of off--diagonal solutions with Killing symmetry on $\partial _{3}$ determined by cosmological
solitonic hierarchies $\widehat{\Im }[\iota ]=[~\ _{h}\widehat{\Im }[\
_{1}\iota ],~\ _{v}\widehat{\Im }[\ _{2}\iota ]]$ (\ref{dsourcparam}) and a
parametric running cosmological constant $\Lambda (\tau ),$
\begin{eqnarray*}
ds^{2}&=& e^{\ \psi (\tau ,x^{k})}[(dx^{1})^{2}+(dx^{2})^{2}] +h_{3}(\tau
)[dy^{3}+(\ _{1}n_{k}+4\ _{2}n_{k}\int dt\frac{(h_{3}^{\ast }{}(\tau ))^{2}}{%
|\int dt\ \ _{v}\widehat{\Im }(\tau )(h_{3}^{\ast }(\tau ))|\ (h_{3}(\tau
))^{5/2}})dx^{k}] \\
&&-\frac{[h_{3}^{\ast }{}(\tau )]^{2}}{|\int dt\ \ _{v}\widehat{\Im }(\tau
)(h_{3}{}^{\ast }(\tau ))|\ h_{3}}[dt+\frac{\partial _{i}(\int dt\ \ _{v}%
\widehat{\Im }(\tau )h_{3}^{\ast }{}])}{\ _{v}\widehat{\Im }(\tau )\
h_{3}^{\ast }{}(\tau )}dx^{i}].
\end{eqnarray*}

Such classes of cosmological solutions involve different types of solitonic
hierarchies and, in general, are with nontrivial nonholonomically induced
torsion (it is possible to impose nonholonomic constraints to
LC-configurations (\ref{lcsolstat})).


{\scriptsize
\begin{eqnarray*}
&&%
\begin{tabular}{l}
\hline\hline
\begin{tabular}{lll}
& {\large \textsf{Table 2:\ Off-diagonal cosmological flows with solitonic
hierarchies}} &  \\
& Exact solutions of $\widehat{\mathbf{R}}_{\mu \nu }(\tau )=\widehat{\Im }%
_{\mu \nu }(\tau )$ (\ref{solitonhierarcheq}) transformed into a system of
nonlinear PDEs (\ref{eq1a})-(\ref{eq4a}) &
\end{tabular}
\\
\end{tabular}
\\
&&%
\begin{tabular}{lll}
\hline\hline
$%
\begin{array}{c}
\mbox{d-metric ansatz with} \\
\mbox{Killing symmetry }\partial _{3}%
\end{array}%
$ &  & $%
\begin{array}{c}
ds^{2}=g_{i}(\tau )(dx^{i})^{2}+g_{a}(\tau )(dy^{a}+N_{i}^{a}(\tau
)dx^{i})^{2},\mbox{ for } \\
g_{i}=e^{\psi {(\tau ,x}^{i}{)}},\,\,\,\,g_{a}=h_{a}(\tau ,x^{i},t),\
\,N_{i}^{3}=n_{i}(\tau ,x^{i},t),N_{i}^{4}=w_{i}(\tau
,x^{i},t),y^{4}=t;\,\,\,%
\end{array}%
$ \\
Effective matter sources &  & $\widehat{\Im }_{\ \nu }^{\mu }(\tau )=[~\ _{h}%
\widehat{\Im }(\tau ,{x}^{i})\delta _{j}^{i},~\ _{v}\widehat{\Im }(\tau
,x^{i},t)\delta _{b}^{a}];\partial _{1}q=q^{\bullet },\partial
_{2}q=q^{\prime },\partial _{3}q=q^{\diamond },\partial _{4}q=q^{\ast }$ \\
\hline
Nonlinear PDEs (\ref{estatsimpl}) &  & $%
\begin{array}{c}
\psi ^{\bullet \bullet }+\psi ^{\prime \prime }=2~\ \ _{h}\widehat{\Im }[\
_{1}\iota ]; \\
\varpi ^{\ast }\ h_{3}^{\ast }=2h_{3}h_{4}~\ _{v}\widehat{\Im }[\ _{2}\iota
]; \\
n_{k}^{\ast \ast }+\gamma n_{k}^{\ast }=0; \\
\beta w_{i}-\alpha _{i}=0;%
\end{array}%
$ for $%
\begin{array}{c}
\varpi {=\ln |h_{3}^{\ast }/\sqrt{|h_{3}h_{4}|}|,} \\
\alpha _{i}=h_{3}^{\ast }\ (\partial _{i}\varpi ),\ \beta =h_{3}^{\ast }\
\varpi ^{\ast }, \\
\ \gamma =\left( \ln |h_{3}|^{3/2}/|h_{4}|\right) ^{\ast },%
\end{array}%
$ \\ \hline
$%
\begin{array}{c}
\mbox{ Generating functions:}\ h_{3}[\ _{3}\iota ], \\
\Psi (\tau ,x^{i},t)=e^{\varpi },\Phi \lbrack \iota ]; \\
\mbox{integration functions:}\ h_{3}^{[0]}(\tau ,x^{k}),\  \\
_{1}n_{k}(\tau ,x^{i}),\ _{2}n_{k}(\tau ,x^{i})%
\end{array}%
$ &  & $%
\begin{array}{c}
\ (\Psi ^{2})^{\ast }=-\int dt\ ~\ _{v}\widehat{\Im }h_{3}^{\ \ast },\Phi
^{2}=-4\Lambda (\tau )h_{3}, \\
\mbox{ see nonlinear symmetries }(\ref{nsym1a}); \\
h_{3}(\tau )=h_{3}^{[0]}-\Phi ^{2}/4\Lambda (\tau ),h_{3}^{\ast }\neq
0,\Lambda (\tau )\neq 0=const%
\end{array}%
$ \\ \hline
Off-diag. solutions, $%
\begin{array}{c}
\mbox{d--metric} \\
\mbox{N-connec.}%
\end{array}%
$ &  & $%
\begin{array}{c}
\ g_{i}(\tau )=e^{\ \psi (\tau ,x^{k})}%
\mbox{ as a solution of 2-d Poisson
eqs. }\psi ^{\bullet \bullet }+\psi ^{\prime \prime }=2~\ _{h}\widehat{\Im }%
(\tau ); \\
h_{3}(\tau )=h_{3}^{[0]}-\int dt(\Psi ^{2})^{\ast }/4\ _{v}\widehat{\Im }%
=h_{3}^{[0]}-\Phi ^{2}/4\Lambda (\tau ); \\
h_{4}(\tau )=-(\Psi ^{\ast })^{2}/4(\ _{v}\widehat{\Im })^{2}h_{3},%
\mbox{
see }(\ref{offcosm1}); \\
n_{k}(\tau )=\ _{1}n_{k}+\ _{2}n_{k}\int dt(\Psi ^{\ast })^{2}/(\ _{v}%
\widehat{\Im })^{2}|h_{3}^{[0]}-\int dt(\Psi ^{2})^{\ast }/4\ _{v}\widehat{%
\Im }|^{5/2}; \\
w_{i}(\tau )=\partial _{i}\ \Psi /\ \Psi ^{\ast }=\partial _{i}\ \Psi ^{2}/\
|(\Psi ^{2})^{\ast }|.%
\end{array}%
$ \\ \hline
LC-configurations (\ref{lccond}) &  & $%
\begin{array}{c}
w_{i}^{\ast }=(\partial _{i}-w_{i}\partial _{t})\ln \sqrt{|h_{4}(\tau )|}%
,(\partial _{i}-w_{i}\partial _{t})\ln \sqrt{|h_{3}(\tau )|}=0, \\
\partial _{k}w_{i}(\tau )=\partial _{i}w_{k}(\tau ),n_{i}^{\ast }(\tau
)=0,\partial _{i}n_{k}(\tau )=\partial _{k}n_{i}(\tau );\Psi =\check{\Psi}%
[\iota ],(\partial _{i}\check{\Psi})^{\ast }=\partial _{i}(\check{\Psi}%
^{\ast }) \\
\mbox{ and }\ _{v}\widehat{\Im }(\tau ,x^{i},t)=\ _{v}\widehat{\Im }[\check{%
\Psi}]=\ _{v}\check{\Im},\mbox{ or }\ _{v}\widehat{\Im }=const.%
\end{array}%
$ \\ \hline
N-connections, zero torsion &  & $n_{k}(\tau )=\check{n}_{k}(\tau )=\partial
_{k}n(\tau ,x^{i})\mbox{ and }w_{i}(\tau )=\partial _{i}\check{A}(\tau
)=\left\{
\begin{array}{c}
\partial _{i}(\int dt\ \check{\Im}\ \check{h}_{3}^{\ast }{}])/\check{\Im}\
\check{h}_{3}^{\ast }{}; \\
\partial _{i}\check{\Psi}/\check{\Psi}^{\ast }; \\
\partial _{i}(\int dt\ \check{\Im}(\check{\Phi}^{2})^{\ast })/(\check{\Phi}%
)^{\ast }\check{\Im};%
\end{array}%
\right. .$ \\ \hline
$%
\begin{array}{c}
\mbox{polarization functions} \\
\mathbf{\mathring{g}}\rightarrow \widehat{\mathbf{g}}\mathbf{=}[g_{\alpha
}=\eta _{\alpha }\mathring{g}_{\alpha },\ \eta _{i}^{a}\mathring{N}_{i}^{a}]%
\end{array}%
$ &  & $%
\begin{array}{c}
ds^{2}=\eta _{1}[\ _{1}\iota ]\mathring{g}_{1}(x^{i})[dx^{1}]^{2}+\eta
_{2}[\ _{2}\iota ]\mathring{g}_{2}(x^{i})[dx^{2}]^{2}+ \\
\eta _{3}[\ _{3}\iota ]\mathring{g}_{3}(x^{i})[dy^{3}+\eta _{i}^{3}[\
_{6}\iota ]\mathring{N}_{i}^{3}(x^{i})dx^{i}]^{2}+\eta _{4}[\ _{4}\iota ]%
\mathring{g}_{4}(x^{i})[dt+\eta _{i}^{4}[\ _{4}\iota ]\mathring{N}%
_{i}^{4}(x^{k})dx^{i}]^{2},%
\end{array}%
$ \\ \hline
Prime metric for a cosm.sol. &  & $%
\begin{array}{c}
\lbrack \mathring{g}_{i}(x^{i}),\mathring{g}_{a}=\mathring{h}_{a}(x^{i});%
\mathring{N}_{k}^{3}=\mathring{n}_{k}(x^{i})],\mathring{N}_{k}^{4}=\mathring{%
w}_{k}(x^{i}), \\
\mbox{diagonalizable by frame/ coordinate transforms.} \\
\end{array}%
$ \\
Example of a prime metric &  & $%
\begin{array}{c}
\mathring{g}_{1}=\mathring{a}^{2}(\varsigma ),\mathring{g}_{2}=\mathring{a}%
^{2}(\varsigma ),\mathring{h}_{3}=\mathring{a}^{2}(\varsigma ),\mathring{h}%
_{4}=-1,\varsigma =\varsigma (t) \\
\mbox{ a FLRW or Biachi type solution };%
\end{array}%
$ \\ \hline
Solutions for polarization funct. &  & $%
\begin{array}{c}
\eta _{i}(\tau )=e^{\ \psi (\tau ,x^{k})}/\mathring{g}_{i};\eta _{4}%
\mathring{h}_{4}=-\frac{4[(|\eta _{3}\mathring{h}_{3}|^{1/2})^{\diamond
}]^{2}}{|\int dt\ _{v}\widehat{\Im }[(\eta _{3}\mathring{h}_{3})]^{\ast }|\ }%
; \\
\eta _{3}(\tau )=\eta _{3}(\tau ,x^{i},t)=\eta _{3}[\ _{3}\iota ]%
\mbox{ as a generating
function}; \\
\eta _{k}^{3}\ (\tau )\mathring{N}_{k}^{3}=\ _{1}n_{k}+16\ \ _{2}n_{k}\int dt%
\frac{\left( [(\eta _{3}\mathring{h}_{3})^{-1/4}]^{\ast }\right) ^{2}}{|\int
dt\ _{v}\widehat{\Im }[(\eta _{3}\ \mathring{h}_{3})]^{\ast }|\ };\ \eta
_{i}^{4}(\tau )\ \mathring{N}_{i}^{4}=\frac{\partial _{i}\ \int dt\ _{v}%
\widehat{\Im }(\eta _{3}\ \mathring{h}_{3})^{\ast }}{\ _{v}\widehat{\Im }\
(\eta _{3}\ \mathring{h}_{3})^{\ast }}%
\end{array}%
$ \\ \hline
Polariz. funct. with zero torsion &  & $%
\begin{array}{c}
\eta _{i}(\tau )=e^{\ \psi (\tau ,x^{k})}/\mathring{g}_{i};\eta _{3}=\check{%
\eta}_{3}(\tau ,x^{i},t)\mbox{ as a generating function}; \\
\eta _{4}(\tau )=-\frac{4[(|\eta _{3}\mathring{h}_{3}|^{1/2})^{\ast }]^{2}}{%
\mathring{g}_{4}|\int dt\ _{v}\check{\Im}[(\check{\eta}_{3}\mathring{h}%
_{3})]^{\ast }|\ };\eta _{k}^{3}(\tau )=\frac{\ \partial _{k}n}{\mathring{n}%
_{k}},\eta _{i}^{4}(\tau )=\frac{\partial _{i}\check{A}}{\mathring{w}_{k}},%
\end{array}%
$ \\ \hline\hline
\end{tabular}%
\end{eqnarray*}%
}

\section{Geometric cosmological flows and solitonic hierarchies}

\label{sec4} In this section, we consider applications of the anholonomic
frame deformation method (AFDM, outlined in Tables 1 and 2) for constructing
in explicit form exact and parametric off-diagonal cosmological solutions
describing solitonic geometric flows.

\subsection{Nonlinear PDEs for geometric flows with cosmological solitonic
hierarchies}

There will be studied solutions of modified Einstein equations (\ref%
{solitonhierarcheq}) transformed into systems of nonlinear PDEs with
decoupling (\ref{estatsimpl}).

\subsubsection{Parametric cosmological solutions with additive solitonic
sources}

Let us introduce such conventions: We shall write that $\ _{0}^{int}\widehat{%
\Im }(\tau )=\ _{v}^{int}\widehat{\Im }[\ _{2}\iota ]$ (i.e. put a left
label "0") for a geometric flow source $~\ _{v}\widehat{\Im }(\tau )$ \ if
it contains a term in $\ \ ^{eff}\widehat{\mathbf{\Upsilon }}_{\ \nu }^{\mu
}(\tau )\ $ (\ref{dsourcparam}) defined as a source functional on a
solitonic hierarchy $[\ _{2}\iota ].$ If it will be written $\ \ _{v}^{int}%
\widehat{\Im }(\tau )$ without a left label "0", we shall consider that such
a term corresponds to a general (effective) $\ ^{eff}\widehat{\mathbf{%
\Upsilon }}_{\ \nu }^{\mu }(\tau )$ (not prescribing any solitonic
configurations) encoding contributions from a distortion tensor $\widehat{%
\mathbf{Z}}$ (\ref{distr}). An effective source term $\ \ _{v}^{fl}\widehat{%
\Im }$ determined by geometric flows (with left label "fl") of the d-metric,
$\partial _{\tau }\mathbf{g}_{\alpha ^{\prime }}(\tau ),$ in (\ref%
{dsourcparam}) is contained. It is of cosmological solitonic character if
the d-metric coefficients are also cosmological solitons. We can consider
cosmological solitonic hierarchies for Ricci soliton configurations, i.e.
nonholonomic Einstein systems, with $\ _{v}^{fl}\widehat{\Im }=0.$

For this class of solutions, we consider a source (\ref{dsourcparam}) with a
left label $a$ is used for "additive functionals"
\begin{equation}
~\ _{v}^{a}\widehat{\Im }(\tau )=\ _{v}^{a}\widehat{\Im }[\iota ]~=\ _{v}^{a}%
\widehat{\Im }(\tau ,{x}^{i},t)=\ _{v}^{fl}\widehat{\Im }[\ _{1}\iota ]+\
_{v}^{int}\widehat{\Im }[\ _{2}\iota ]+\ _{v}^{int}\widehat{\Im }[\
_{3}\iota ].  \label{adsourcosm}
\end{equation}%
In such source functionals, it is considered that we prescribe an effective
cosmological solitonic hierarchy for matter fields even, in general, such
gravitational and matter field interactions can be of non--solitonic type.
The second equation (\ref{estatsimpl}) with cosmological source $\ _{v}%
\widehat{\Im }[\iota ]$ = $\ _{v}^{a}\widehat{\Im }[\iota ]$ can be
integrated on time like coordinate $y^{4}=t.$ This allows us to construct
off-diagonal cosmological metrics and generalized connections encoding
solitonic hierarchies determined by a generating function $h_{3}(\tau ,{x}%
^{i},t)$ with Killing symmetry on $\partial _{3},$ by effective sources $\
^{a}\widehat{\Im }[\iota ]~=(\ _{h}^{a}\widehat{\Im }[\ _{1}\iota ],\ \
_{v}^{a}\widehat{\Im }[\ _{2}\iota ])$ and an effective cosmological constant%
\begin{equation}
\ ^{a}\Lambda (\tau )=\ ^{fl}\Lambda (\tau )+\ ^{m}\Lambda (\tau )+\
_{0}^{int}\Lambda (\tau ).  \label{adcosmconst2}
\end{equation}%
This constant is related to $\ _{v}^{a}\widehat{\Im }[\iota ]~$ (\ref%
{adsourcosm}) via nonlinear symmetry transforms (\ref{nsym1a}).

Following the AFDM\ summarized in Table 2, we construct such a class of
quadratic line elements for generic off-diagonal cosmological solutions
determined by effective sources encoding solitonic hierarchies,
\begin{eqnarray}
ds^{2} &=&e^{\ \psi \lbrack \ _{1}\iota ]}[(dx^{1})^{2}+(dx^{2})^{2}]-\frac{%
[h_{3}^{\ast }{}(\tau )]^{2}}{|\int dt\ \ _{v}^{a}\widehat{\Im }[\iota
]h_{3}^{\ast }{}(\tau )|\ h_{3}(\tau )}\left[ dt+\frac{\partial _{i}(\int
dt\ \ _{v}^{a}\widehat{\Im }[\iota ]\ h_{3}^{\ast }{}(\tau ))}{\ \ \ _{v}^{a}%
\widehat{\Im }[\iota ]\ h_{3}^{\ast }{}(\tau )}dx^{i}\right]  \notag \\
&&+h_{3}(\tau )\left[ dy^{3}+\left( \ _{1}n_{k}(\tau )+4\ _{2}n_{k}(\tau
)\int dt\frac{[h_{3}^{\ast }{}(\tau )]^{2}}{|\int dt\ \ \ _{v}^{a}\widehat{%
\Im }[\iota ]h_{3}^{\ast }(\tau )|\ [h_{3}(\tau )]^{5/2}}\right) dx^{k}%
\right] .  \label{cosmdm2}
\end{eqnarray}%
Such solutions can be nonholonomically constrained\ in order to extract
LC-configurations. The formulas (\ref{cosmdm2}) can be re-defined
equivalently in terms of generating functions $\Psi (\tau ,{x}^{i},t)$ or $%
\Phi (\tau ,{x}^{i},t)$ which can be of a non--solitonic character.

\subsubsection{Einstein gravity with cosmological solitonic generating
functions}

Another class of cosmological NESs can be generated as (off-) diagonal
cosmological solutions using generating functionals encoding cosmological
solitonic hierarchies $\Phi (\tau )=\Phi \lbrack i].$ Such functionals are
subjected to nonlinear symmetries of type (\ref{nsym1a}) and general
effective sources $\ _{v}\widehat{\Im }(\tau )$ which can be of
non-solitonic character. The second equation into (\ref{estatsimpl})
transforms into
\begin{equation*}
\varpi ^{\ast }(\tau )[\ \Phi \lbrack i],\Lambda (\tau )]\ h_{3}^{\ast
}(\tau )[\Phi \lbrack i],\Lambda (\tau )]=2h_{4}(\tau )[\Phi \lbrack
i],\Lambda (\tau )]h_{3}(\tau )[\Phi \lbrack i],\Lambda (\tau )]\ \ _{v}%
\widehat{\Im }(\tau ).
\end{equation*}%
This equation can be solved together with other equations (\ref{eq1a})-(\ref%
{eq4a}) following the AFDM, see Table 2.

The solutions for such cosmological configurations determined by general
nonlinear functionals for generating functions can be written in all forms (%
\ref{gencosm1}). We present \ here the quadratic line element corresponding
only to solutions of third type parametrization
\begin{eqnarray}
ds^{2} &=&e^{\ \psi (\tau
,x^{k})}[(dx^{1})^{2}+(dx^{2})^{2}]+(h_{3}^{[0]}(\tau ,x^{k})-\frac{(\Phi
\lbrack i])^{2}}{4\Lambda (\tau )})  \label{cosm2dmnf} \\
&&\lbrack dy^{3}+(_{1}n_{k}(\tau ,x^{k})+\ _{2}n_{k}(\tau ,x^{k})\int dt%
\frac{\ [(\Phi \lbrack i])^{2}]^{\ast }}{|\Lambda (\tau )\int dt\ \ _{v}%
\widehat{\Im }(\tau )\ [(\Phi \lbrack i])^{2}]^{\ast }|}|h_{3}^{[0]}(\tau
,x^{k})-\frac{\ (\Phi \lbrack i])^{2}}{4\Lambda (\tau )}|^{-5/2})dx^{k}]
\notag \\
&&-\frac{(\Phi \lbrack i])^{2}[(\Phi \lbrack i])^{2}]^{\ast }}{|\Lambda
(\tau )\int dt\ _{v}\widehat{\Im }(\tau )[(\Phi \lbrack i])^{2}]^{\ast }|\
(h_{3}^{[0]}-\frac{(\Phi \lbrack i])^{2}}{4\Lambda (\tau )})}[dt+\frac{%
\partial _{i}\left( \int dt\ _{v}\widehat{\Im }(\tau )[(\Phi \lbrack
i])^{2}]^{\ast }\right) }{\ _{v}\widehat{\Im }(\tau )\ [(\Phi \lbrack
i])^{2}]^{\ast }}dx^{i}].  \notag
\end{eqnarray}%
The zero torsion constraints (\ref{lccond}) allow us to extract
LC--configurations parameterized as a subclass of (\ref{cosm2dmnf}),%
\begin{eqnarray}
ds^{2} &=&e^{\psi (\tau )}[(dx^{1})^{2}+(dx^{2})^{2}]+(h_{3}^{[0]}(\tau
,x^{k})-\frac{(\check{\Phi}[i])^{2}}{4\Lambda (\tau )})\left[
dy^{3}+(\partial _{k}n(\tau ))dx^{k}\right]  \notag \\
&&-\frac{(\check{\Phi}[i])^{2}[(\check{\Phi}[i])^{2}]^{\ast }}{|\Lambda
(\tau )\int dt\ _{v}\widehat{\Im }(\tau )[(\check{\Phi}[i])^{2}]^{\ast }|\
(h_{3}^{[0]}-\frac{(\check{\Phi}[i])^{2}}{4\Lambda (\tau )})}[dt+(\partial
_{i}\ \check{A}(\tau ))dx^{i}],  \label{cosm2dmnflc}
\end{eqnarray}%
where $\check{A}(\tau )$ and $n(\tau )$ are also generating functions.
Dualizing such solutions and their symmetries, we can generate stationary
configurations.

\subsubsection{ Small N-adapted cosmological solitonic flow deformations}

\label{assedef}Let us analyze some classes of cosmological solitonic flow
solutions with small parametric deformations for a well-known cosmological
metric in GR (for instance, a FLRW or Bianchi one as in Appendix \ref%
{appendixb}) $\mathbf{\mathring{g}}=[\mathring{g}_{i},\mathring{g}_{a},%
\mathring{N}_{b}^{j}]$ (\ref{primedm}) when $\partial _{4}\mathring{g}_{3}=%
\mathring{g}_{3}^{\ast }\neq 0.$ We formulate a geometric formalism for
small generic off--diagonal parametric deformations of $\mathbf{\mathring{g}}
$ into certain target cosmological solitonic metrics of type $\mathbf{g}$ (%
\ref{dme}) when%
\begin{eqnarray}
ds^{2} &=&\eta _{i}(\varepsilon ,\tau )\mathring{g}_{i}(dx^{i})^{2}+\eta
_{a}(\varepsilon ,\tau )\mathring{g}_{a}(\mathbf{e}^{a})^{2},  \label{targm}
\\
\mathbf{e}^{3} &=&dy^{3}+\ ^{n}\eta _{i}(\varepsilon ,\tau )\mathring{n}%
_{i}dx^{i},\mathbf{e}^{4}=dt+\ ^{w}\eta _{i}(\varepsilon ,\tau )\mathring{w}%
_{i}dx^{i}.\   \notag
\end{eqnarray}%
The coefficients $[g_{\alpha }=\eta _{\alpha }\mathring{g}_{\alpha },\
^{w}\eta _{i}\mathring{w}_{i},\ ^{n}\eta _{i}n_{i}]$ in these formulas
depend on a small parameter $\varepsilon ,$ $0\leq \varepsilon \ll 1,$ on
evolution parameter $\tau $ and on coordinates $x^{i}$ and $t.$ We suppose
that a family of (\ref{targm}) define a solution of cosmological solitonic
flow equations described by a system of nonlinear PDEs with decoupling (\ref%
{estatsimpl}). The $\varepsilon $ -deformations are parameterised in the
form
\begin{eqnarray}
&&\eta _{i}(\varepsilon ,\tau )=1+\varepsilon \upsilon _{i}(\tau
,x^{k}),\eta _{a}=1+\varepsilon \upsilon _{a}(\tau ,x^{k},t)%
\mbox{  for
the coefficients of d-metrics };  \label{smpolariz} \\
&&^{w}\eta _{i}(\varepsilon ,\tau )=1+\varepsilon \ ^{w}\upsilon _{i}(\tau
,x^{k},t),\ \ ^{n}\eta _{i}(\tau ,x^{k},t)=1+\varepsilon \ \ ^{n}\upsilon
_{i}(\tau ,x^{k},t)\mbox{ for the coefficients of N-connection },  \notag
\end{eqnarray}%
where a generating function can be given by $g_{3}(\tau )=\eta _{3}(\tau )%
\mathring{g}_{3}=\ \eta _{3}(\tau ,x^{i},t)\mathring{g}_{3}(\tau
,x^{i},t)=[1+\varepsilon \upsilon (\tau ,x^{i},t)]\mathring{g}_{3},$ for $%
\upsilon =\upsilon _{3}(\tau ,x^{i},t).$

The deformations of $h$-components of a prime cosmological d-metric are $\
_{\varepsilon }g_{i}=\mathring{g}_{i}(1+\varepsilon \upsilon _{i})=e^{\psi
(\tau ,x^{k})}$ for a solution of the 2-d Laplace equation in (\ref%
{estatsimpl}). For parameterizations
\begin{equation*}
\psi (\tau )=\ \ ^{0}\psi (\tau ,x^{k})+\varepsilon \ ^{1}\psi (\tau ,x^{k})%
\mbox{ and }\ _{h}\widehat{\Im }(\tau )(\tau )=\ _{h}^{0}\widehat{\Im }(\tau
,x^{k})+\varepsilon \ _{h}^{1}\widehat{\Im }(\tau ,x^{k}),
\end{equation*}
we compute the deformation polarization functions in the form $\upsilon
_{i}=e^{\ ^{0}\psi }\ ^{1}\psi /\mathring{g}_{i}\ \ _{h}^{0}\widehat{\Im }.$
The horizontal generating and source functions are solutions of $\ ^{0}\psi
^{\bullet \bullet }+\ ^{0}\psi ^{\prime \prime }=\ _{h}^{0}\widehat{\Im }$
and $\ ^{1}\psi ^{\bullet \bullet }+\ ^{1}\psi ^{\prime \prime }=\ \ _{h}^{1}%
\widehat{\Im }.$ Using $\varepsilon $-decompositions (\ref{smpolariz}) and
similar formulas for $v$-components, we compute $\varepsilon $-decomposition
of the target cosmological solitonic d-metric and N-connection coefficients
\begin{eqnarray*}
\mathring{g}_{i}\eta _{i}(\tau ) &=&e^{\ \psi (\tau ,x^{k})}%
\mbox{ as a
solution of 2-d Poisson equations } \\
\ _{\varepsilon }g_{i}(\tau ) &=&[1+\varepsilon e^{\ ^{0}\psi }\ ^{1}\psi /%
\mathring{g}_{i}\ \ _{h}^{0}\widehat{\Im }]\mathring{g}_{i},%
\mbox{ also constructed as a
solution of 2-d Poisson equations for }\ ^{1}\psi \\
\mathring{g}_{4}\eta _{4}(\tau ) &=&-\frac{4[(|\eta _{3}(\tau )\mathring{g}%
_{3}|^{1/2})^{\ast }]^{2}}{|\int dt\ \ _{v}\widehat{\Im }(\tau )[\eta
_{3}(\tau )\mathring{g}_{3}]^{\ast }|\ } \\
\mbox{ i.e. }\ _{\varepsilon }g_{4}(\tau ) &=&[1+\varepsilon \ \upsilon _{4}]%
\mathring{g}_{4}\mbox{ for }\upsilon _{4}(\tau ,x^{i},t)=2\frac{(\upsilon
\mathring{g}_{3})^{\ast }}{\mathring{g}_{3}^{\ast }}-\frac{\int dt\ \ _{v}%
\widehat{\Im }(\tau )(\upsilon \mathring{g}_{3})^{\ast }}{\int dt\ \ _{v}%
\widehat{\Im }(\tau )\mathring{g}_{3}^{\ast }}.
\end{eqnarray*}%
For such formulas, the system of coordinates is chosen in such a form that
there are satisfied the condition $(\mathring{g}_{3}^{\ast })^{2}=\mathring{g%
}_{4}|\int dt\ _{v}\widehat{\Im }(\tau )\mathring{g}_{3}^{\ast }|,$ which
allow to find $\mathring{g}_{4}$ for any prescribed values $\mathring{g}_{3}$
and $\ _{v}\widehat{\Im }(\tau ).$

The $\varepsilon $ -deformations of N-connection coefficients are computed%
\begin{eqnarray*}
\eta _{k}^{3}(\tau )\mathring{n}_{k} &=&\ _{1}n_{k}(\tau )+16\
_{2}n_{k}(\tau )\int dt\frac{\left( [(\eta _{3}(\tau )\mathring{g}%
_{3})^{-1/4}]^{\ast }\right) ^{2}}{|\int dt\ _{v}\widehat{\Im }(\tau )(\eta
_{3}\ (\tau )\mathring{g}_{3})^{\ast }|\ } \\
\mbox{ i.e. }\ _{\varepsilon }n_{i}(\tau ) &=&[1+\varepsilon \ ^{n}\upsilon
_{i}(\tau )]\mathring{n}_{k}=0\mbox{ for }\ ^{n}\upsilon _{i}(\tau
,x^{i},t)=0,
\end{eqnarray*}%
if the integration functions are chosen $_{1}n_{k}(\tau )=0$ and $%
_{2}n_{k}(\tau )=0,$ and
\begin{eqnarray*}
\eta _{i}^{4}(\tau )\mathring{w}_{i} &=&\frac{\partial _{i}\ \int dt\ \ _{v}%
\widehat{\Im }(\tau )[\eta _{3}(\tau )\ \mathring{g}_{3}]^{\ast }}{\ \ \ _{v}%
\widehat{\Im }(\tau )\ [\eta _{3}(\tau )\ \mathring{g}_{3}]^{\ast }} \\
\mbox{ i.e. }\ _{\varepsilon }w_{i}(\tau ) &=&[1+\varepsilon \ ^{w}\upsilon
_{i}(\tau )]\mathring{w}_{i}\mbox{ for }\ ^{w}\upsilon _{i}(\tau ,x^{i},t)=%
\frac{\partial _{i}\ \int dt\ \ _{v}\widehat{\Im }(\tau )(\upsilon \mathring{%
g}_{3})^{\ast }}{\partial _{i}\ \int dt\ \ \ _{v}\widehat{\Im }(\tau )%
\mathring{g}_{3}^{\ast }}-\frac{(\upsilon \mathring{g}_{3})^{\ast }}{%
\mathring{g}_{3}^{\ast }},
\end{eqnarray*}%
when a prime $\mathring{w}_{i}=\partial _{i}\ \int dt\ _{v}\widehat{\Im }%
(\tau )\mathring{g}_{3}^{\ast }/\ _{v}\widehat{\Im }(\tau )\mathring{g}%
_{3}^{\ast }$ is well defined for some prescribed $\ _{v}\widehat{\Im }(\tau
)$ and $\mathring{g}_{3}^{\ast }.$

Finally, we conclude that $\varepsilon $--deformed quadratic line elements
for cosmological flow deformations can be written in a general form {\small
\begin{equation*}
ds_{\varepsilon t}^{2}=\ _{\varepsilon }g_{\alpha \beta }(\tau
,x^{k},t)du^{\alpha }du^{\beta }=\ _{\varepsilon }g_{i}(\tau
,x^{k})[(dx^{1})^{2}+(dx^{2})^{2}]+\ _{\varepsilon }g_{3}(\tau
,x^{k},t)(dy^{3})^{2}+\ _{\varepsilon }h_{4}(\tau ,x^{k},t)\ [dt+\
_{\varepsilon }w_{i}(\tau ,x^{k},t)dx^{i}]^{2}.
\end{equation*}%
} We can impose additional nonholonomic constraints (\ref{lcsolstat}) in
order to extract LC--configurations for $\varepsilon $--deformati\-ons with
zero torsion.

\subsection{FLRW metrics in (off-) diagonal cosmological media with
solitonic hierarchies}

Various classes of generic off-diagonal cosmological solitonic solutions can
be constructed and parameterized in terms of $\eta $--polarization functions
introduced in formulas (\ref{dme}) and applying the AFDM summarized in
Tables 1 and 2. A primary cosmological d-metric can be parameterized in a
necessary form as in Appendix (\ref{appendixb}) when $\mathbf{\mathring{g}=}[%
\mathring{g}_{i}(x^{i},t),\mathring{g}_{a}=\mathring{h}_{a}(x^{i},t);%
\mathring{N}_{k}^{3}=\mathring{n}_{k}(x^{i},t),\mathring{N}_{k}^{4}=%
\mathring{w}_{k}(x^{i},t)]$ (\ref{primedm}) which for a FLRW configuration
can be diagonalized by frame/ coordinate transforms. A cosmological
solitonic target metric $\mathbf{g}$ can be generated by nonholonomic $\eta $%
--deformations, $\mathbf{\mathring{g}}\rightarrow \mathbf{g}(\tau )\mathbf{=}%
[g_{i}(\tau ,x^{k})=\eta _{i}(\tau )\mathring{g}_{i},g_{b}(\tau
,x^{k},t)=\eta _{b}(\tau )\mathring{g}_{b},N_{i}^{a}(\tau ,x^{k},t)=\ \eta
_{i}^{a}(\tau )\mathring{N}_{i}^{a}],$ and constrained to the conditions to
define exact and parametric solutions of the system of nonlinear PDEs with
decoupling (\ref{estatsimpl}). A corresponding quadratic line element for $%
\mathbf{g}$ can be parameterized in a form (\ref{dme}),
\begin{equation}
ds^{2}=\eta _{i}(\tau ,x^{i},t)\mathring{g}_{i}[dx^{i}]^{2}+\eta _{a}(\tau
,x^{i},t)\mathring{g}_{a}[dy^{a}+\eta _{k}^{a}(\tau ,x^{i},t)\mathring{N}%
_{k}^{a}dx^{k}]^{2},  \label{cosm3gf}
\end{equation}
with summation on repeating contracted low-up indices. The polarizaton
values $\eta _{\alpha }(\tau )$ and $\eta _{i}^{a}(\tau )$ are determined by
geometric and cosmological solitonic flows and nonlinear interactions.

\subsubsection{Cosmological solutions generated by solitonic sources}

We prescribe that the effective $v$--source is determined by a solitonic
hierarchy $\widehat{\mathbf{\Upsilon }}(\tau ,x^{i},t)=\ _{v}\widehat{\Im }%
[\ _{2}\iota ]$ (\ref{dsourcparam}) and compute the coefficients for a
target d-metric (\ref{cosm3gf}) following formulas summarized in Table 2,%
\begin{eqnarray}
\eta _{i}(\tau ) &=&\frac{e^{\ \psi (\tau ,x^{k})}}{\mathring{g}}_{i};\ \eta
_{3}(\tau )=\eta _{3}(\tau ,x^{i},t)\mbox{ as a generating
function};\ \eta _{4}(\tau )=-\frac{4[(|\eta _{3}(\tau )\mathring{h}%
_{3}|^{1/2})^{\ast }]^{2}}{\mathring{h}_{4}|\int dt\ \ _{v}\widehat{\Im }[\
_{2}\iota ](\eta _{3}(\tau )\mathring{h}_{3})^{\ast }|};  \notag \\
\ \eta _{k}^{3}\ (\tau ) &=&\frac{\ _{1}n_{k}}{\mathring{n}_{k}}+16\ \ \frac{%
\ _{2}n_{k}}{\mathring{n}_{k}}\int dt\frac{\left( [(\eta _{3}(\tau )%
\mathring{h}_{3})^{-1/4}]^{\ast }\right) ^{2}}{|\int dt\ _{v}\widehat{\Im }%
[\ _{2}\iota ](\eta _{3}\ (\tau )\mathring{h}_{3})^{\ast }|};\eta
_{i}^{4}(\tau )=\frac{\partial _{i}\ \int dt\ \ _{v}\widehat{\Im }[\
_{2}\iota ](\eta _{3}(\tau )\mathring{h}_{3})^{\ast }}{\mathring{w}_{i}\ \ \
_{v}\widehat{\Im }[\ _{2}\iota ]\ (\eta _{3}(\tau )\mathring{h}_{3})^{\ast }}%
,  \label{cosm3gfa}
\end{eqnarray}
with integration functions $\ _{1}n_{k}(\tau ,x^{i})$ and $\ _{2}n_{k}(\tau
,x^{i}).$

In (\ref{cosm3gfa}), the gravitational polarization $\eta _{3}(x^{i},t)$ is
taken as a (non) singular generating function subjected to nonlinear
symmetries of type (\ref{nsym1a}) which can be written in the form
\begin{equation*}
\Phi ^{2} = -4\ \Lambda (\tau )h_{3}(\tau )=-4\ \Lambda \eta _{3}(\tau ,{x}%
^{i},t)\mathring{h}_{3}(\tau ,x^{i},t) \mbox{ and } (\Psi ^{2})^{\ast } =
-\int dt\ \ \ _{v}\widehat{\Im }[\ _{2}\iota ][\eta _{3}(\tau ,{x}^{i},t)%
\mathring{h}_{4}(\tau ,{x}^{i},t)]^{\ast }.
\end{equation*}%
In this section, the values $\Phi ,h_{3}$ and $\eta _{3}$ may not encode
soltionic hierarchies but $\Psi $ and other coefficients of such target
cosmological d-metric are solitonic ones if they are computed using $\ _{v}%
\widehat{\Im }[\ _{2}\iota ].$ We can constrain the coefficients (\ref%
{cosm3gfa}) to a subclass of data generating target LC-configurations when
the d-metrics satisfy the constraints (\ref{lcsolstat}) for zero torsion.
The nonlinear functionals for the soliton v-source and (effective)
cosmological constant can be changed into additive functionals $\ _{v}%
\widehat{\Im }\ \rightarrow $ $\ _{v}^{a}\widehat{\Im }$ and $\Lambda
\rightarrow $ $\ ^{a}\Lambda $ as $\ _{v}^{a}\widehat{\Im }[\iota ]$ (\ref%
{adsourcosm}) and $\ ^{a}\Lambda $ (\ref{adcosmconst2}).

\subsubsection{FLRW metrics deformed by solitonic generating functions}

Solutions with geometric flow and cosmological $\eta $--polarizations (\ref%
{targm}) can be constructed with coefficients of the d-metrics determined by
nonlinear generating functionals $\Phi \lbrack i],$ or prescribed additive
functionals $\ ^{a}\Phi \lbrack i]$ corresponding to (\ref{adsourcosm}).
This includes terms with integration functions$\ h_{3}^{[0]}(\tau ,{x}^{i})$
for $h_{3}[i]$ with explicit dependence on a time like variable $t.$ Such
configurations can be generated also by some prescribed cosmological data $\
\ _{v}\widehat{\Im }\ (\tau ,{x}^{i},t)$ and $\Lambda (\tau ),$ which are
not obligatory of solitonic nature. We can compute corresponding nonlinear
functionals $\ \eta _{3}(\tau ,{x}^{i},t)$ (we omit here similar formulas
for additive functionals $\ ^{a}\eta _{3}(\tau ,{x}^{i},t))$ using nonlinear
symmetries (\ref{nsym1a}) and related polarization functions,
\begin{equation*}
\ \eta _{3}[i]=-\Phi ^{2}[i]/4\Lambda (\tau )\mathring{h}_{3}({x}^{i},t),\
[\Psi ^{2}(\tau )]^{\ast }=-\int dt\ \ \ _{v}\widehat{\Im }(\tau ,{x}%
^{i},t)h_{3}^{\ast }(\tau )=-\int dt\ \ _{v}\widehat{\Im }(\tau ,{x}%
^{i},t)[\ \eta _{3}[i]\mathring{h}_{3}({x}^{i},t)]^{\ast }.
\end{equation*}%
We apply these formulas for the AFDM outlined in Table 2 and compute the
coefficients of a cosmological solitonic d-metric of type (\ref{cosm3gf}),
{\small
\begin{eqnarray}
\eta _{i}(\tau ) &=&\frac{e^{\ \psi (\tau ,x^{k})}}{\mathring{g}_{i}};\eta
_{3}(\tau )=\eta _{3}(\tau ,x^{i},t)=\eta _{3}[i]%
\mbox{ as a generating
function};\eta _{4}=-\frac{4[(|\ \eta _{3}[i]\mathring{h}_{3}|^{1/2})^{\ast
}]^{2}}{\mathring{h}_{3}|\int dt\ \ _{v}\widehat{\Im }(\tau )\ \eta _{3}[i]%
\mathring{h}_{3})^{\ast }|\ };  \label{cosm3fp1} \\
\eta _{k}^{3}(\tau ) &=&\frac{\ _{1}n_{k}(\tau )}{\mathring{n}_{k}}+16\ \
\frac{\ _{2}n_{k}(\tau )}{\mathring{n}_{k}}\int dt\frac{\left( [(\eta _{3}[i]%
\mathring{h}_{3})^{-1/4}]^{\ast }\right) ^{2}}{|\int dt\ _{v}\widehat{\Im }%
(\tau )(\eta _{3}[i]\mathring{h}_{3})^{\ast }|\ };\eta _{i}^{4}(\tau )=\frac{%
\partial _{i}\ \int dt\ \ _{v}\widehat{\Im }(\tau )(\eta _{3}[i]\ \mathring{h%
}_{3})^{\ast }}{\mathring{w}_{i}\ \ _{v}\widehat{\Im }(\tau )\ (\eta
_{3}[i]\ \mathring{h}_{3})^{\ast }},  \notag
\end{eqnarray}%
} for integrating functions $\ _{1}n_{k}\tau ,{x}^{i}{)}$ and $\
_{2}n_{k}\tau ,{x}^{i}).$

Using (\ref{cosm3fp1}), target cosmological solitonic off-diagonal metrics
with zero torsion which solve (\ref{lccond}) can be generated by
polarization functions subjected to additional nonholonomic constraints,
\begin{eqnarray*}
\eta _{i}(\tau ) &=&\frac{e^{\ \psi (\tau ,x^{k})}}{\mathring{g}_{i}};\eta
_{3}(\tau )=\check{\eta}_{3}(\tau ,x^{i},t)=\check{\eta}_{3}[i]%
\mbox{ as a
generating function }; \\
\ \eta _{4}(\tau ) &=&-\frac{4[(|\ \check{\eta}_{3}[i]\mathring{h}%
_{3}|^{1/2})^{\ast }]^{2}}{\mathring{h}_{4}|\int dt\ \ _{v}\check{\Im}(\tau
)(\check{\eta}_{3}[i]\mathring{h}_{3})^{\ast }|\ };\eta _{k}^{3}(\tau )=%
\frac{\ \partial _{k}n(\tau )}{\mathring{n}_{k}},\eta _{i}^{4}(\tau )=\frac{%
\partial _{i}\ \check{A}(\tau )}{\mathring{w}_{k}},
\end{eqnarray*}%
for an integrating function $n(\tau ,{x}^{i})$ and a generating function $\
\check{A}(\tau ,{x}^{i},t).$

The solutions constructed in this subsection describe certain
nonholonomically deformed cosmological geometric flow configurations
self-consistently imbedded into a solitonic gravitational evolution media
which can model nonholonomic dark energy and dark energy configurations. A
compatibility with observational data can be chosen for respective
integration functions and constants and corresponding classes of solitonic
hierarchies.

\subsubsection{Cosmological deformations by solitonic sources \& solitonic
generating functions}

General classes of cosmological solutons and nonholonomic deformations can
be constructed with nonlinear solitonic functionals both for generating
functions and generating sources. Nonlinear superpositions of solutions of
type (\ref{cosm3gfa}) and (\ref{cosm3fp1}) can be performed if the
coefficients of d-metric are computed {\small
\begin{eqnarray}
\eta _{i}(\tau ) &=&\frac{e^{\ \psi (\tau ,x^{k})}}{\mathring{g}_{i}};\eta
_{3}(\tau )=\eta _{3}(\tau ,x^{i},t)=\ \eta _{3}[\ _{3}\iota ]%
\mbox{ as a generating
function};\eta _{4}(\tau )=-\frac{4[(|\eta _{3}[\ _{3}\iota ]\mathring{h}%
_{3}|^{1/2})^{\ast }]^{2}}{\mathring{h}_{4}|\int dt\ \ _{v}\widehat{\Im }%
[\iota ](\eta _{3}[\ _{3}\iota ]\mathring{h}_{3})^{\ast }|\ };  \notag \\
\ \eta _{k}^{3}(\tau ) &=&\frac{\ _{1}n_{k}(\tau )}{\mathring{n}_{k}}+16\ \
\frac{\ _{2}n_{k}(\tau )}{\mathring{n}_{k}}\int dt\frac{\left( [(\eta _{3}[\
_{3}\iota ]\mathring{h}_{3})^{-1/4}]^{\ast }\right) ^{2}}{|\int dt\ \ _{v}%
\widehat{\Im }[\iota ](\eta _{3}[\ _{3}\iota ]\ \mathring{h}_{3})^{\ast }|\ }%
;\eta _{i}^{4}(\tau )=\frac{\partial _{i}\ \int dt\ \ _{v}\widehat{\Im }%
[\iota ](\eta _{3}[\ _{3}\iota ]\ \mathring{h}_{3})^{\ast }}{\mathring{w}%
_{i}\ \ \ _{v}\widehat{\Im }[\iota ]\ (\eta _{3}[\ _{3}\iota ]\ \mathring{h}%
_{3})^{\ast }},  \label{cosm312}
\end{eqnarray}%
} where $\ _{1}n_{k}(\tau ,x^{k})$ and $\ _{2}n_{k}(\tau ,x^{k})$ are
integration functions. In formulas (\ref{cosm312}), we consider two
different prescribed a nonlinear generating functional, $\Phi \lbrack \
_{3}\iota ],$ and a nonlinear functional for source,$\ \ _{v}\widehat{\Im }%
[\iota ].$ and running constant $\Lambda (\tau )$ related via nonlinear
symmetries of type (\ref{nsym1a}). This allows us to compute a corresponding
nonlinear functional $\ \eta _{3}(\tau ,{x}^{i},t)=\ \eta _{3}[\iota ,\
_{4}\iota ,...]$ and a polarization function,
\begin{equation}
\ \eta _{3}(\tau )=-\ \Phi ^{2}[\ _{3}\iota ]/4\ \Lambda (\tau )\mathring{h}%
_{3}({x}^{i},t),\ (\ \Psi ^{2}(\tau ))^{\ast }=-\int dt\ \ _{v}\widehat{\Im }%
[\iota ]h_{3}^{\ast }[\ _{3}\iota ]=-\int dt\ \ _{v}\widehat{\Im }[\iota ][\
\eta _{3}(\tau )\mathring{h}_{3}({x}^{i},t)]^{\ast }.  \label{nsym1b}
\end{equation}%
Imposing additional constraints (\ref{lcsolstat}) for a zero torsion,
LC-cosmological solitonic metrics with geometric flows are generated.

In result, the quadratic line element corresponding to such classes of
cosmological solutions (\ref{cosm312}) can be written for generating data $%
\left( \Phi \lbrack \ _{3}\iota ],\Lambda (\tau )\right) :$
\begin{eqnarray}
ds^{2} & = & e^{\ \psi (\ _{3}\iota )}[(dx^{1})^{2}+(dx^{2})^{2}]
+(h_{3}^{[0]}(\tau ,x^{k}) -\frac{(\Phi \lbrack \ _{3}\iota ])^{2}}{4\Lambda
(\tau )})  \notag \\
&& [dy^{3}+ (_{1}n_{k}(\tau ,x^{k})+\ _{2}n_{k}(\tau ,x^{k})\int dt\frac{\
[(\Phi \lbrack \ _{3}\iota ])^{2}]^{\ast }}{|\Lambda (\tau )\int dt\ \ _{v}%
\widehat{\Im }[\iota ]\ [(\Phi \lbrack \ _{3}\iota ])^{2}]^{\ast }|}%
|h_{3}^{[0]}(\tau ,x^{k})-\frac{\ (\Phi \lbrack \ _{3}\iota ])^{2}}{4\Lambda
(\tau )}|^{-5/2})dx^{k}]  \notag \\
&& -\frac{(\Phi \lbrack \ _{3}\iota ])^{2}[(\Phi \lbrack \ _{3}\iota
])^{2}]^{\ast }}{|\Lambda (\tau )\int dt\ _{v}\widehat{\Im }[\iota ][(\Phi
\lbrack \ _{3}\iota ])^{2}]^{\ast }|\ (h_{3}^{[0]}-\frac{(\Phi \lbrack \
_{3}\iota ])^{2}}{4\Lambda (\tau )})}[dt+\frac{\partial _{i}\left( \int dt\
_{v}\widehat{\Im }[\iota ][(\Phi \lbrack \ _{3}\iota ])^{2}]^{\ast }\right)}{%
\ _{v}\widehat{\Im }[\iota ]\ [(\Phi \lbrack \ _{3}\iota ])^{2}]^{\ast }}%
dx^{i}]  \label{cosm312nqle}
\end{eqnarray}

The data for a primary cosmological solution can be extracted using
nonlinear symmetries (\ref{nsym1b}), when $\mathring{g}_{i}=e^{\ \psi (\tau
,x^{k})}/\eta _{i}(\tau )$ and $\mathring{h}_{3}({x}^{i},t)=-\ \Phi ^{2}[\
_{3}\iota ]/4\ \Lambda (\tau )\eta _{3}(\tau )$ are considered for certain
values which for a $\tau _{0}$ are prescribed in some forms that the
integration functions $h_{4}^{[0]}(\tau ,x^{k}),_{1}n_{k}(\tau ,x^{k})$ and $%
_{2}n_{k}(\tau ,x^{k})$ encode a prime d-metric $\mathbf{\mathring{g}}=[%
\mathring{g}_{i},\mathring{g}_{a},\mathring{N}_{b}^{j}]$ (\ref{primedm}).
Such cosmological scenarios (\ref{cosm312nqle}) describe geometric evolution
for $\tau >$ $\tau _{0}$ self-consistently imbedded into solitonic
gravitational (dark energy) backgrounds and solitonic dark and/or standard
matter.

\subsection{Off--diagonal deformations of FLRW metrics by solitonic flow
sources}

We study how effective sources for geometric flows with cosmological
solitonic hierarchies flows result in generic off--diagonal deformations and
generalizations of a FLRW metric. Such nonholonomic deformations of
geometric objects define new classes of exact solutions of systems of
nonlinear PDEs (\ref{estatsimpl}). To apply the AFDM\ is necessary to define
\ some nonholonomic variables which allow decoupling and integration of
corresponding systems of equations describing N-adapted nonholonomic
deformations of a prime d-metric $\mathbf{\mathring{g}=}[\mathring{g}_{i},%
\mathring{g}_{a}=\mathring{h}_{a};\mathring{N}_{k}^{3}=\mathring{n}_{k},%
\mathring{N}_{k}^{4}=\mathring{w}_{k},]$ (\ref{primedm}), see formulas for
parameterizations of cosmological d-metrics in Appendix \ref{appendixb}. We
cite \cite{misner} as a standard monograph on GR with necessary details on
geometry of cosmological spaces and \cite%
{gheorghiu14,bubuianu18,bub19,vacaru18tc,vacaru19b} for examples of
nonholonomic deformations of BH and cosmological solutions in geometric
flows and gravity theories.

\subsubsection{Nonholonomic evolution of FLRW metrics with induced (or zero)
torsion}

We consider a primary cosmological d-metric parameterized as in Appendix (%
\ref{appendixb}) when $\mathbf{\mathring{g}=}[\mathring{g}_{i}(x^{i},t),%
\mathring{g}_{a}=\mathring{h}_{a}(x^{i},t);\mathring{N}_{k}^{3}=\mathring{n}%
_{k}(x^{i},t),\mathring{N}_{k}^{4}=\mathring{w}_{k}(x^{i},t),]$ (\ref%
{primedm}) with $\mathring{g}_{3}^{\ast }\neq 0,$ which for a FLRW
configuration can be diagonalized by frame/ coordinate transforms. \ This
allows us to construct nonholonomic cosmological deformations following the
geometric formalism outlined in section \ref{assedef} and Table 2.

For general $\eta $--deformations (\ref{targm}) and constraints $n_{i}=0,$
the solitonic flow modifications of the FLRW metric are computed
\begin{eqnarray}
ds^{2} &=&e^{\ \psi (\tau ,x^{k})}[(dx^{1})^{2}+(dx^{2})^{2}]+\eta _{3}[\
_{3}\iota ]\mathring{h}_{3}(\mathbf{e}^{3})^{2}-\frac{4[(|\eta _{3}[\
_{3}\iota ]\mathring{h}_{3}|^{1/2})^{\ast }]^{2}}{|\int dt\ _{v}\widehat{\Im
}[\iota ][\eta _{3}[\ _{3}\iota ]\mathring{h}_{3}]^{\ast }|\ }\mathring{h}%
_{4}(\mathbf{e}^{4})^{2},  \notag \\
\mathbf{e}^{3} &=&dy^{3},\ \mathbf{e}^{4}=dt+\frac{\partial _{i}\ \int dt\
_{v}\widehat{\Im }[\iota ][\eta _{3}[\ _{3}\iota ]\mathring{h}_{3}]^{\ast }}{%
\ _{v}\widehat{\Im }[\iota ]\ [\eta _{3}[\ _{3}\iota ]\mathring{h}%
_{3}]^{\ast }}dx^{i^{\prime }},  \label{ofindtmga}
\end{eqnarray}%
where $\eta _{3}(\tau )=\eta _{3}(\tau ,x^{k},t)=\eta _{3}[\ _{3}\iota ]$ is
a generating function and $\ _{v}\widehat{\Im }(\tau )=\ \ _{v}\widehat{\Im }%
[\iota ]$ is a flow generating source as in (\ref{cosm312}) and $\psi (\tau
,x^{k})$ is a solution of a 2-d Poisson equation.

\subsubsection{Small parametric modifications of FLRW metrics and effective
flow sources}

Let us elaborate on models of geometric cosmological flows for nonholonomic
distributions describing $\varepsilon $-deformations (\ref{smpolariz}) for
for $\mathring{g}_{3}^{\ast }\neq 0$ in target metrics of type (\ref{targm}%
). The corresponding quadratic line elements are
\begin{eqnarray}
ds^{2} &=&[1+\varepsilon e^{\ ^{0}\psi }\frac{\ ^{1}\psi }{\mathring{g}_{i}\
}\ _{h}^{0}\widehat{\Im }]\mathring{g}_{i}(dx^{i})^{2}+[1+\varepsilon
\upsilon ]\mathring{g}_{3}(\mathbf{e}^{3})^{2}+\left[ 1+\varepsilon \ (2%
\frac{[\upsilon \mathring{g}_{3}]^{\ast }}{\mathring{g}_{3}^{\ast }}-\frac{%
\int dt\ _{v}\widehat{\Im }([\upsilon \mathring{g}_{3}])^{\ast }}{\int
dy^{3}\ _{v}\widehat{\Im }\mathring{g}_{3}^{\ast }})\right] \mathring{g}_{4}(%
\mathbf{e}^{4})^{2},  \notag \\
\mathbf{e}^{3} &=&dx^{3},\mathbf{e}^{4}=dt+[1+\varepsilon (\frac{\partial
_{i}\ \int dt\ _{v}\widehat{\Im }(\upsilon \mathring{g}_{3})^{\ast }}{%
\partial _{i}\ \int dt\ _{v}\widehat{\Im }\mathring{g}_{3}^{\ast }}-\frac{%
(\upsilon \mathring{g}_{3})^{\ast }}{\mathring{g}_{3}^{\ast }})]\ \mathring{w%
}_{i}dx^{i}.  \label{smalparstat}
\end{eqnarray}%
In these formulas, $\ ^{0}\psi (\tau )=\ ^{0}\psi (\tau ,x^{k})$ and $\
^{1}\psi (\tau )=\ ^{1}\psi (\tau ,x^{k})$ are solutions of 2-d Poisson
equations with a generating h-source $\ _{h}\widehat{\Im }(\tau )=\ _{h}%
\widehat{\Im }(\tau ,x^{k})=\ _{h}^{0}\widehat{\Im }(\tau
,x^{k})+\varepsilon \ _{h}^{1}\widehat{\Im }(\tau ,x^{k})$ as described in
section \ref{assedef}; $\ _{v}\widehat{\Im }(\tau )=\ \ _{v}\widehat{\Im }%
[\iota ]$ is a generating v-source with $\varepsilon $-decomposition where $%
\upsilon _{3}=\upsilon (\tau )=\upsilon (\tau ,x^{k},t)=\upsilon \lbrack \
_{3}\iota ]$ is a generating function. The cosmological flow solution (\ref%
{smalparstat}) is for a N-adapted system of references and space coordinates
$[x^{i},y^{3}]$ for which the condition $(\mathring{g}_{3}^{\ast })^{2}=%
\mathring{g}_{4}|\int dt\ _{v}\widehat{\Im }\mathring{g}_{3}^{\ast }|$
allows us to compute well-defined coefficients $\mathring{g}_{3}$ and $%
\mathring{w}_{i}=\partial _{i}\ [dt\ \ _{v}\widehat{\Im }\mathring{g}%
_{3}^{\ast }]/\ _{v}\widehat{\Im }\mathring{g}_{3}^{\ast }$ when there are
prescribed certain values$\ _{v}\widehat{\Im }$ and $\mathring{g}_{3}^{\ast
}\neq 0.$ We can fix the conditions $_{1}n_{k}(\tau )=0$ and $_{2}n_{k}(\tau
)=0$ for which $N_{i}^{3}=n_{i}=0$ but even in such cases a non-zero
coefficient $N_{i}^{4}=w_{i}(\varepsilon ,\tau ,x^{k},t)$ results in
nontrivial nonholonomic torsion and anholonomy coefficients. To extract
LC-configurations we can impose on $\upsilon (\tau )$ and sources additional
zero torsion constraints (\ref{lcsolstat}).

Considering for nonlinear symmetries of type (\ref{nsym1b}) the formula $%
\eta _{3}(\tau )=-\ \Phi ^{2}[\ _{3}\iota ]/4\ \Lambda (\tau )\mathring{g}%
_{3}$ for $\ $(\ref{smalparstat}), we conclude that as a generating function
(including solitonic hierarchies) can be used the value
\begin{equation}
\varepsilon \upsilon \lbrack \ _{3}\iota ]=-\left( 1+\Phi ^{2}[\ _{3}\iota
]/4\ \Lambda (\tau )\mathring{g}_{3}\right) \mbox{ or }\Phi \lbrack \
_{3}\iota ]\simeq 2\sqrt{|\Lambda (\tau )\mathring{g}_{3}|}(1-\frac{%
\varepsilon }{2}\upsilon \lbrack \ _{3}\iota ]).  \label{infgenerf}
\end{equation}%
Other types of geometric flow and cosmologic solitonic hierarchies can be
encoded into generated sources $\ _{h}\widehat{\Im }(\tau ,x^{k})=\ _{h}^{0}%
\widehat{\Im }(\tau ,x^{k})+\varepsilon \ _{h}^{1}\widehat{\Im }(\tau
,x^{k}),$ $\ _{v}\widehat{\Im }(\tau )=\ \ _{v}\widehat{\Im }[\iota ]$ and $%
(\Lambda (\tau ),\upsilon \lbrack \ _{4}\iota ],\ _{h}^{0}\widehat{\Im }%
+\varepsilon \ _{h}^{1}\widehat{\Im },_{v}\widehat{\Im }[\iota ]).$

\subsection{W--entropy and thermodynamics of cosmological solitonic solutions%
}

\label{sec5a} We provide explicit examples how G. Perelman's W-entropy and
related thermodynamic values can be computed for cosmological solitonic
solutions under geometric flow evolution. To simplify formulas we fix
certain values for normalization and integration functions corresponding to
a constant normalizing function, $\ \widehat{f}(\tau )=\ \widehat{f}%
_{0}=const=0,$ in (\ref{normalizfunct}).\footnote{%
Having defined such values in a convenient system of reference/coordinates,
we can consider changing to any system of reference.} In result, the F- and
W-functionals (\ref{ffcand}) are written
\begin{eqnarray}
\widehat{\mathcal{F}} &=&\frac{1}{8\pi ^{2}}\int \tau ^{-2}\sqrt{|\mathbf{g}%
[\Phi (\tau ,x^{i},t)]|}\delta ^{4}u[\ _{h}\Lambda (\tau )+\Lambda (\tau
)],\   \label{wn} \\
\widehat{\mathcal{W}} &=&\frac{1}{4\pi ^{2}}\int \tau ^{-2}\sqrt{|\mathbf{g}%
[\Phi (\tau ,x^{i},t)]|}\delta ^{4}u(\tau \left[ \ _{h}\Lambda (\tau
)+\Lambda (\tau )\right] ^{2}-1),  \notag
\end{eqnarray}%
\begin{equation*}
\mbox{ where }\sqrt{|\mathbf{g}[\Phi (\tau ,x^{i},t)\mathbf{]}|}=\sqrt{%
|q_{1}q_{2}\mathbf{q}_{3}(_{q}N)|}=2e^{\ \psi (\tau ,x^{k})}\left\vert \Phi
(\tau ,x^{i},t)\right\vert \sqrt{\frac{\left\vert [\Phi ^{2}(\tau
,x^{i},t)]^{\ast }\right\vert }{|\Lambda (\tau )\int dt\ _{v}\widehat{\Im }%
(\tau ,x^{i},t)[\Phi ^{2}(\tau ,x^{i},t)]^{\ast }|\ }}
\end{equation*}
is computed for d-metrics parameterized in the form (\ref{decomp31}) with
\begin{equation*}
q_{1}(\tau ) = q_{2}(\tau )=e^{\ \psi (\tau ,x^{k})},\mathbf{q}_{3}(\tau )=-%
\frac{\Phi ^{2}(\tau ,x^{i},t)}{4\Lambda (\tau )},\ \lbrack \ _{q}N(\tau
)]^{2} = h_{4}(\tau ,x^{k},t)=-\frac{4[\Phi ^{2}(\tau ,x^{i},t)]^{\ast }}{%
|\int dt\ _{v}\widehat{\Im }(\tau ,x^{i},t)[\Phi ^{2}(\tau ,x^{i},t)]^{\ast
}|\ }
\end{equation*}%
for $h_{4}^{[0]}=0.$ The N-adapted differential $\delta ^{4}u=dx^{1}dx^{2}%
\mathbf{e}^{3}\mathbf{e}^{4}=dx^{1}dx^{2}[dy^{3}+n_{i}(\tau
)dx^{i}][dt+w_{i}(\tau )dx^{i}]$ is for N-connection coefficients with fixed
integration functions $_{1}n_{k}(\tau )=0$ and $_{2}n_{k}(\tau )=0$, when
\begin{equation*}
N_{i}^{a}=[n_{i}(\tau )=0,w_{i}(\tau )=\frac{\partial _{i}\left( \int dt\
_{v}\widehat{\Im }(\tau ,x^{i},t)[\Phi ^{2}(\tau ,x^{i},t)]^{\ast }\right) }{%
\ _{v}\widehat{\Im }(\tau ,x^{i},t)[\Phi ^{2}(\tau ,x^{i},t)]^{\ast }}].
\end{equation*}

The statistical thermodynamic values can be computed using the thermodynamic
generating function (\ref{genfcanv}) corresponding to $\widehat{\mathcal{W}}$
(\ref{wn}) (for simplicity, with fixed normalization)
\begin{equation}
\widehat{\mathcal{Z}}[\mathbf{g}(\tau )]=\frac{1}{4\pi ^{2}}\int \tau ^{-2}d%
\mathcal{V}(\tau ).  \label{genfn}
\end{equation}%
The effective integration volume functional $d\mathcal{V}(\tau )=d\mathcal{V(%
}\psi (\tau ,x^{k}),\Phi (\tau ,x^{i},t),\ _{v}\widehat{\Im }(\tau
,x^{i},t),\Lambda (\tau ))$ is determined by data $(\psi (\tau ),\Phi (\tau
),\ _{v}\widehat{\Im }(\tau ),\Lambda (\tau ))$ and computed%
\begin{equation}
d\mathcal{V}(\tau )=e^{\ \psi (\tau )}\left\vert \Phi (\tau )\right\vert
\sqrt{\frac{\left\vert [\Phi ^{2}(\tau )]^{\ast }\right\vert }{|\Lambda
(\tau )\int dt\ _{v}\widehat{\Im }(\tau )[\Phi ^{2}(\tau )]^{\ast }|\ }}%
dx^{1}dx^{2}dy^{3}\left[ dt+\frac{\partial _{i}\left( \int dy^{3}\ _{v}%
\widehat{\Im }(\tau )[\Phi ^{2}(\tau )]^{\ast }\right) }{\ _{v}\widehat{\Im }%
(\tau )\ [\Phi ^{2}(\tau )]^{\ast }}dx^{i}\right] .  \label{eiv}
\end{equation}%
These formulas allow us to compute thermodynamic values for cosmological
solitonic geometric flows,%
\begin{equation}
\widehat{\mathcal{E}}\ (\tau )=-\frac{1}{4\pi ^{2}}\int \left( \left[ \
_{h}\Lambda (\tau )+\Lambda (\tau )\right] -\frac{2}{\tau }\right) d\mathcal{%
V}(\tau ),\ \widehat{\mathcal{S}}(\tau )\ =-\frac{1}{4\pi ^{2}\tau ^{2}}\int
\left( \tau \left[ \ _{h}\Lambda (\tau )+\Lambda (\tau )\right] -2\right) d%
\mathcal{V}(\tau ).  \label{thvcann}
\end{equation}

\subsubsection{Thermodynamic values for cosmological solitonic generating
functions and sources}

We can consider Perelman and/or Carath\'{e}odory thermodynamic values are
computed for a 3+1 spitting (\ref{decomp31}) determined by a cosmological
solitonic d-metric (\ref{cosm312nqle}) (for LC-configurations, we can
consider (\ref{lccond})), when
\begin{eqnarray*}
q_{1} &=&q_{2}[\ _{h}i]=e^{\ \psi \lbrack \ _{h}i]},\mathbf{q}_{3}[\ _{3}i]=-%
\frac{(\Phi \lbrack \ _{3}i])^{2}}{4\Lambda (\tau )},\ [\ _{q}N(\tau
)]^{2}=h_{4}[\ _{4}i]=-\frac{4[(\Phi \lbrack \ _{3}i])^{2}]^{\ast }}{|\int
dt\ _{v}\widehat{\Im }[i][(\Phi \lbrack \ _{3}i])^{2}]^{\ast }|\ } \\
\mbox{ and }N_{i}^{a} &=&[n_{i}(\tau )=0,w_{i}(\tau )=\frac{\partial
_{i}\left( \int dt\ _{v}\widehat{\Im }[i][\Phi ^{2}[\ _{3}i]]^{\ast }\right)
}{\ _{v}\widehat{\Im }[i][\Phi ^{2}[\ _{3}i]]^{\ast }}].
\end{eqnarray*}%
For such coefficients and prescribed solitonic hierarchies for the effective
volume (\ref{eiv}), we obtain%
\begin{equation*}
d\mathcal{V}[\ _{h}i,\ _{3}i,i] = e^{\ \psi \lbrack \ _{h}i]}\left\vert \Phi
\lbrack \ _{3}i]\right\vert \sqrt{\frac{\left\vert [\Phi ^{2}[\
_{3}i]]^{\ast }\right\vert }{|\Lambda (\tau )\int dt\ _{v}\widehat{\Im }%
[i][\Phi ^{2}[\ _{3}i]]^{\ast }|\ }}dx^{1}dx^{2}dy^{3}\ [ dt+\frac{\partial
_{i}\left( \int dt\ _{v}\widehat{\Im }[i][\Phi ^{2}[\ _{3}i]]^{\ast }\right)
}{\ _{v}\widehat{\Im }[i]\ [\Phi ^{2}[\ _{3}i]]^{\ast }}dx^{i}] .
\end{equation*}%
This volume element allows us to define the thermodynamic generating
function (\ref{genfn}) and compute respective thermodynamic values (\ref%
{thvcann}) for geometric flows of such solitonic hierarchies,
\begin{eqnarray*}
\widehat{\mathcal{Z}}[\ _{h}i,\ _{3}i,i]&=&\frac{1}{4\pi ^{2}}\int \tau
^{-2}d\mathcal{V}[\ _{h}i,\ _{3}i,i] \mbox{ and } \\
\widehat{\mathcal{E}}\ [\ _{h}i,\ _{3}i,i] &=&-\frac{1}{4\pi ^{2}}\int
\left( \left[ \ _{h}\Lambda (\tau )+\Lambda (\tau )\right] -\frac{2}{\tau }%
\right) d\mathcal{V}[\ _{h}i,\ _{3}i,i], \\
\widehat{\mathcal{S}}[\ _{h}i,\ _{3}i,i]\ &=& -\frac{1}{4\pi ^{2}}\int
\left( \tau \left[ \ _{h}\Lambda (\tau )+\Lambda (\tau )\right] -2\right)
\tau ^{-2}d\mathcal{V}[\ _{h}i,\ _{3}i,i].
\end{eqnarray*}

\subsubsection{Small parametric cosmological solitonic and geometric flow
thermodynamics}

The d-metrics for such parametric solutions are described by quadratic
elements (\ref{smalparstat}) \ and generating functions $\Phi \lbrack \
_{3}\iota ,\mathring{h}_{3}]\simeq 2\sqrt{|\Lambda (\tau )\mathring{h}_{3}|}%
(1-\frac{\varepsilon }{2}\upsilon \lbrack \ _{3}\iota ])$ (\ref{infgenerf})
and a primary cosmological metric. The $\varepsilon $--decomposition for the
respective effective volume form is{\small
\begin{equation*}
d\mathcal{V}[\ _{h}i,\ \varepsilon \upsilon \lbrack \ _{3}\iota ],i,%
\mathring{h}_{3}] = 2e^{\ \psi \lbrack \ _{h}i]}\left\vert (1-\frac{%
\varepsilon }{2}\upsilon \lbrack \ _{3}\iota ])\right\vert \sqrt{\frac{%
\left\vert \ \mathring{h}_{3}|\ \upsilon \lbrack \ _{3}\iota ]|^{\ast
}\right\vert }{|\int dt\ _{v}\widehat{\Im }[i]|\ \upsilon \lbrack \
_{3}\iota ]|^{\ast }|\ }}dx^{1}dx^{2}dy^{3} [dt+\frac{\partial _{i}\left(
\int dt\ _{v}\widehat{\Im }[i]|\ \upsilon \lbrack \ _{3}\iota ]|^{\ast
}\right) }{\ _{v}\widehat{\Im }[i]\ |\ \upsilon \lbrack \ _{3}\iota ]|^{\ast
}}dx^{i}] ,
\end{equation*}%
} which allows to compute corresponding thermodynamic generating function (%
\ref{genfn}) and canonical energy and entropy (\ref{thvcann}) for geometric
flow cosmological solitonic flow parametric $\varepsilon$--deformations.

Finally, we emphasize that it is not possible to define and compute the
Bekenstein--Hawking entropy for locally anisotropic cosmological solutions
constructed in this section.

\section{Conclusions, Discussion, and Perspectives}

\label{sec5} The axiomatic side of thermodynamics due to  Constantin Carath%
\'{e}odory \cite{carath09,carath25} and, further, the axiomatic treatment of
physics were of constant and deep interest both to mathematicians and
physicists (including D. Hilbert, W. Pauli, M. Born, and, in a critical
sense, M. Planck)  \cite{giles64,tisza70,redlich70,bunge73}. Relativistic
generalizations of Grigory Perelman geometric flow thermodynamics \cite%
{perelman1} allows us to include in the scheme and find geometric
connections to mathematical physics, (modified) gravity theories, cosmology
and (quantum) information theory \cite%
{vacaru08,vacaru13,vacaru16,vacaru17,vacaru19a,vacaru19b}.

In this article, we developed Carath\'{e}odory's axiomatic approach to
foundations of thermodynamics and statistical physics (considering Pfaff
forms but also certain work on measure theory) and demonstrated also that his
methods are useful for research and applications in modern
gravity and cosmology theories. Although C. Carath\'{e}odory and G. Perelman
geometric thermodynamic construction played a strategic role in finding
solutions of most important and difficult problems in geometry and physics,
their contributions have not yet properly appreciated by many physicists and
mathematicians. This is probably due to the multi- and inter-disciplinary
character of their works when further applications request both a  "deep
physical intuition" and "advanced mathematical education and very
sophisticate geometric methods". In our works, we try to establish a bridge
between different communities of researchers.

Let us speculate on certain further perspectives on elaborating unified geometric methods to
thermodynamics of geometric flows, gravity \& cosmology, quantum information \
etc.:

\begin{enumerate}
\item Carath\'{e}odory's works on measure theory \cite{carath27,carath86},
see further developments in \cite{royden88} and (on symbolic shifts and
ergodic theory) \cite{confeld82,bedford91}, seem to be useful in modeling
information sources and processing \cite{khinchin57,kakihara99} and (recent applications)  quantum information theory \cite{vacaru19,vacaru19a,vacaru19b}.

\item The extended spectral decompositions are applicable for various
conservative systems, complex systems and their macroscopic descriptions,
with new possibilities for probabilistic prediction and control \cite%
{ant83,ant98,misra83,ant00,vacaru00,vacaru10,vacaru12}, in the theory of
locally anisotropic kinetic processes and diffusion \cite%
{vacaru00,vacaru10,vacaru12}, see also applications in noncommutative
geometric flow theory and physics \cite{vacaru08}.

\item P. Finsler  elaborated his geometry as a postgraduate of C. Carath\'{e}%
odory. Such  Finsler-Lagrange-Hamilton theories and their modifications of
Einstein gravity, geometric flows theories, cosmology and astrophysics have
been recently axiomatized for nonholonomic Lorentz manifolds and (co)
tangent bundles, see recent reviews in \cite{vacaru18axiom,vacarububax}. \
This paper should be considered as a cosmological partner of the works \cite%
{bub19,bub20} on Finsler and other types modified black hole configurations and their generalized Perelman thermodynamics.
\end{enumerate}

\vskip3pt

\textbf{Acknowledgments:} This research develops former programs partially
supported by IDEI, PN-II-ID-PCE-2011-3-0256, CERN and DAAD and extended to
collaborations at California State University at Fresno, the USA, and Yu.
Fedkovych Chernivtsi National University, Ukraine. Author S. Vacaru is
grateful to Prof.  P. Stavrinos for his former support and collaboration. He
thanks Prof. I. Antoniou for providing very important references on
Carath\'{e}odory research in mathematics and physics.

\appendix

\setcounter{equation}{0} \renewcommand{\theequation}
{A.\arabic{equation}} \setcounter{subsection}{0}
\renewcommand{\thesubsection}
{A.\arabic{subsection}}

\section{ Pfaffian differential equations}

\label{appendixa} Let us provide a brief introduction into the theory of
Pfaff forms and thermodynamics, see details and references in \cite%
{pog00,ant02,bel02}. A Pfaff differential form is $\ \delta \phi
=\sum\nolimits_{I}X_{I}dz^{I},$ where $I$ runs integer values (for
simplicity, we consider $I=1,2$) and $\delta f$ is differential 1-form but
may be not a differential of a real valued function $\phi (z^{I})$ of real
variables $z^{I},$ where $\partial _{I}:=\partial /\partial z^{I}$. An
equation
\begin{equation}
\delta \phi =0  \label{pfaff1}
\end{equation}%
is called a non-exact Pfaff equation. If $\delta \phi =d\phi =(\partial
_{I}\phi )dz^{I}$ is an exact differential of a function $\phi (z^{I}),$
i.e. we have an exact Pfaff equation, it is possible to integrate (\ref%
{pfaff1}) along a path $C$ connecting two points $z_{[1]}^{I}$ and $%
z_{[2]}^{I}$ (when $\phi $ is path--independent) and express the solution in
the form
\begin{equation*}
\phi =\int\nolimits_{C}d\phi =\phi (z_{[2]}^{I})-\phi (z_{[1]}^{I})=const.
\end{equation*}%
The H. A. Schwarz criterium is the necessary and sufficient condition to
detect a total differential equation%
\begin{equation}
\partial _{I}X_{J}=\partial _{J}X_{I},\mbox{ for }I\neq J,\mbox{ i.e. }\frac{%
\partial ^{2}\phi }{\partial x^{1}\partial x^{2}}=\frac{\partial ^{2}\phi }{%
\partial x^{2}\partial x^{1}}.  \label{schw1}
\end{equation}

In many cases, a non-exact Pfaffian with $\partial _{I}X_{J}\neq \partial
_{J}X_{I}$ can be transformed into an exact one by the aid of an integating
factor $K(z^{I}),$ when the coefficients of $\sum\nolimits_{I}KX_{I}dz^{I}$
satisfy the Schwarz condition%
\begin{equation}
\partial _{I}(KX_{J})=\partial _{J}(KX_{I}),\mbox{ for }I\neq J.
\label{schw2}
\end{equation}%
In such a case, the equation
\begin{equation}
K\delta \phi =d(K\phi )=0  \label{pfaff2}
\end{equation}%
can be integrated in an explicit form which allows us to find $\phi $ for
any prescribed $K$ satisfying (\ref{schw2}).

In a more general context, if we are not able to transform (\ref{pfaff1})
into a (\ref{pfaff2}), we can additionally add to
\begin{equation*}
\delta (K\phi )=\sum\nolimits_{I}KX_{I}dz^{I}\neq d(K\phi )
\end{equation*}%
a differential of a new function $B(z^{I}),dB=(\partial _{I}B)dz^{I}$ and
search for such $K$ and $B$ when
\begin{equation*}
\partial _{I}(KX_{J}+B)=\partial _{J}(KX_{I}+B),\mbox{ for }I\neq J%
\mbox{
and }\delta (K\phi )+dB=d(K\phi +B).
\end{equation*}%
In such a case, we can integrate
\begin{equation}
d(K\phi +B)=0  \label{pfaff3}
\end{equation}%
for any suitable $K$ and $B$ and find $\phi $ in nonexplicit form from a
so-called nonholonomic (non-integrable) function $F(\phi ,z^{I})=const.$
Usually, in thermodynamics we deal with equations of type (\ref{pfaff1})
into a (\ref{pfaff2}), but on nonholonomic manifolds, equations of type (\ref%
{pfaff3}) are involved.

\setcounter{equation}{0} \renewcommand{\theequation}
{B.\arabic{equation}} \setcounter{subsection}{0}
\renewcommand{\thesubsection}
{B.\arabic{subsection}}

\section{Parameterizatons for families of cosmological d-metrics}

\label{appendixb} We consider basic notations for quadratic line elements
describing geometric flow evolutions and nonholonomic deformations of prime
metrics into target cosmological ones.

\subsection{Target d-metrics with geometric evolution of polarization
functions}

Families of target quadratic line elements can be represented in
off-diagonal form, $\mathbf{g}_{\alpha \beta }=[g_{i},h_{a},n_{i},w_{i}],$
and/or using $\eta $-polarization functions,
\begin{eqnarray}
ds^{2}(\tau ) &=&g_{i}(\tau ,x^{k})[dx^{i}]^{2}+h_{3}(\tau
,x^{k},t)[dy^{3}+n_{i}(\tau ,x^{k},t)dx^{i}]^{2}+h_{4}(\tau
,x^{k},t)[dt+w_{i}(\tau ,x^{k},t)dx^{i}]^{2}  \label{targ1} \\
&=&\eta _{i}(\tau ,x^{k},t)\mathring{g}_{i}(x^{k},t)[dx^{i}]^{2}+\eta
_{3}(\tau ,x^{k},t)\mathring{h}_{3}(x^{k},t)[dy^{3}+\eta _{i}^{3}(\tau
,x^{k},t)\mathring{N}_{i}^{3}(x^{k},t)dx^{i}]^{2}  \notag \\
&&+\eta _{4}(\tau ,x^{k},t)\mathring{h}_{4}(x^{k},t)[dt+\eta _{i}^{4}(\tau
,x^{k},t)\mathring{N}_{i}^{4}(x^{k},t)dx^{i}]^{2}  \notag \\
&=&\eta _{i}(\tau )\mathring{g}_{i}[dx^{i}]^{2}+\eta _{3}(\tau )\mathring{h}%
_{3}[dy^{3}+\eta _{k}^{3}(\tau )\mathring{N}_{k}^{3}dx^{k}]^{2}+\eta
_{4}(\tau )\mathring{h}_{4}[dt+\eta _{k}^{4}(\tau )\mathring{N}%
_{k}^{4}dx^{k}]^{2},  \label{targ3}
\end{eqnarray}%
where $\tau $ is a temperature like geometric evolution parameter and, for
simplicity, we consider that prime metrics do not depend on such a
parameter. There will be stated dependencies of type $\eta _{a}(\tau )=\eta
_{a}(\tau ,x^{k},t)$ if such not notations do not result in ambiguities. We
consider a coordinate transform to a new time like coordinate $%
y^{4}=t\rightarrow \varsigma $ when $t=t(x^{i},\varsigma ),$
\begin{equation*}
dt=\partial _{i}tdx^{i}+(\partial t/\partial \varsigma )d\varsigma
;d\varsigma =(\partial t/\partial \varsigma )^{-1}(dt-\partial _{i}tdx^{i}),%
\mbox{  i.e. }(\partial t/\partial \varsigma )d\varsigma =(dt-\partial
_{i}tdx^{i}),
\end{equation*}%
and rewrite the target d-metric using the new time variable $\varsigma $ $.$
For instance, the 4th term in (\ref{targ3}) is computed
\begin{eqnarray*}
&&\eta _{4}(\tau )\mathring{h}_{4}[dt+\eta _{k}^{4}(\tau )\mathring{N}%
_{k}^{4}dx^{k}]^{2}=\eta _{4}(\tau )\mathring{h}_{4}[\partial
_{k}tdx^{k}+(\partial t/\partial \varsigma )d\varsigma +\eta _{k}^{4}(\tau )%
\mathring{N}_{k}^{4}dx^{k}]^{2} \\
&=&\eta _{4}(\tau )\mathring{h}_{4}[(\partial _{k}t)dx^{k}+(\partial
t/\partial \varsigma )d\varsigma +\eta _{k}^{4}(\tau )\mathring{N}%
_{k}^{4}dx^{k}]^{2}=\eta _{4}(\tau )\mathring{h}_{4}[(\partial t/\partial
\varsigma )d\varsigma +(\partial _{k}t+\eta _{k}^{4}(\tau )\mathring{N}%
_{k}^{4})dx^{k}]^{2} \\
&=&\mathring{h}_{4}[\eta _{4}(\tau )(\partial t/\partial \varsigma
)d\varsigma +\eta _{4}(\tau )(\partial _{k}t+\eta _{k}^{4}(\tau )\mathring{N}%
_{k}^{4})dx^{k}]^{2}=\mathring{h}_{4}[\eta _{4}(\tau )(\partial t/\partial
\varsigma )d\varsigma +\eta _{4}(\tau )(\partial _{k}t/\mathring{N}%
_{k}^{4}+\eta _{k}^{4}(\tau ))\mathring{N}_{k}^{4}dx^{k}]^{2}
\end{eqnarray*}%
If \ $\eta _{4}\partial t/\partial \varsigma =1,$ when $\partial t/\partial
\varsigma =(\eta _{4})^{-1}$ is introduced for $dt=\partial
_{i}tdx^{i}+(\partial t/\partial \varsigma )d\varsigma ,$ we obtain
\begin{equation*}
dt=(\partial _{i}t)dx^{i}+(\eta _{4})^{-1}d\varsigma \mbox{ for }\check{\eta}%
_{k}^{4}=\eta _{4}(\partial _{k}t+\eta _{k}^{4}\mathring{N}_{k}^{4}).
\end{equation*}%
In result, \ a new time coordinate $\varsigma $ can be found from $\partial
t/\partial \varsigma =(\eta _{4})^{-1}$ which results in
\begin{equation*}
d\varsigma =\eta _{4}(x^{k},t)dt;\varsigma =\int \eta
_{4}(x^{k},t)dt+\varsigma _{0}(x^{k}).
\end{equation*}%
Such coordinates with flow parameter $\tau $ and time like $\varsigma $ are
useful for computations of geometric evolution and nonholonomic deformations
of the FLRW metrics.

\subsection{Off-diagonal and diagonal parameterizations of prime d-metrics}

Let us consider a target line quadratic element for an off-diagonal
cosmological solution written in the form (\ref{targ3}). We can introduce an
effective target locally anisotropic cosmological scaling factor $\check{a}%
^{2}(\tau ,x^{k},\varsigma ):=\eta (\tau ,x^{k},\varsigma )\mathring{a}%
^{2}(x^{i},\varsigma )$ with gravitational polarization $\eta (\tau
,x^{k},\varsigma )$ and prime cosmological scaling factor $\mathring{a}%
^{2}(\tau ,x^{i},\varsigma ),$ which allows to consider limits $\mathring{a}%
(\tau ,x^{i},\varsigma )\rightarrow \mathring{a}(\varsigma )$ with typical \
FLRW configurations. This can be performed following formulas{\small
\begin{eqnarray}
ds^{2} &=&\eta _{3}(\tau )\{\frac{\eta _{i}(\tau )}{\eta _{3}(\tau )}%
\mathring{g}_{i}[dx^{i}]^{2}+\mathring{h}_{3}[dy^{3}+\eta _{k}^{3}\mathring{N%
}_{k}^{3}dx^{k}]^{2}\}+\mathring{h}_{4}[d\tau +\check{\eta}_{k}^{4}\mathring{%
N}_{k}^{4}dx^{k}]^{2}  \label{targ4} \\
&=&\check{a}^{2}(\tau ,x^{k},\varsigma )\{\check{\eta}_{i}(\tau
,x^{k},\varsigma )\mathring{g}_{i}[dx^{i}]^{2}+\mathring{h}_{3}[dy^{3}+%
\check{\eta}_{k}^{3}(\tau ,x^{k},\varsigma )\mathring{N}_{k}^{3}dx^{k}]^{2}%
\}+\mathring{h}_{4}[d\varsigma +\check{\eta}_{k}^{4}(\tau ,x^{k},\varsigma )%
\mathring{N}_{k}^{4}dx^{k}]^{2}  \notag \\
&=&\eta (\tau ,x^{k},\varsigma )\mathring{a}^{2}(\tau ,x^{i},\varsigma )\{%
\check{\eta}_{i}(\tau ,x^{k},\varsigma )\mathring{g}_{i}[dx^{i}]^{2}+%
\mathring{h}_{3}[dy^{3}+\check{\eta}_{k}^{3}(\tau ,x^{k},\varsigma )%
\mathring{N}_{k}^{3}dx^{k}]^{2}\}+\mathring{h}_{4}[d\varsigma +\check{\eta}%
_{k}^{4}(\tau ,x^{k},\varsigma )\mathring{N}_{k}^{4}dx^{k}]^{2},  \notag
\end{eqnarray}%
}
\begin{eqnarray*}
\mbox{where }\check{a}^{2}(\tau ,x^{k},\varsigma ):= &&\eta _{3}(\tau
,x^{k},t(x^{i},\varsigma ))=\eta (\tau ,x^{k},t(x^{i},\varsigma ))\mathring{a%
}^{2}(x^{k},t(x^{i},\varsigma ))=\eta (\tau ,x^{k},\varsigma )\mathring{a}%
^{2}(x^{i},\varsigma ); \\
\check{\eta}_{i}(\tau ,x^{k},\varsigma ):= &&\frac{\eta _{i}(\tau
,x^{k},t(x^{i},\varsigma ))}{\eta (\tau ,x^{k},t(x^{i},\varsigma ))};\
\check{\eta}_{k}^{3}(\tau ,x^{k},\varsigma ):=\eta _{k}^{3}(\tau
,x^{k},t(x^{i},\varsigma )); \\
\check{\eta}_{k}^{4}(\tau ,x^{k},\varsigma ):= &&\eta _{4}\{\tau ,,\partial
_{k}t(x^{i},\varsigma )[\mathring{N}_{k}^{4}(x^{i},t(x^{i},\varsigma
))]^{-1}+\eta _{k}^{4}(\tau ,x^{i},t(x^{i},\varsigma ))\}\mathring{N}%
_{k}^{4}(x^{i},t(x^{i},\varsigma )).
\end{eqnarray*}

Considering a prime d-metric as a flat FLRW metric written in local
coordinates \newline
$\overline{u}=\{\overline{u}^{\alpha }(x^{i},y^{3},\varsigma )=(\overline{x}%
^{1}(x^{i},y^{3},\varsigma ),\overline{x}^{2}(x^{i},y^{3},\varsigma ),%
\overline{y}^{3}(x^{i},y^{3},\varsigma ),\overline{y}^{4}(x^{i},y^{3},%
\varsigma ))\},$ a d-metric (\ref{targ1}) can be written in curved
coordinate form $\mathring{a}^{2}(\overline{u}),$ with local coordinated $%
\overline{u}^{\alpha }$ using a prime cosmological scaling factor $\mathring{%
a}^{2}(\varsigma ),$
\begin{eqnarray}
d\mathring{s}^{2} &=&\mathring{a}^{2}(\overline{u})\{\mathring{g}_{i}(%
\overline{u})[d\overline{x}^{i}]^{2}+\mathring{h}_{3}(\overline{u})[d%
\overline{y}^{3}+\mathring{N}_{k}^{3}(\overline{u})d\overline{x}^{k}]^{2}\}+%
\mathring{h}_{4}(\overline{u})[d\overline{y}^{4}+\mathring{N}_{k}^{4}(%
\overline{u})d\overline{x}^{k}]^{2}\rightarrow \mathring{a}^{2}(\varsigma
)[dx^{\check{i}}]^{2}-d\varsigma ^{2},  \notag \\
\mbox{ for }\overline{u}^{\alpha } &\rightarrow &(x^{i},y^{3},\varsigma ),%
\mathring{g}_{i}\rightarrow 1,\mathring{h}_{3}\rightarrow 1,\mathring{h}%
_{4}\rightarrow -1,\mathring{N}_{k}^{a}(\overline{u})\rightarrow 0%
\mbox{ and
}\mathring{a}^{2}(\overline{u})\rightarrow \mathring{a}^{2}(\varsigma ).
\notag
\end{eqnarray}%
By definition, a quasi FLRW configuration is stated by a diagonalized
solution for a d-metric is of type (\ref{targ4}) when the integration
functions and coordinates result in $\check{\eta}_{k}^{a}(\tau
,x^{k},\varsigma )=0,$%
\begin{equation}
ds^{2}=\eta (\tau ,x^{k},\varsigma )\mathring{a}^{2}\{\check{\eta}_{i}(\tau
,x^{k},\varsigma )\mathring{g}_{i}[dx^{i}]^{2}+\mathring{h}%
_{3}[dy^{3}]^{2}\}+\mathring{h}_{4}[d\varsigma ]^{2}.  \label{qflrwm}
\end{equation}%
Small nonholonomic deformations of such d-metrics can be parameterized $%
\check{\eta}_{i}\tau )\simeq $ $1+\varepsilon \check{\chi}_{i}(\tau
,x^{k},\varsigma )$ (see below formulas relevant to (\ref{targdm4e})) by the
polarization of the target cosmological factor, $\eta (\tau ,x^{k},\varsigma
)$ can be arbitrary one and not a value of $1+\varepsilon \chi (\tau
,x^{k},\varsigma )$ with a small parameter $\varepsilon .$ We can consider a
resulting scaling factor $a^{2}(\tau ,x^{k},\varsigma )=\eta (\tau
,x^{k},\varsigma )\mathring{a}^{2}(x^{k},\varsigma ),$ with possible further
re-parametrizations or limits to $a^{2}(\tau ,\varsigma )=\eta (\tau
,\varsigma )\mathring{a}^{2}(\varsigma )$ encoding possible nonlinear
off-diagonal and parametric interactions determined by systems of nonlinear
PDEs.

\subsection{Approximations for flows of target d-metrics}

To study nonlinear properties of cosmological models is convenient to
consider different types of parameterizations and approximations for
nonholonomic deformations of a prime metric to a target d-metric (\ref{targ4}%
) being under geometric flow evolution. For our purposes, there are
important six classes of exact, or parametric, solutions which can be
generated by a respective subclass of generating functions and/or generating
sources and, for certain cases, making some diagonal approximations, or by
introducing small $\varepsilon $-parameters.

\begin{enumerate}
\item We can chose mutual re--parametrization of generating functions $(\Psi
,\Upsilon )\iff (\Phi ,\Lambda =const)$ and integrating functions when the
coefficients of a family of target d-metric $\widehat{\mathbf{g}}_{\alpha
\beta }(\tau ,\varsigma )$ depend only a time like coordinate $\varsigma ,$
when $\eta (\tau ,x^{k},\varsigma )\rightarrow $ $\widetilde{\eta }(\tau
,\varsigma )$ and $a(\tau ,x^{k},\varsigma )\rightarrow \widetilde{a}%
^{2}(\tau ,\varsigma )=$ $\widetilde{\eta }(\tau ,\varsigma )\mathring{a}%
^{2}(\varsigma ).$ Respective families of linear quadratic elements (\ref%
{targ4}) can be represented in the form
\begin{equation}
ds^{2}(\tau )=\eta (\tau ,\varsigma )\mathring{a}^{2}(\varsigma )\{\check{%
\eta}_{i}(\tau ,\varsigma )\mathring{g}_{i}[dx^{i}]^{2}+\mathring{h}%
_{3}[dy^{3}+\check{\eta}_{k}^{3}(\tau ,\varsigma )\mathring{N}%
_{k}^{3}dx^{k}]^{2}\}+\mathring{h}_{4}[d\varsigma +\check{\eta}_{k}^{4}(\tau
,\varsigma )\mathring{N}_{k}^{4}dx^{k}]^{2}.  \label{targdm4a}
\end{equation}%
With respect to coordinate bases, such families of cosmological solutions
can be generic off-diagonal and could be chosen in some forms describing
nonholonomic deformations of Bianchi cosmological models.

\item For FLRW prime configurations, we can consider families of generation
functions and integration functions which result in zero values of the
target N-connection coefficients under geometric flow evolutons and/or
consider limits $\mathring{N}_{k}^{a}\rightarrow 0.$ For such cases, we can
transform families (\ref{targdm4a}) into families of diagonal metrics
\begin{equation}
ds^{2}(\tau )=\eta (\tau ,\varsigma )\mathring{a}^{2}(\varsigma )\{\check{%
\eta}_{i}(\tau ,\varsigma )\mathring{g}_{i}[dx^{i}]^{2}+\mathring{h}%
_{3}(dy^{3})^{2}\}+\mathring{h}_{4}(d\tau )^{2}  \label{targdm4b}
\end{equation}%
modeling locally anisotropic interactions with a "memory" of nonholonomic/
off-diagonal structures.

\item Flow evolution with small parametric nonholonomic deformations of a
prime metric into families of target off-diagonal cosmological solutions (%
\ref{targ4}) can be approximated
\begin{equation*}
\check{\eta}_{i}(\tau ,x^{k},\varsigma )\simeq 1+\varepsilon _{i}\check{\chi}%
_{i}(\tau ,x^{k},\varsigma ),\eta (\tau ,x^{k},\varsigma )\simeq
1+\varepsilon _{3}\chi (\tau ,x^{k},\varsigma ),\check{\eta}_{k}^{a}(\tau
,x^{k},\varsigma )\simeq 1+\varepsilon _{k}^{a}\check{\chi}_{k}^{a}(\tau
,x^{k},\varsigma ),
\end{equation*}%
where small parameters $\varepsilon _{i},\varepsilon _{3},\varepsilon
_{k}^{a}$ satisfy conditions of type $0\leq |\varepsilon _{i}|,|\varepsilon
_{3}|,|\varepsilon _{k}^{a}|\ll 1$ and, for instance, $\chi (\tau
,x^{k},\varsigma )$ is taken as a generating function. Such approximations
restrict the class of generating functions subjected to nonlinear symmetries
and may impose certain relations between such $\varepsilon $-constants and $%
\chi $-functions. Corresponding quadratic line elements can be parameterized%
\begin{eqnarray}
ds^{2}(\tau ) &=&[1+\varepsilon _{3}\chi (\tau ,x^{k},\varsigma )]\mathring{a%
}^{2}(x^{i},\varsigma )\{[1+\varepsilon _{i}\check{\chi}_{i}(\tau
,x^{k},\varsigma )]\mathring{g}_{i}[dx^{i}]^{2}+  \label{targdm4c} \\
&&\mathring{h}_{3}[dy^{3}+(1+\varepsilon _{k}^{3}\check{\chi}_{k}^{3}(\tau
,x^{k},\varsigma ))\mathring{N}_{k}^{3}dx^{k}]^{2}\}+\mathring{h}%
_{4}[d\varsigma +(1+\varepsilon _{k}^{4}\check{\chi}_{k}^{4}(\tau
,x^{k},\varsigma ))\mathring{N}_{k}^{4}dx^{k}]^{2}.  \notag
\end{eqnarray}%
Such $\tau $-families of off-diagonal solutions define cosmological metrics
with certain small independent fluctuations, for instance, a FLRW embedded
self-consistently into a locally anisotropic background under geometric flow
evolution.

\item We can consider also families of off-diagonal cosmological solutions
with small parameters $\varepsilon _{i},\varepsilon _{3},\varepsilon
_{k}^{a} $ when the generating functions and d-metric and N-connection
coefficients do not depend on space like coordinates, which is typical for a
number of cosmological models. For such approximations, the family of
quadratic line element (\ref{targdm4c}) transforms into%
\begin{eqnarray}
ds^{2}(\tau ) &=&[1+\varepsilon _{3}\chi (\tau ,\varsigma )]\mathring{a}%
^{2}(\varsigma )\{[1+\varepsilon _{i}\check{\chi}_{i}(\tau ,\varsigma )]%
\mathring{g}_{i}[dx^{i}]^{2}+  \label{targdm4d} \\
&&\mathring{h}_{3}[dy^{3}+(1+\varepsilon _{k}^{3}\check{\chi}_{k}^{3}(\tau
,\varsigma ))\mathring{N}_{k}^{3}dx^{k}]^{2}\}+\mathring{h}_{4}[d\varsigma
+(1+\varepsilon _{k}^{4}\check{\chi}_{k}^{4}(\tau ,\varsigma ))\mathring{N}%
_{k}^{4}dx^{k}]^{2}.  \notag
\end{eqnarray}

\item There are off-diagonal deformations, for instance, of a FLRW metric
into a family of locally anisotropic cosmological solutions which can be
constructed using only one small parameter $\varepsilon =\varepsilon
_{i}=\varepsilon _{3}=\varepsilon _{k}^{a},$ and when the formulas (\ref%
{targdm4d}) transform into%
\begin{eqnarray}
ds^{2}(\tau ) &=&[1+\varepsilon \chi (\tau ,x^{k},\varsigma )]\mathring{a}%
^{2}(x^{i},\varsigma )\{[1+\varepsilon \check{\chi}_{i}(\tau
,x^{k},\varsigma )]\mathring{g}_{i}[dx^{i}]^{2}+  \label{targdm4e} \\
&&\mathring{h}_{3}[dy^{3}+(1+\varepsilon \check{\chi}_{k}^{3}(\tau
,x^{k},\varsigma ))\mathring{N}_{k}^{3}dx^{k}]^{2}\}+\mathring{h}%
_{4}[d\varsigma +(1+\varepsilon \check{\chi}_{k}^{4}(\tau ,x^{k},\varsigma ))%
\mathring{N}_{k}^{4}dx^{k}]^{2}.  \notag
\end{eqnarray}%
Such flows with $\varepsilon $-deformations can be generated by
corresponding small $\varepsilon $--deformations of flows generating
functions.

\item We can impose on families (\ref{targdm4e}) the condition that the $%
\varepsilon $--deformations depend only on evoluiton temperature like
parameter and a time like coordinate. This results in d-metrics
\begin{eqnarray*}
ds^{2}(\tau ) &=&[1+\varepsilon \chi (\tau ,\varsigma )]\mathring{a}%
^{2}(\varsigma )\{[1+\varepsilon \check{\chi}_{i}(\tau ,\varsigma )]%
\mathring{g}_{i}[dx^{i}]^{2}+  \label{targdm4f} \\
&&\mathring{h}_{3}[dy^{3}+(1+\varepsilon \check{\chi}_{k}^{3}(\tau
,\varsigma ))\mathring{N}_{k}^{3}dx^{k}]^{2}\}+\mathring{h}_{4}[d\varsigma
+(1+\varepsilon \check{\chi}_{k}^{4}(\tau ,\varsigma ))\mathring{N}%
_{k}^{4}dx^{k}]^{2}  \notag
\end{eqnarray*}%
which can be considered as some ansatz used, for instance, for describing
geometric evolution of quantum fluctuations of FLRW metrics.
\end{enumerate}

In various classes of cosmological models with families of solutions with
parametric $\varepsilon $-decompositions can be performed in a
self-consistent form by omitting quadratic and higher order terms after a
class of locally anisotropic solutions have been found for some general data
$(\eta _{\alpha },\eta _{i}^{a}).$ They are more general than approximate
solutions found, for instance, for classical and quantum fluctuations of
standard FLRW metrics and may involve flow evolution parameters of
cosmological constants and generating functions and sources. For certain
subclasses of generic off-diagonal solutions, we can consider that $%
\varepsilon _{i},\varepsilon _{a},\varepsilon _{i}^{a}\sim \varepsilon ,$
when only one small parameter is considered for all coefficients of
nonholonomic deformations.

\end{document}